\documentclass[aps,prd,reprint,twocolumn,superscriptaddress,longbibliography,nofootinbib,floatfix,showpacs]{revtex4-1}
\usepackage{epsfig}
\usepackage{amsmath,amssymb,amsfonts}
\usepackage{hyperref}
\usepackage{mathrsfs}
\usepackage{bbm}
\usepackage{slashed}
\usepackage{graphicx}
\usepackage{verbatim}

\usepackage{bm}

\usepackage[usenames]{color}

\usepackage{tikz}
\usetikzlibrary{calc,decorations.markings,decorations.pathmorphing,positioning,shapes,snakes,arrows}

\definecolor{darkgreen}{rgb}{0.2,0.6,0}

\newcommand{\be}{\begin{equation}}
\newcommand{\ee}{\end{equation}}
\newcommand{\bw}{\begin{widetext}}
\newcommand{\ew}{\end{widetext}}
\newcommand{\bi}{\begin{itemize}}
\newcommand{\ei}{\end{itemize}}
\newcommand{\bea}{\begin{eqnarray}}
\newcommand{\eea}{\end{eqnarray}}
\newcommand{\ud}{\mathrm{d}}

\newcommand{\LCm}{{\scriptscriptstyle -}} %LC supersripts
\newcommand{\LCp}{{\scriptscriptstyle +}}
\newcommand{\LCpm}{{\scriptscriptstyle \pm}}
\newcommand{\LCmp}{{\scriptscriptstyle \mp}}

\newcommand{\LCperp}{{\scriptscriptstyle \perp}}

\newcommand{\LCpara}{{\scriptscriptstyle \parallel}}

\usepackage[T1]{fontenc} \usepackage[latin1]{inputenc}

\begin{document}

\title{Loops and polarization in strong-field QED}

\author{Greger Torgrimsson}
\email{g.torgrimsson@hzdr.de}
\affiliation{Helmholtz-Zentrum Dresden-Rossendorf, Bautzner Landstra{\ss}e 400, 01328 Dresden, Germany}

\begin{abstract}

In a previous paper we showed how higher-order strong-field-QED processes in long laser pulses can be approximated by multiplying sequences of ``strong-field Mueller matrices''. We obtained expressions that are valid for arbitrary field shape and polarization. In this paper we derive practical approximations of these Mueller matrices in the locally-constant- and the locally-monochromatic-field regimes. We allow for arbitrary laser polarization as well as arbitrarily polarized initial and final particles.  
The spin and polarization can also change due to loop contributions (the mass operator for electrons and the polarization operator for photons). We derive Mueller matrices for these as well.     
	
\end{abstract}	

\maketitle

\section{Introduction}

A single particle colliding with a high-intensity laser can lead to the production of many particles in cascade processes~\cite{Bell:2008zzb,Elkina:2010up,Nerush:2010fe}. These processes are challenging to calculate. In fact, even the simplest nontrivial cascades, nonlinear trident~\cite{Dinu:2017uoj,King:2018ibi,Mackenroth:2018smh,Acosta:2019bvh,Krajewska15,Hu:2014ooa,King:2013osa,Ilderton:2010wr,Hu:2010ye,Bamber:1999zt,Ritus:1972nf,Baier,Dinu:2019wdw}, double nonlinear Compton scattering~\cite{Morozov:1975uah,Lotstedt:2009zz,Loetstedt:2009zz,Seipt:2012tn,Mackenroth:2012rb,King:2014wfa,Dinu:2018efz,Wistisen:2019pwo} and photon trident~\cite{Torgrimsson:2020mto,MorozovNarozhnyiPhTr}, are challenging to calculate exactly, even after modeling the laser as a plane wave, or even after approximating the plane wave as a constant crossed field.
One therefore needs a good approximation for studying higher-order cascades. For sufficiently high intensity, which here means\footnote{We use units with $m_e=1$ and absorb $e$ into the field $eE\to E$.} $a_0=E/\omega\gg1$, one can use the locally-constant-field (LCF) approximation, where higher-order processes are approximated by an incoherent product of the probabilities of nonlinear Compton scattering and nonlinear Breit-Wheeler pair production, and the short formation length means that the field is locally constant during these first-order processes. This approximation is implemented in particle-in-cell (PIC) codes~\cite{RidgersCode,Gonoskov:2014mda,Osiris,Smilei,King:2013zw,Gelfer:2015ora,Li:2018fcz}. However, in most codes so far, the spin and polarization of intermediate particles have been neglected, i.e. the first-order probabilities are summed/averaged over spin/polarization before multiplying them together. 
Some recent papers, though, have started taking spin/polarization into account~\cite{Li:2018fcz,Chen:2019vly,Seipt:2019ddd,Wan:2019gow,Li:2019oxr,Seipt:2020uxv,King:2013zw} (see also~\cite{CAIN}). This is often done in order to study whether high-intensity lasers can be used to generate polarized particle beams~\cite{DelSorbo:2017fod,DelSorboPlasmaJ,Seipt:2018adi,Li:2018fcz,Chen:2019vly,Seipt:2019ddd,Wan:2019gow,Li:2019oxr}. 
Overall, there seems to be a great deal of interest at the moment in studying spin and polarization effects in strong-field QED, see also~\cite{Ilderton:2020gno,Seipt:2020diz,Titov:2020taw,Wistisen:2020rsq,Kohlfurst:2018kxg,Al-Naseri:2020dxl} for more recent papers.

Moreover, even if one does not measure the spin/polarization of the initial and final particles, one still has to sum over the spin/polarization of the intermediate particles in order to obtain the full approximation of the probabilities for higher-order processes. 
For trident and double Compton scattering in a constant field and for the probability summed/averaged over the spin/polarization of initial and final particles, it was shown in~\cite{Ritus:1972nf,Baier,King:2013osa,Morozov:1975uah,King:2014wfa} how to perform the spin sums for intermediate particles.  
For example, the LCF version of the two-step part of trident is obtained by summing the incoherent product of nonlinear Compton scattering and Breit-Wheeler pair production over two orthogonal polarization vectors of the intermediate photon, rather than summing/averaging before multiplying. Note that on the probability level one cannot simply sum over an arbitrary spin/polarization basis, but at least in LCF there is a basis which does give the correct result. In~\cite{Dinu:2019pau} we showed that for $a_0\sim1$ and fields that do not have linear polarization, one in general does not have such simple sums. It is of course always true that one can sum over any basis on the amplitude level, but on the probability level this gives in general a double sum, where the spin from the amplitude does not have to be the same as the spin from its complex conjugate. In LCF (summed over all the external spins/polarizations) there is a basis where the off-diagonal terms vanish. That is also the case for $a_0\sim1$ if the field has linear polarization. In the general case, where there is no simple basis for which the off-diagonal terms vanish, we have found a way to treat these double spin sums by expressing spin/polarization in terms of Stokes vectors and spin transitions in terms of strong-field-QED Mueller matrices~\cite{Dinu:2019pau}.
Thus, in~\cite{Dinu:2018efz,Dinu:2019pau} we showed how to obtain approximations of general higher-order tree processes using the $\mathcal{O}(\alpha)$ Mueller matrices as building blocks. This generalizes the LCF approximation to fields with intermediate intensities $a_0\gtrsim1$, arbitrary field polarization and field shape, and for arbitrarily polarized initial and final particles. 

In addition to LCF, another case for which one can expect to find simple results is for a circularly polarized field with long pulse length, where one can use a locally monochromatic field (LMF) approximation~\cite{NarozhnyiLMF,Seipt:2010ya,Heinzl:2020ynb}. Since our gluing approximation is valid for long pulses, it is therefore natural to derive LMF approximations of all the Mueller matrices.

In addition to the tree processes, nonlinear Compton and Breit-Wheeler, loop diagrams can also contribute to the changes in spin and polarization~\cite{Ilderton:2020gno,Meuren:2011hv,Dinu:2013gaa}. Here we will derive Mueller matrices for these loop contributions and study their role in the gluing/incoherent-product approach.   

So, the aims of this paper are:

\bi
\item Derive LCF and LMF approximations for all components of the Mueller matrices of all $\mathcal{O}(\alpha)$ processes. 
\item Derive the full Mueller matrices for the loop contributions to $e^\LCm\to e^\LCm$ and $\gamma\to\gamma$ (at $\mathcal{O}(\alpha)$). These include both diagonal and off-diagonal terms, related to e.g. spin flip and spin rotation, respectively. 
\item Show that, despite the vanishing contribution to spin flip at $\mathcal{O}(\alpha)$, the $\mathcal{O}(\alpha)$ Mueller matrices for the loops contain all the necessary information to approximate higher orders. We show in particular how to recover the exact spin-flip probability at $\mathcal{O}(\alpha^2)$ from the product of two Mueller matrices, and the solution to the BMT equation and the Sokolov-Ternov effect from resummations of series of Mueller matrices.     
\ei

This paper is organized as follows. In Sec.~\ref{Definitions} we give definitions and summarize some results from~\cite{Dinu:2019pau}. In Sec.~\ref{LMFsection} we derive the LMF approximations of the Mueller matrices for nonlinear Compton scattering and nonlinear Breit-Wheeler pair production for a circularly polarized laser. In Sec.~\ref{LMFtridentSection} we show that this LMF approximation agrees well with the exact result for nonlinear trident. In Sec.~\ref{LCF building blocks} we derive the LCF version of these Mueller matrices. In Sec.~\ref{MassOperatorSection} we first present the general $\mathcal{O}(\alpha)$ Mueller matrix for spin change due to the electron mass operator loop. In Sec.~\ref{loop circular polarization} we consider a circularly polarized field in LMF. In Sec.~\ref{LCF loop and C} we study the loop in LCF and combine it with the contribution from Compton scattering, in Sec.~\ref{compare Sokolov-Ternov} we consider the low-$\chi$ limit and recover literature results for the Sokolov-Ternov effect, and in Sec.~\ref{LCF larger chi} we discuss what happens at larger $\chi$. In Sec.~\ref{loop low energy limit} we consider the low-energy limit and compare with the solution to the BMT equation. In Sec.~\ref{Electrons with negligible recoil} we consider electrons with negligible recoil, which allows us to neglect Compton scattering and resum the Mueller-matrix series.
In Sec.~\ref{PolarizationOperatorSection} we derive the general $\mathcal{O}(\alpha)$ Mueller matrix for polarization change due to the polarization-operator loop. We conclude in Sec.~\ref{conclusions}. There are several appendices where we collect most of the derivations.

\section{Definitions}\label{Definitions}

We use lightfront coordinates $v^\LCpm=2v_\LCmp=v^0\pm v^3$ and $v_\LCperp=\{v_1,v_2\}$. The plane wave is given by a potential in lightfront gauge $a_\LCperp(\phi)$, where $\phi=kx=\omega x^\LCp$ and $\omega$ is a frequency scale.
For a photon (that is not part of the laser) with momentum $l_\mu$, an arbitrary polarization vector $\epsilon_\mu$ is given in the lightfront gauge $\epsilon_\LCm=0$, $\epsilon_\LCp=l_\LCperp\epsilon_\LCperp/(2l_\LCm)$ by
\be\label{epsilonDefinition}
\epsilon_\LCperp=\left\{\cos\left(\frac{\rho}{2}\right),\sin\left(\frac{\rho}{2}\right)e^{i\lambda}\right\} \;,
\ee
where $\rho$ and $\lambda$ are two constants. 
The corresponding Stokes vector is
\be\label{Stokes3D}
{\bf n}=\{\cos\lambda\sin\rho,\sin\lambda\sin\rho,\cos\rho\} \;.
\ee
For electrons the Stokes vector is given by
\be
{\bf n}=\frac{1}{2}u^\dagger{\bf \Sigma}u({\bf p}=0) \;,
\ee
where ${\bf\Sigma}=i\{\gamma^2\gamma^3,\gamma^3\gamma^1,\gamma^1\gamma^2\}$,
and similarly for positrons. 
Another, equivalent definition of ${\bf n}$ is via
\be
u\bar{u}=\frac{1}{2}(\slashed{p}+1)(1+\gamma^5\slashed{\alpha}) \;,
\ee
where 
\be\label{Stokes4vec}
\alpha_\mu=\sum_{i=1}^3n_i\alpha_\mu^{(i)}
\ee 
and (cf.~\cite{Seipt:2018adi})
\be\label{alphaidefin}
\alpha_\mu^{(1,2)}=-\delta_\mu^{1,2}-\frac{p_{1,2}}{kp}k_\mu
\quad
\alpha_\mu^{(3)}=p_\mu-\frac{1}{kp}k_\mu \;.
\ee

The probability of nonlinear Compton scattering by an electron or a positron, or nonlinear Breit-Wheeler pair production, can now be expressed as (cf.~\cite{Misaki2000,Ivanov:2004fi,Ivanov:2004vh,Galynskii:2000fk,Grinchishin:1984aw,Galynskii:1992tm})
\be\label{Pnnn}
\begin{split}
\mathbb{P}=&\langle\mathbb{P}\rangle+{\bf n}_\gamma\!\cdot\!{\bf P}_\gamma+{\bf n}_1\!\cdot\!{\bf P}_1+{\bf n}_0\!\cdot\!{\bf P}_0 \\
&+{\bf n}_\gamma\!\cdot\!{\bf P}_{\gamma1}\!\cdot\!{\bf n}_1+{\bf n}_\gamma\!\cdot\!{\bf P}_{\gamma0}\!\cdot\!{\bf n}_0+{\bf n}_1\!\cdot\!{\bf P}_{10}\!\cdot\!{\bf n}_0 \\
&+{\bf P}_{\gamma10,ijk}{\bf n}_{\gamma i}{\bf n}_{1j}{\bf n}_{0k}\;,
\end{split}
\ee 
where ${\bf n}_\gamma$ is the Stokes vector for the photon, and ${\bf n}_{1,0}$ are the Stokes vectors for the fermions. Spin up and down along some direction ${\bf n}_r$ (e.g. $\{0,1,0\}$) corresponds to ${\bf n}=\pm{\bf n}_r$. With~\eqref{Pnnn} we can also study e.g. rotation from ${\bf n}_r$ to some orthogonal spin.

Similar expressions in QED without a background field can be found in~\cite{Fano,LippsTolhoekI,LippsTolhoekII,McMasterRevModPhys,QED-book}, and~\cite{Misaki2000,Ivanov:2004fi,Ivanov:2004vh} derived such representations for nonlinear Compton scattering and nonlinear Breit-Wheeler pair production. Our main focus here is how to use the $\langle\mathbb{P}\rangle$ and ${\bf P}$'s in~\eqref{Pnnn} as building blocks for higher-order processes. The vectors and matrices ${\bf P}$ are given by double $\phi$ integrals which depend on the longitudinal momenta but not on the spins and polarizations. 

In~\cite{Dinu:2019pau} we presented two equivalent ways of how to glue together a sequence of first-order building blocks, each on the form~\eqref{Pnnn}, to construct the ``N-step'' part of  higher-order processes. In the ``averaging'' approach, we write 
\be
\mathbb{P}_{\rm N-step}=\frac{2^n}{m}\langle\mathbb{P}_1\mathbb{P}_2\dots\mathbb{P}_N\rangle \;,
\ee  
where $n$ is the number of particles for which there is a sum rather than an average over spin/polarization (this includes all the intermediate particles), $m$ is an integer that prevents double counting due to identical particles in the final state, and $\mathbb{P}_i$ gives~\eqref{Pnnn} for step $i$ (i.e. emission of a photon or pair production). The bracket ``operator'' $\langle...\rangle$ is defined by (for each ${\bf n}$ separately)
\be
\langle1\rangle=1 \qquad \langle{\bf n}\rangle=0 \qquad \langle{\bf n}{\bf n}\rangle={\bf 1} \;,
\ee  
where ${\bf 1}$ is the unit matrix in 3D. The first two formulas are just what one would expect by averaging over any basis ${\bf n}=\pm{\bf n}_{\rm r}$ with arbitrary ${\bf n}_{\rm r}$. The third formula is the nontrivial one, since clearly $\frac{1}{2}\sum{\bf n}{\bf n}$ cannot be equal to ${\bf 1}$ for any basis. The reason that one can nevertheless sum over a certain basis in the LCF case or for linear polarization, is due to vanishing elements of the vectors and matrices that form products with the matrix $\frac{1}{2}\sum{\bf n}{\bf n}$, i.e. the nonzero elements in $\frac{1}{2}\sum{\bf n}{\bf n}-{\bf 1}$ would multiply zeroes and then it does not matter whether one uses ${\bf 1}$ or $\frac{1}{2}\sum{\bf n}{\bf n}$. However, for the general case we need $\langle{\bf n}{\bf n}\rangle={\bf 1}$.

In the second approach we replace the $\langle...\rangle$ operator with Mueller matrices. For this we use 4D Stokes vectors 
\be\label{Stokes4D}
{\bf N}=\{1,{\bf n}\} \;.
\ee 
The first-order probabilities can be expressed as
\be
\mathbb{P}={\bf M}_{ijk}{\bf N}_{1i}{\bf N}_{0j}{\bf N}_{\gamma k} \;,
\ee
where ${\bf M}$ is a $4\times4\times4$ matrix and $i,j,k=1,...,4$. The ``N-step'' can now be obtained by matrix multiplication. For example, if a photon is emitted at step $m$ and decays at step $n$, then the sum over its polarization is included via ${\bf M}_{i_mj_mk}^{(m)}{\bf M}_{i_nj_nk}^{(n)}$. A similar matrix approach exists for QED in the absence of a strong field~\cite{McMasterRevModPhys}. One can also compare this with the use of Mueller matrices for the propagation of light in optics. 
       
In~\cite{Dinu:2019pau} we presented general results for the Mueller matrices of nonlinear Compton scattering and Breit-Wheeler pair production, which are valid for any polarization or field shape. The field could for example have $a_0\sim1$ and elliptical polarization, or some sort of asymmetric structure. Of course, for this to give a good approximation one has to assume that the field is sufficiently long or intense. If the field is intense, i.e. if $a_0$ is sufficiently large, then it is useful to have a LCF approximation of these Mueller matrices, which we will derive in the following. If $a_0\sim1$ one can find simple expressions for a circularly polarized field, which we now turn to.

\section{The locally-monochromatic-field approximation}\label{LMFsection}

The expressions for ${\bf P}$ and ${\bf M}$ are given in~\cite{Dinu:2019pau}. In this section we will consider fields with long pulse and circular polarization, which is a case where one can expect to find simpler results.
So, we consider fields on the form
\be
{\bf a}(\phi)=a_0h\left(\frac{\phi}{\mathcal{T}}\right)\{\sin\phi,\cos\phi,0\} \;,
\ee
where $h(x)$ gives the pulse envelope; it could for example be a Gaussian pulse $h(x)=e^{-x^2}$, but we will keep it general. For large $\mathcal{T}$ one can obtain LMF approximations, as in~\cite{NarozhnyiLMF,Seipt:2010ya,Heinzl:2020ynb} for first-order processes. Large $\mathcal{T}$ is experimentally relevant, and it also means that one can approximate higher-order processes with our gluing method even for $a_0\gtrsim1$ (in contrast to the standard LCF version of the N-step part).
Since the building blocks in the gluing method are first order and since they all have similar structure, one can expect that parts of the calculation will be similar to~\cite{Heinzl:2020ynb}. However, here we calculate all terms that are needed for a general higher-order process.

In all terms we have two lightfront time integration variables, $\phi_1$ and $\phi_2$. We change variables to $\sigma=(\phi_1+\phi_2)/2$ and $\theta=\phi_2-\phi_1$ and then to $u=\sigma/\mathcal{T}$. The integrand can now be expanded to leading order in $\mathcal{T}$. The exponent of each term is (before making any approximation) expressed solely in terms of the effective mass 
\be
M^2_{ij}=1+\langle{\bf a}^2\rangle_{ij}-\langle{\bf a}\rangle_{ij}^2 \;,
\ee
where
\be
\langle F\rangle_{ij}=\frac{1}{\theta_{ij}}\int_{\phi_j}^{\phi_i}\!\ud\phi\; F(\phi) \;.
\ee
In the LMF limit this becomes
\be\label{MLMF}
\Theta=\theta M^2=\theta\left[1+a_0^2(u)\left(1-\text{sinc}^2\frac{\theta}{2}\right)\right] \;,
\ee
where $a_0(u)=a_0h(u)$.
The field enters the prefactor via
\be
{\bf w}_1={\bf\Delta}_{12} \qquad {\bf w}_1={\bf\Delta}_{21} \;,
\ee
where ${\bf\Delta}_{ij}$ is given by
\be\label{DeltaDefinition}
{\bf\Delta}_{ij}={\bf a}(\phi_i)-\langle{\bf a}\rangle_{ij} \;.
\ee
We also use
\be\label{XVdefinition}
{\bf X}=\frac{1}{2}({\bf w}_2+{\bf w}_1) \qquad {\bf V}=\frac{1}{2}{\bm \sigma}_2^{(3)}\cdot({\bf w}_2-{\bf w}_1) \;,
\ee
where ${\bm \sigma}_i^{(3)}$ are the Pauli matrices with a trivial third component added (recall ${\bf e}_3\cdot{\bf a}(\phi)=0$)
\be\label{sigmaDefinition}
{\bm\sigma}_i^{(3)}=\begin{pmatrix} {\bm\sigma}_i & 0 \\ 0&0\end{pmatrix} \;.
\ee
In the LMF case we have
\be\label{XLMF}
{\bf X}=a_0(u)\{\sin(\mathcal{T}u),\cos(\mathcal{T}u),0\}\left(\cos\frac{\theta}{2}-\text{sinc}\frac{\theta}{2}\right)
\ee
and
\be\label{VLMF}
{\bf V}=i a_0(u)\{\sin(\mathcal{T}u),\cos(\mathcal{T}u),0\}\sin\frac{\theta}{2} \;.
\ee
Note that the exponential part of the integrand has a $u$ dependence given by~\eqref{MLMF}, which is a smooth function and varies on the scale $u\sim1$. We see from~\eqref{XLMF} and~\eqref{VLMF} that some terms in the prefactor are proportional e.g. to $\sin(\mathcal{T}u)$. For large $\mathcal{T}$ these terms oscillate rapidly and can be neglected. Consequently, several elements of the ${\bf P}$ vectors/matrices in~\eqref{Pnnn} are negligible. But terms with e.g. ${\bf X}\cdot{\bf V}$ remain.

We have in mind using the following first-order results as building blocks for higher-order processes. So, for example, the electron in the following photon-emission results could have emitted other photons before or itself been produced at an earlier step in the cascade. We use $b_0=kp$ to denote the longitudinal momentum of the original particle that entered the laser. All the other longitudinal momenta are expressed as ratios, $s_i=kp_i/b_0$ for fermions and $q_i=kl_i/b_0$ for photons. 
   
For photon emission by an electron we find
\be
\{\langle\mathbb{P}\rangle^{\rm C},{\bf P}_0^{\rm C},...\}=\frac{\alpha\mathcal{T}}{4 b_0s_0^2}\int\ud u\{\langle\mathbb{R}\rangle^{\rm C},{\bf R}_0^{\rm C},...\} \;,
\ee
where $s_0$ and $s_1$ are the momentum ratios for the electron before and after emitting a photon with momentum ratio $q_1$
\be\label{aveQCcirc}
\langle\mathbb{R}\rangle^{\rm C}=\frac{\kappa}{2}\mathcal{J}_1-\mathcal{J}_0
\ee
\be
{\bf R}_0^{\rm C}=\frac{1}{2}\frac{q_1}{s_0}\left(1+\frac{s_0}{s_1}\right)\mathcal{J}_2{\bf e}_3
\ee
\be
{\bf R}_1^{\rm C}=\frac{1}{2}\frac{q_1}{s_1}\left(1+\frac{s_1}{s_0}\right)\mathcal{J}_2{\bf e}_3
\ee
\be
{\bf R}_\gamma^{\rm C}=\frac{\kappa}{2}\mathcal{J}_2{\bm \epsilon}_2
\ee
\be\label{QC10circ}
{\bf R}_{01}^{\rm C}=-\frac{q_1^2}{2s_0s_1}\mathcal{J}_0{\bf 1}_\LCpara+(\mathcal{J}_1-\mathcal{J}_0)\left({\bf 1}_\LCperp+\frac{\kappa}{2}{\bf 1}_\LCpara\right)
\ee
\be
{\bf R}_{\gamma0}^{\rm C}=\frac{q_1}{s_0}\left[\frac{1}{2}\left(1+\frac{s_0}{s_1}\right)\mathcal{J}_1-\mathcal{J}_0\right]{\bf e}_3{\bm\epsilon_2}
\ee
\be
{\bf R}_{\gamma1}^{\rm C}=\frac{q_1}{s_1}\left[\frac{1}{2}\left(1+\frac{s_1}{s_0}\right)\mathcal{J}_1-\mathcal{J}_0\right]{\bf e}_3{\bm\epsilon_2}
\ee
\be
{\bf R}_{\gamma01}^{\rm C}=\mathcal{J}_2\left({\bf1}_\LCperp+\frac{\kappa}{2}{\bf1}_\LCpara\right){\bm\epsilon}_2-\frac{q_1^2\mathcal{J}_0}{2s_0s_1}({\bm\sigma}_1^{(3)}{\bm\epsilon}_1+{\bm\sigma}_3^{(3)}{\bm\epsilon}_3)
\ee
where $\mathcal{J}_i$ denote three integrals that can be expressed in terms of sums of Bessel functions as in~\eqref{J0Bessel}, \eqref{J1Bessel} and~\eqref{J2Bessel}, $\kappa=(s_0/s_1)+(s_1/s_0)$, ${\bf e}_3=\{0,0,1\}$ and ${\bm\epsilon}_2=\{0,1,0\}$ etc., ${\bf 1}_\LCpara={\bf e}_3{\bf e}_3$ and ${\bf 1}_\LCperp={\bf 1}-{\bf 1}_\LCpara$.
%\be
%{\bf 1}_\LCperp=\begin{pmatrix} 1&0&0 \\ 0&1&0 \\ 0&0&0 \end{pmatrix}
%\qquad
%{\bf 1}_\LCpara=\begin{pmatrix} 0&0&0 \\ 0&0&0 \\ 0&0&1 \end{pmatrix} \;.
%\ee
${\bf e}_i$, ${\bf 1}_\LCperp$ and ${\bf 1}_\LCpara$ form dot products with the fermions' Stokes vectors, and ${\bm\epsilon}_i$ with the photon Stokes vector.
Photon emission by a positron is described in general (i.e. not just in LMF) by the same expressions but with the replacements ${\bf a}\to-{\bf a}$ and ${\bf n}_e\to-{\bf n}_p$.

For pair production we have similar expressions
\be
\{\langle\mathbb{P}\rangle^{\rm BW},{\bf P}_0^{\rm BW},...\}=\frac{\alpha\mathcal{T}}{4 b_0q_1^2}\int\ud u\{\langle\mathbb{R}\rangle^{\rm BW},{\bf R}_0^{\rm BW},...\} \;,
\ee
where
\be
\langle\mathbb{R}\rangle^{\rm BW}=\frac{\kappa}{2}\mathcal{J}_1+\mathcal{J}_0
\ee
\be
{\bf R}_2^{\rm BW}=\frac{1}{2}\frac{q_1}{s_2}\left(1-\frac{s_2}{s_3}\right)\mathcal{J}_2{\bf e}_3
\ee
\be
{\bf R}_3^{\rm BW}=-\frac{1}{2}\frac{q_1}{s_3}\left(1-\frac{s_3}{s_2}\right)\mathcal{J}_2{\bf e}_3
\ee
\be\label{QBWgamma}
{\bf R}_\gamma^{\rm BW}=-\frac{\kappa}{2}\mathcal{J}_2{\bm \epsilon}_2
\ee
\be
{\bf R}_{23}^{\rm BW}=-\frac{q_1^2}{2s_2s_3}\mathcal{J}_0{\bf 1}_\LCpara+(\mathcal{J}_1-\mathcal{J}_0)\left({\bf 1}_\LCperp+\frac{\kappa}{2}{\bf 1}_\LCpara\right)
\ee
\be
{\bf R}_{\gamma2}^{\rm BW}=-\frac{q_1}{s_2}\left[\frac{1}{2}\left(1-\frac{s_2}{s_3}\right)\mathcal{J}_1-\mathcal{J}_0\right]{\bf e}_3{\bm\epsilon_2}
\ee
\be
{\bf R}_{\gamma3}^{\rm BW}=-\frac{q_1}{s_3}\left[-\frac{1}{2}\left(1-\frac{s_3}{s_2}\right)\mathcal{J}_1+\mathcal{J}_0\right]{\bf e}_3{\bm\epsilon_2}
\ee
\be
{\bf R}_{\gamma23}^{\rm BW}=-\mathcal{J}_2\left({\bf1}_\LCperp+\frac{\kappa}{2}{\bf1}_\LCpara\right){\bm\epsilon}_2+\frac{q_1^2\mathcal{J}_0}{2s_2s_3}({\bm\sigma}_1^{(3)}{\bm\epsilon}_1+{\bm\sigma}_3^{(3)}{\bm\epsilon}_3)
\ee
where $\kappa=(s_2/s_3)+(s_3/s_2)$ and $s_2$ and $s_3$ are the longitudinal-momentum ratios of the electron and positron, respectively. The notation $s_{2,3}$ rather than e.g. $s_{0,1}$ is due to the comparison with trident, where $s_{0,1}$ would be used in the first, Compton step and $s_{2,3}$ for the second, pair-production step. But this is just notation and we are considering any sequence of photon emission and pair production, so at some later step we would have e.g. $s_n$ and $s_{n+1}$.

Note that if we sum over the spins of all the final-state fermions then effectively ${\bf n}\to0$ and so any multiplication of fermion matrices (${\bf 1}_\LCperp$, ${\bf 1}_\LCpara$ and ${\bm\sigma}_{i}^{(3)}$) ends with a dot product with ${\bf e}_3$, coming e.g. from ${\bf R}_1$. Since ${\bf 1}_\LCperp\!\cdot{\bf e}_3={\bm\sigma}_{i}^{(3)}\!\cdot{\bf e}_3=0$ and ${\bf 1}_\LCpara\!\cdot{\bf e}_3={\bf e}_3$ the terms with ${\bf 1}_\LCperp$ and ${\bm\sigma}_{i}^{(3)}$ drop out and for the remaining terms the matrix multiplication becomes trivial. So, we see that in this case it is not necessary to have $\langle{\bf n}{\bf n}\rangle={\bf 1}$ for intermediate fermions; it is enough to have $\langle{\bf n}{\bf n}\rangle={\bf 1}_\LCpara$. This is something that can be obtained with a single (rather than double) sum $\frac{1}{2}\sum{\bf n}{\bf n}$ by summing over ${\bf n}=\pm{\bf e}_3$. This corresponds to a basis with spin down and up along the laser propagation direction ($-\hat{\bf k}=-{\bf e}_3$). 

For the photon part, note that the only terms that involve ${\bm\epsilon}_1$ and ${\bm\epsilon}_3$ are the ones that couple all three Stokes vectors, i.e. the terms in ${\bf R}_{\gamma01}^{\rm C}$ and ${\bf R}_{\gamma23}^{\rm BW}$ with ${\bm\sigma}_i^{(3)}$, but, since we effectively have ${\bm\sigma}_i^{(3)}\to0$ in the case of unpolarized fermions, this means that ${\bm\epsilon}_1$ and ${\bm\epsilon}_3$ also drop out. So, for the intermediate photons we again do not need $\langle{\bf n}{\bf n}\rangle={\bf 1}$, but just $\langle{\bf n}{\bf n}\rangle_{ij}=\delta_{i2}\delta_{j2}$, which can be obtained with a single sum over polarization vectors with ${\bf n}=\pm{\bm\epsilon}_2$. From~\eqref{epsilonDefinition} we see that this is, as expected, a basis of circular polarization.

Thus, for the probability summed over all final-state spins, there is a basis for the spin and polarization of intermediate particles which allows one to obtain the full result using single spin/polarization sums, i.e. a basis for which the off-diagonal terms in the double spin/polarization sums vanish. However, if one is interested in the spin of one of the particles in the final state, then one needs in general the full gluing method with $\langle{\bf n}{\bf n}\rangle={\bf 1}$.

Since the above LMF approximations are exactly linear in the pulse length $\mathcal{T}$, we can see explicitly the volume scaling $\mathcal{T}^N$ of the N-step. Corrections to the N-step approximation have a subdominant scaling with respect to $\mathcal{T}$. In comparison, the dominance of the N-step in the LCF case for large $a_0$ is due to the $a_0^N$ scaling (with $\chi$ as independent). For example the two-step part of trident scales as $a_0^2$ in LCF or $\mathcal{T}^2$ in LMF.

We have performed the oscillating $\theta$ integrals in terms of sums over Bessel functions, see~\eqref{J0Bessel}, \eqref{J1Bessel} and~\eqref{J2Bessel}. This has a huge numerical advantage, because theses sums converge quickly. To obtain the spectrum we now only have the $u$ integrals left, but these are relatively easy to perform numerically since their integrands are determined by the envelope function $h(u)$, which has a simple shape (e.g. Gaussian $e^{-u^2}$). The $u$ integrals can not be performed at this stage anyway, because when gluing together the above first-order results we should include step functions to ensure lightfront-time ordering $u_1<u_2<...$, with $u_i$ corresponding to step $i$.

\subsection{Trident}\label{LMFtridentSection}

\begin{figure}
\vspace{-.5cm}
\makebox[\linewidth][c]{
\hspace{-0.5cm}
\includegraphics[width=0.5\linewidth]{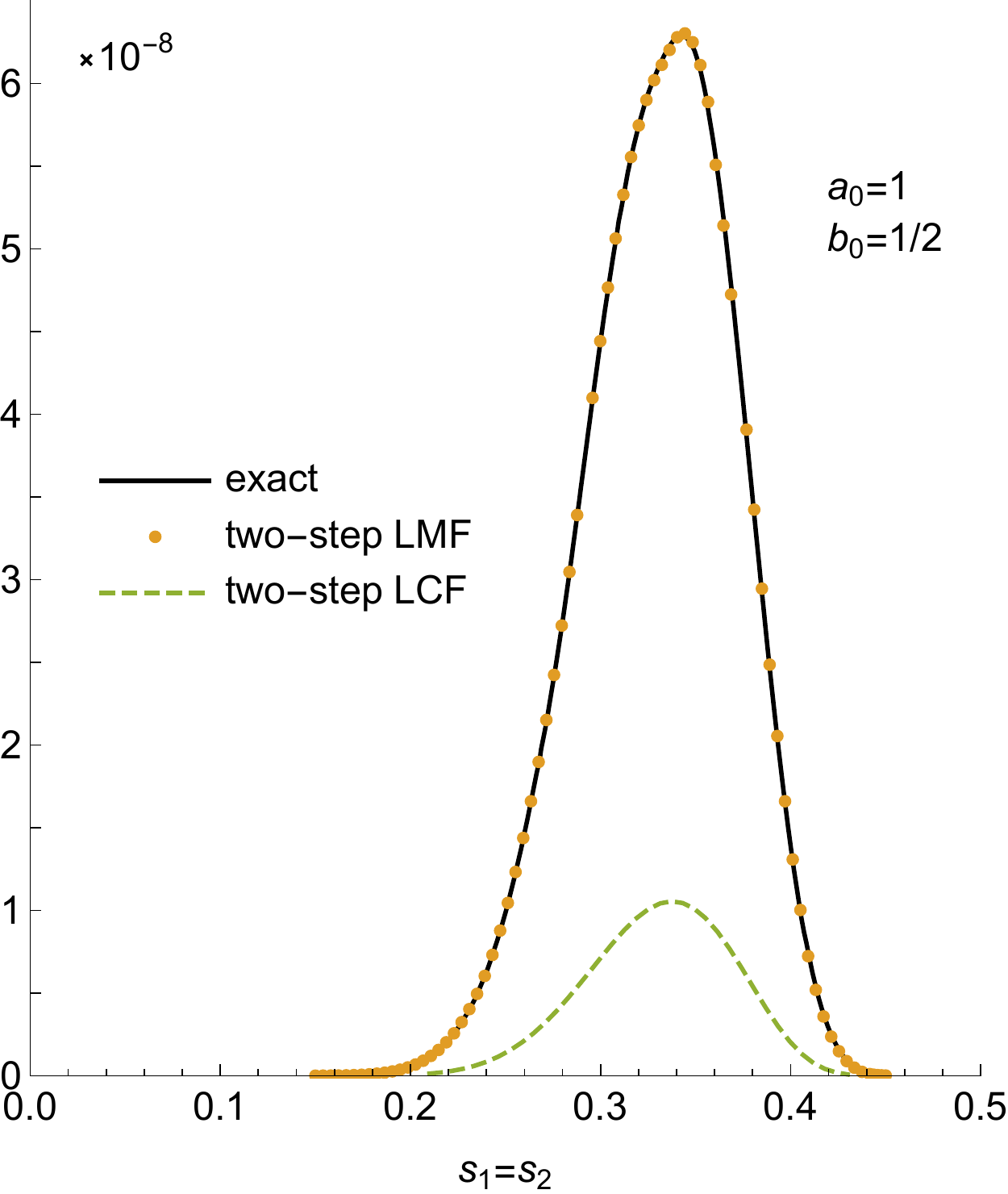}
\includegraphics[width=0.5\linewidth]{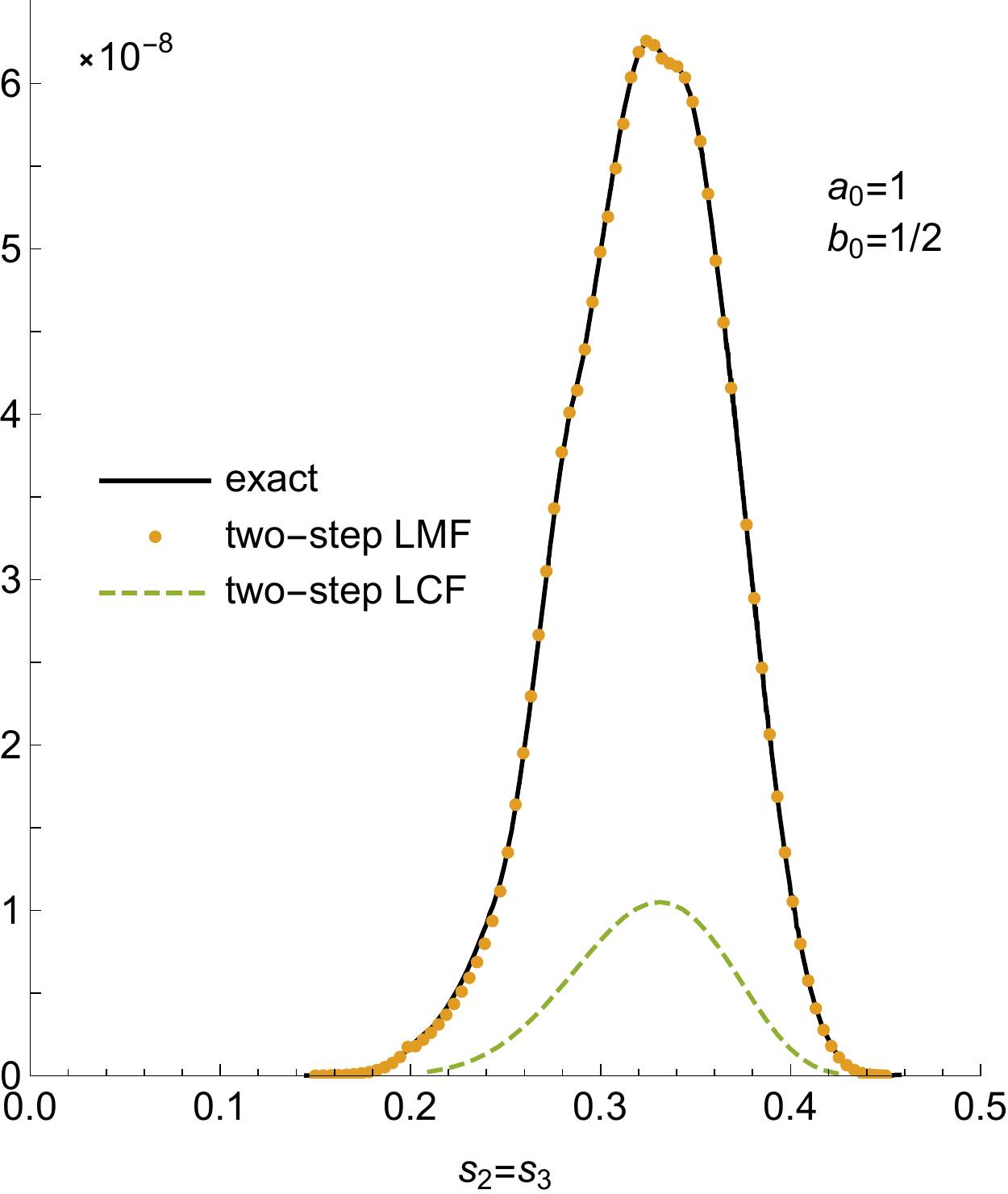}} \\
\makebox[\linewidth][c]{
\hspace{-0.5cm}
\includegraphics[width=0.5\linewidth]{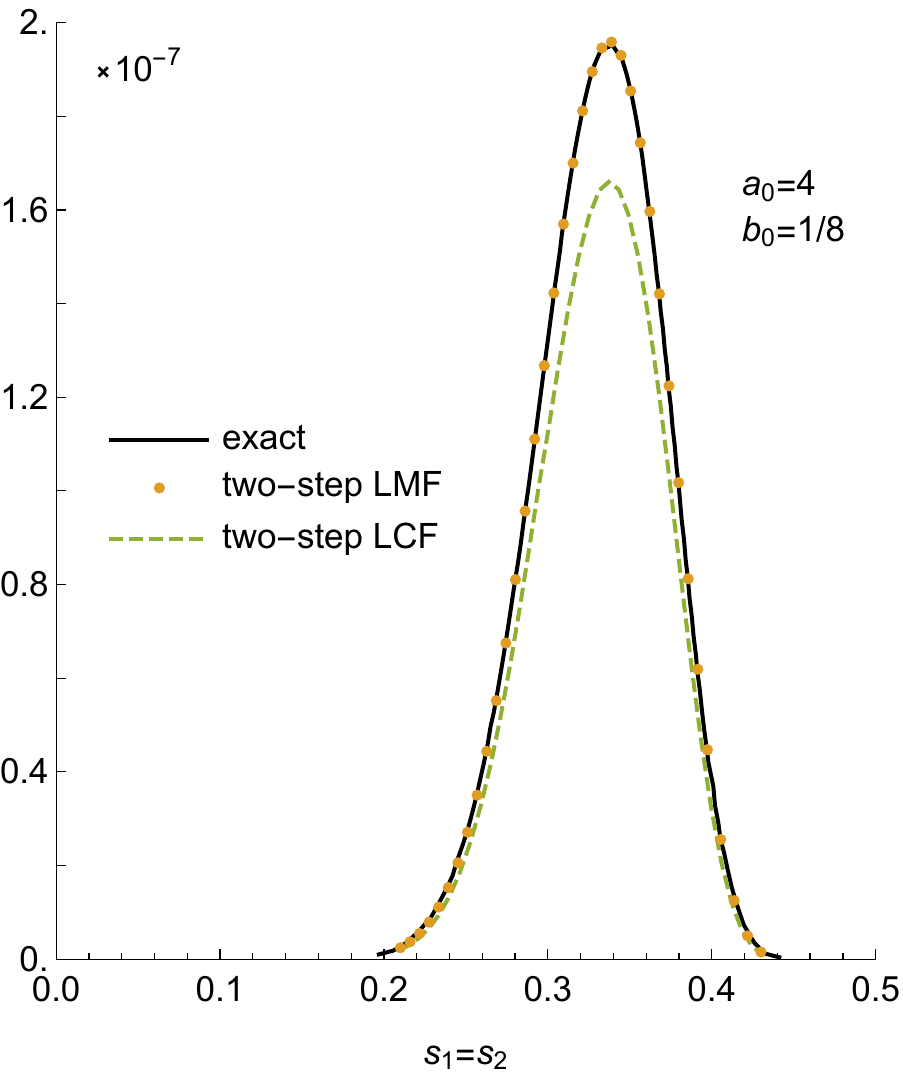}
\includegraphics[width=0.5\linewidth]{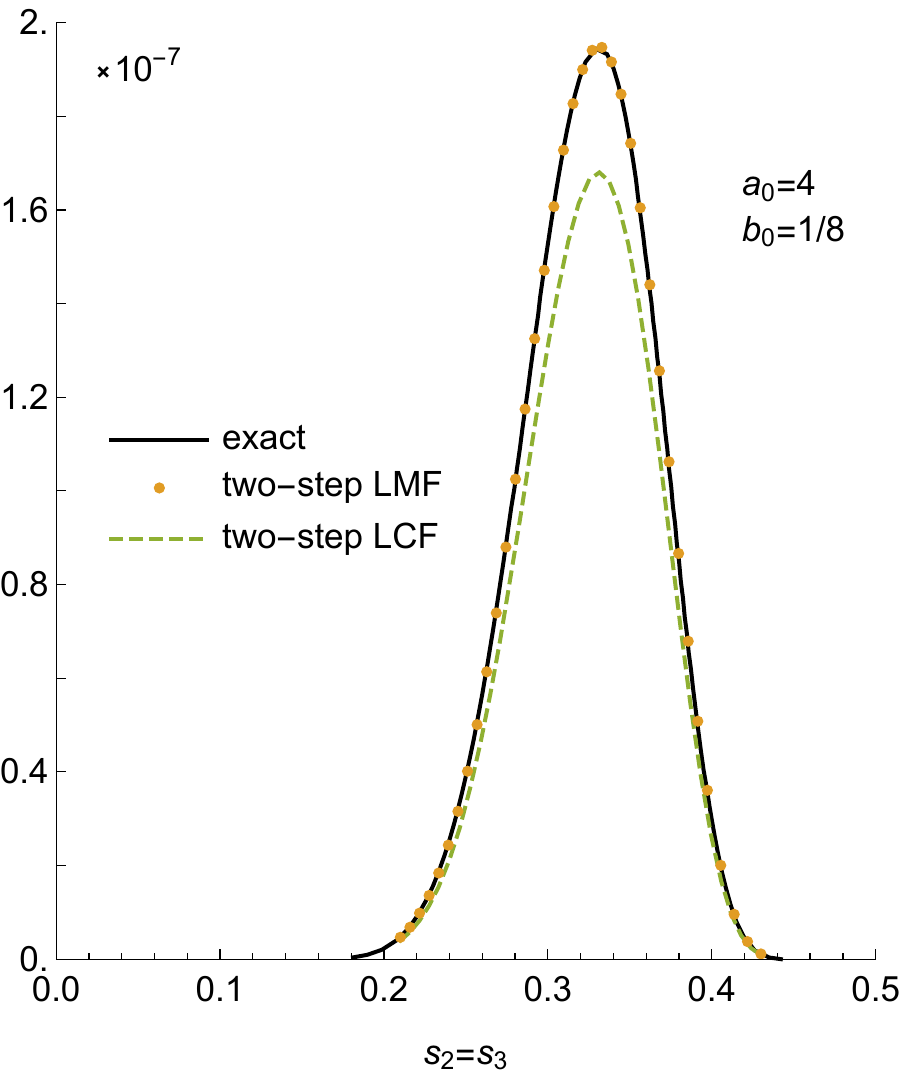}}
\caption{Comparison of the LMF approximation of the two-step with the exact results for all terms (i.e. two-step $+$ one-step), for the longitudinal-momentum spectrum in trident. The exact results are taken from~\cite{Dinu:2019pau} and the LMF results are derived here.}
\label{tridentSpectrum}
\end{figure} 

In this section we will benchmark the LMF approximation with trident as an example. 
Here $s_1$, $s_2$, $s_3=1-s_1-s_2$ and $q_1=1-s_1$ are the longitudinal momenta of the two electrons, the positron and the intermediate photon, respectively, divided by the the initial longitudinal momentum $b_0=kp$.
Using either the gluing method with the LMF results presented above, or by applying the same LMF treatment directly to the exact expressions in~\cite{Dinu:2017uoj} for the full probability, we find to leading order in LMF
\be
\begin{split}
\mathbb{P}(s)=&\frac{\alpha^2\mathcal{T}^2}{2b_0^2q_1^2}\int\ud u_2\ud u_1\theta(u_2-u_1)\bigg\{-\frac{\kappa_1\kappa_2}{4}\mathcal{J}_2^{(1)}\mathcal{J}_2^{(2)} \\
+&\left[\frac{\kappa_1}{2}\mathcal{J}_1^{(1)}-\mathcal{J}_0^{(1)}\right]\left[\frac{\kappa_2}{2}\mathcal{J}_1^{(2)}+\mathcal{J}_0^{(2)}\right]\bigg\}+(s_1\leftrightarrow s_2) \;,
\end{split}
\ee 
where $\kappa_1=(1/s_1)+s_1$, $\kappa_2=(s_2/s_3)+(s_3/s_2)$, and $\mathcal{J}^{(i)}$ is obtained from $\mathcal{J}$ in~\eqref{J0Bessel}, \eqref{J1Bessel} and~\eqref{J2Bessel} by replacing $u\to u_i$ and $r\to r_i$ with $r_1=(1/s_1)-1$ and $r_2=(1/s_2)+(1/s_3)$.

In~\cite{Dinu:2019pau} we presented sections of the spectrum with $s_1=s_2$ and $s_2=s_3$ for several different values of $a_0$ and $b_0$ for a circularly polarized field, and there we showed that our full gluing approximation agrees well with the exact result. Here we compare the LMF approximation of the gluing/Mueller-matrix approximation with the exact result. In Fig.~\ref{tridentSpectrum} we have chosen the $a_0$ and $b_0$ values from~\cite{Dinu:2019pau} that are closest to the parameter values that are planned for the LUXE experiment~\cite{Abramowicz:2019gvx}. We can see that, even after approximating the full gluing approximation with its LMF approximation, we still have a very good agreement with the exact results. We can also see that this is in a regime where the LCF approximation of the gluing approximation is not great.  

The LMF approximation looks indistinguishable from the full result in~\cite{Dinu:2019pau}, but for higher energies $b_0$ one will start to see a difference between the full version of the two-step part and its LMF approximation. However, as seen in the plots in~\cite{Dinu:2019pau}, for larger $b_0$ the one-step terms will also become non-negligible, which means that one will also start to see a difference between full two-step part and the exact probability (two-step $+$ one-step).

\section{LCF building blocks}\label{LCF building blocks}

In this section we will obtain the LCF approximation of the Mueller matrices. This can be obtained from the large $a_0$ limit of the general expressions in~\cite{Dinu:2019pau}. As usual, the results are obtained by rescaling $\theta\to\theta/a_0$ and expanding to leading order in $1/a_0$. All $\theta$ integrals can be expressed in terms of the Airy function $\text{Ai}$, its derivative $\text{Ai}'$ and the integral
\be
\text{Ai}_1(\xi)=\int_\xi^\infty\ud t\text{Ai}(t) \;.
\ee 
For nonlinear Compton we find
\be
\{\langle\mathbb{P}^{\rm C}\rangle,{\bf P}_0^{\rm C},\dots\}=\frac{\alpha}{4s_0^2}\int\frac{\ud\sigma}{b_0}\{\langle\hat{\mathbb{R}^{\rm C}}\rangle,\hat{\bf R}_0^{\rm C},\dots\} \;,
\ee  
where 
\be\label{LCFRCaveDefinition}
\langle\hat{\mathbb{R}}^{\rm C}\rangle=-\text{Ai}_1(\xi)-\kappa\frac{\text{Ai}'(\xi)}{\xi} \;,
\ee
\be\label{LCFRC0Definition}
\hat{\bf R}_0^{\rm C}=\frac{q_1}{s_0}\frac{\text{Ai}(\xi)}{\sqrt{\xi}}\hat{\bf B} \;,
\ee
\be
\hat{\bf R}_1^{\rm C}=\frac{q_1}{s_1}\frac{\text{Ai}(\xi)}{\sqrt{\xi}}\hat{\bf B} \;,
\ee
\be
\hat{\bf R}_\gamma^{\rm C}=-\frac{\text{Ai}'(\xi)}{\xi}\hat{\bf E}\cdot{\bf S}\cdot\hat{\bf E} \;,
\ee
\be\label{LCFRC10Definition}
\hat{\bf R}_{01}^{\rm C}=-\text{Ai}_1(\xi)({\bf 1}_\LCperp+[\kappa-1]{\bf 1}_\LCpara)-\frac{\text{Ai}'(\xi)}{\xi}(2{\bf 1}_\LCperp+\kappa{\bf 1}_\LCpara) \;,
\ee
\be
\begin{split}
\hat{\bf R}_{\gamma0}^{\rm C}=&-\frac{q_1}{s_1}\frac{\text{Ai}(\xi)}{\sqrt{\xi}}{\bf S}\cdot\hat{\bf B} \\
&-\frac{q_1}{s_0}\left(\text{Ai}_1(\xi)+\left[1+\frac{s_0}{s_1}\right]\frac{\text{Ai}'(\xi)}{\xi}\right){\bm\epsilon}_2\hat{\bf k} \;,
\end{split}
\ee
\be
\begin{split}
\hat{\bf R}_{\gamma1}^{\rm C}=&-\frac{q_1}{s_0}\frac{\text{Ai}(\xi)}{\sqrt{\xi}}{\bf S}\cdot\hat{\bf B} \\
&-\frac{q_1}{s_1}\left(\text{Ai}_1(\xi)+\left[1+\frac{s_1}{s_0}\right]\frac{\text{Ai}'(\xi)}{\xi}\right){\bm\epsilon}_2\hat{\bf k} \;,
\end{split}
\ee
\be
\begin{split}
\hat{\bf R}_{\gamma01}^{\rm C}=&-\frac{q_1^2}{2s_0s_1}\text{Ai}_1(\xi){\bf S}-\frac{\text{Ai}'(\xi)}{\xi}\hat{\bf E}\cdot{\bf S}\cdot\hat{\bf E}\left(\frac{\kappa}{2}{\bf1}_\LCperp+{\bf1}_\LCpara\right)  \\
&-\frac{\tilde{\kappa}}{2}\frac{\text{Ai}'(\xi)}{\xi}\hat{\bf B}\cdot{\bf S}\cdot\hat{\bf E}\; i{\bm\sigma}_2^{(3)} \\
&+q_1\frac{\text{Ai}(\xi)}{\sqrt{\xi}}{\bm\epsilon}_2\left(\frac{\hat{\bf k}\,\hat{\bf B}}{s_1}-\frac{\hat{\bf B}\,\hat{\bf k}}{s_0}\right) \;,
\end{split}
\ee
where $\kappa=(s_0/s_1)+(s_1/s_0)$, $\tilde{\kappa}=(s_0/s_1)-(s_1/s_0)$, ${\bf S}={\bm\epsilon}_1{\bm\sigma}_1^{(3)}+{\bm\epsilon}_3{\bm\sigma}_3^{(3)}$, $\xi=(r/\chi(\sigma))^{2/3}$ with $\chi(\sigma)=|{\bf a}'(\sigma)|b_0$ being the local version of $\chi=a_0b_0$, $r=(1/s_1)-(1/s_0)$, $\hat{\bf E}(\sigma)$ and $\hat{\bf B}(\sigma)$ are unit vectors parallel\footnote{$\hat{\bf E}$ and $\hat{\bf B}$ are actually anti-parallel to the electric and magnetic field, because we have absorbed the charge into the background field, i.e. $e{\bf a}\to{\bf a}$, and $e<0$. The laser travels in the $-\hat{\bf k}$ direction.} to the local electric and magnetic fields
\be\label{EBdefinition}
\hat{\bf E}(\sigma)=\frac{{\bf a}'(\sigma)}{|{\bf a}'(\sigma)|} \qquad \hat{\bf B}(\sigma)=\hat{\bf E}(\sigma)\times\hat{\bf k} \;,
\ee
and the ${\bm\epsilon}$ vectors only form dot products with themselves or with Stokes vectors for (initial or final) photons. 

In order to replace the constant vectors ${\bm\epsilon}_1$ and ${\bm\epsilon}_3$ with ones that are related to the local field polarization, we write 
\be
\hat{\bf E}(\sigma)=:\{\cos\Omega,\sin\Omega,0\} \qquad \hat{\bf B}(\sigma)=\{\sin\Omega,-\cos\Omega,0\} \;.
\ee
Then, a photon with linear polarization parallel to $\hat{\bf E}(\sigma)$ corresponds to the following Stokes vector 
\be\label{photonStokesE}
{\bm\epsilon}_{\rm E}(\sigma)=\{\sin(2\Omega),0,\cos(2\Omega)\} \;,
\ee
and $-{\bm\epsilon}_{E}$ corresponds to polarization parallel to $\hat{\bf B}$. Diagonal linear polarization lying between $\hat{\bf E}$ and $\hat{\bf B}$, i.e. $\epsilon_\LCperp=\{\cos[\Omega-\pi/4],\sin[\Omega-\pi/4]\}$, corresponds to the following Stokes vector 
\be
{\bm\epsilon}_{\rm EB}(\sigma)=\{-\cos(2\Omega),0,\sin(2\Omega)\} \;.
\ee
${\bm\epsilon}_{\rm E}(\sigma)$, ${\bm\epsilon}_{\rm EB}(\sigma)$ and ${\bm\epsilon}_2$ form a local basis for linear parallel (or orthogonal), linear diagonal and circular photon polarization.
Using
\be
{\bf S}={\bm\epsilon}_{\rm E}(\hat{\bf E}\hat{\bf E}-\hat{\bf B}\hat{\bf B})+{\bm\epsilon}_{\rm EB}(\hat{\bf E}\hat{\bf B}+\hat{\bf B}\hat{\bf E})
\ee  
we can now express also the photonic parts of the ${\bf P}$'s in terms of the local direction of the field. For example, $\hat{\bf E}\cdot{\bf S}\cdot\hat{\bf E}={\bm\epsilon}_{\rm E}$, so ${\pm}\hat{\bf R}_\gamma^{\rm C}$ corresponds to a photon emitted with polarization parallel to $\hat{\bf E}$ or $\hat{\bf B}$.  
Since we also have ${\bf 1}_\LCperp={\bf 1}-\hat{\bf k}\hat{\bf k}$ and $(i{\bm\sigma}_2^{(3)})_{ij}=\hat{\bf k}_l \varepsilon_{ijl}$, we can write all terms in a frame independent way.

For nonlinear Breit-Wheeler we find
\be
\{\langle\mathbb{P}^{\rm BW}\rangle,{\bf P}_0^{\rm BW},\dots\}=\frac{\alpha}{4q_1^2}\int\frac{\ud\sigma}{b_0}\{\langle\hat{\mathbb{R}^{\rm BW}}\rangle,\hat{\bf R}_0^{\rm BW},\dots\} \;,
\ee  
where
\be
\langle\hat{\mathbb{R}}^{\rm BW}\rangle=\text{Ai}_1(\xi)-\kappa\frac{\text{Ai}'(\xi)}{\xi} \;,
\ee
\be
\hat{\bf R}_2^{\rm BW}=\frac{q_1}{s_2}\frac{\text{Ai}(\xi)}{\sqrt{\xi}}\hat{\bf B} \;,
\ee
\be
\hat{\bf R}_3^{\rm BW}=\frac{q_1}{s_3}\frac{\text{Ai}(\xi)}{\sqrt{\xi}}\hat{\bf B} \;,
\ee
\be
\hat{\bf R}_\gamma^{\rm BW}=\frac{\text{Ai}'(\xi)}{\xi}\hat{\bf E}\cdot{\bf S}\cdot\hat{\bf E} \;,
\ee
\be
\hat{\bf R}_{23}^{\rm BW}=-\text{Ai}_1(\xi)({\bf 1}_\LCperp+[\kappa+1]{\bf 1}_\LCpara)-\frac{\text{Ai}'(\xi)}{\xi}(2{\bf 1}_\LCperp+\kappa{\bf 1}_\LCpara) \;,
\ee
\be
\begin{split}
\hat{\bf R}_{\gamma2}^{\rm BW}=&\frac{q_1}{s_3}\frac{\text{Ai}(\xi)}{\sqrt{\xi}}{\bf S}\cdot\hat{\bf B} \\
&+\frac{q_1}{s_2}\left(\text{Ai}_1(\xi)+\left[1-\frac{s_2}{s_3}\right]\frac{\text{Ai}'(\xi)}{\xi}\right){\bm\epsilon}_2\hat{\bf k} \;,
\end{split}
\ee
\be
\begin{split}
\hat{\bf R}_{\gamma3}^{\rm C}=&\frac{q_1}{s_2}\frac{\text{Ai}(\xi)}{\sqrt{\xi}}{\bf S}\cdot\hat{\bf B} \\
&-\frac{q_1}{s_3}\left(\text{Ai}_1(\xi)+\left[1-\frac{s_3}{s_2}\right]\frac{\text{Ai}'(\xi)}{\xi}\right){\bm\epsilon}_2\hat{\bf k} \;,
\end{split}
\ee
\be
\begin{split}
\hat{\bf R}_{\gamma23}^{\rm BW}=&\frac{q_1^2}{2s_2s_3}\text{Ai}_1(\xi){\bf S}+\frac{\text{Ai}'(\xi)}{\xi}\hat{\bf E}\cdot{\bf S}\cdot\hat{\bf E}\left(\frac{\kappa}{2}{\bf1}_\LCperp+{\bf1}_\LCpara\right)  \\
&+\frac{\tilde{\kappa}}{2}\frac{\text{Ai}'(\xi)}{\xi}\hat{\bf B}\cdot{\bf S}\cdot\hat{\bf E}\; i{\bm\sigma}_2^{(3)} \\
&+q_1\frac{\text{Ai}(\xi)}{\sqrt{\xi}}{\bm\epsilon}_2\left(\frac{\hat{\bf k}\,\hat{\bf B}}{s_3}-\frac{\hat{\bf B}\,\hat{\bf k}}{s_2}\right) \;,
\end{split}
\ee
where $\kappa=(s_2/s_3)+(s_3/s_2)$, $\tilde{\kappa}=(s_2/s_3)-(s_3/s_2)$, and $\xi=(r/\chi(\sigma))^{2/3}$ with $r=(1/s_2)+(1/s_3)$.

When gluing together these first-order building blocks to approximate higher-order processes, one finds terms with e.g. $\hat{\bf B}(\sigma_1)\cdot\hat{\bf B}(\sigma_2)$ which, for a rotating field, range from $1$ to $-1$ since $\sigma_1$ and $\sigma_2$ are not forced to be within the same formation length, i.e. they can be e.g. at different field maxima.

Spin and polarization of all three particles in nonlinear Compton and Breit-Wheeler have recently been studied in LCF in~\cite{Seipt:2020diz}. The spin and polarization states considered in~\cite{Seipt:2020diz} correspond to the ${\bm\epsilon}_3$ components for the photon and to the ${\bf e}_2$ components for the fermions, for a field with ${\bf a}$ polarized in the $x$ direction. We have checked that the corresponding components of our LCF expressions above agree with those in~\cite{Seipt:2020diz}. However, the full Mueller matrices contain additional nonzero elements. There are two reasons for this: 1) We allow the field to rotate. 2) We allow for arbitrary polarization of the initial and final particles. 

Consider for example an electron that emits several photons, which do not decay into pairs. If we sum over the polarization states of all these photons, then we only need $\langle\hat{\mathbb{R}}\rangle$, $\hat{\bf R}_0$, $\hat{\bf R}_1$ and $\hat{\bf R}_{01}$. If we either average and sum over the spins of the initial and final electron or if we only consider initial and final electrons with ${\bf e}_3\cdot{\bf n}=0$, then the ${\bf 1}_\LCpara$ terms in $\hat{\bf R}_{01}$ drop out and the matrix multiplications reduce from 3 to 2 dimensions. If in addition the field has linear polarization and if we either average and sum over the spins of the initial and final electron or if we only consider initial and final electrons with Stokes vector parallel to the magnetic field, ${\bf n}=\pm\hat{\bf B}$, then the matrix multiplication reduces to a one-dimensional problem. In this case it is not necessary to have $\langle{\bf n}{\bf n}\rangle={\bf 1}$ in the gluing approach; it is enough to have $\langle{\bf n}{\bf n}\rangle=\hat{\bf B}\,\hat{\bf B}$, which one can achieve by simply summing over spin states for the intermediate electrons with ${\bf n}=\pm\hat{\bf B}$. So, if we sum (average) over all the spins/polarizations and if the field has linear polarization, then it is enough to know the probability for nonlinear Compton with initial and final spin parallel and antiparallel to the magnetic field. However, for the general case where the field is rotating or if one is interested in the spin/polarization of initial and final particles, there are more relevant terms and we need to use $\langle{\bf n}{\bf n}\rangle={\bf 1}$.     

We consider again trident as an example and for simplicity we sum and average over all the external spins. Compton scattering and Breit-Wheeler pair production are glued together according to $\mathbb{P}_{\rm glue}=(2^4/2)\langle\mathbb{P}_{\rm C}\mathbb{P}_{\rm BW}\rangle+(1\leftrightarrow2)$ (cf. Eq.~(44) in~\cite{Dinu:2019pau}), which gives us
\be
\begin{split}
\frac{\alpha^2}{2q_1^2b_0^2}&\int\ud\sigma_2\ud\sigma_1\theta(\sigma_2-\sigma_1)\Big[\langle\hat{\mathbb{R}}^{\rm BW}\rangle(\sigma_2)\langle\hat{\mathbb{R}}^{\rm C}\rangle(\sigma_1) \\
&+\hat{\bf R}_\gamma^{\rm BW}(\sigma_2)\cdot\hat{\bf R}_\gamma^{\rm C}(\sigma_1)\Big]+(1\leftrightarrow2) \;.
\end{split}
\ee 
For a linearly polarized field with $\hat{\bf E}={\bf e}_1$, $\hat{\bf E}\cdot{\bf S}\cdot\hat{\bf E}={\bm\epsilon}_3$ is independent of $\sigma$, so $\hat{\bf R}_\gamma^{\rm BW}(\sigma_2)\cdot\hat{\bf R}_\gamma^{\rm C}(\sigma_1)=-(\text{Ai}'(\xi_2)/\xi_2)(\text{Ai}'(\xi_1)/\xi_1)$ only depends on $\sigma_1$ and $\sigma_2$ via $\chi(\sigma_1)$ and $\chi(\sigma_2)$. For a circularly polarized field with $\hat{\bf E}=\cos(\sigma){\bf e}_1+\sin(\sigma){\bf e}_2$ we have $\hat{\bf E}\cdot{\bf S}\cdot\hat{\bf E}=\sin(2\sigma){\bm\epsilon}_1+\cos(2\sigma){\bm\epsilon}_3$, which means $\hat{\bf R}_\gamma^{\rm BW}(\sigma_2)\cdot\hat{\bf R}_\gamma^{\rm C}(\sigma_1)=-(\text{Ai}'(\xi_2)/\xi_2)(\text{Ai}'(\xi_1)/\xi_1)\cos[2(\sigma_2-\sigma_1)]$ is now an oscillating term and will therefore tend to average out. So, although the field and therefore its polarization is locally constant, the two steps can occur at macroscopically separated $\sigma_1$ and $\sigma_2$ and therefore see a different polarization, which leads to a qualitative difference between linear and circular polarization.

Consider again trident in a linearly polarized field. We just saw that for the probability summed over all the external spins, we could replace the general gluing prescription $\langle{\bf n}{\bf n}\rangle={\bf 1}$ with a sum over intermediate photons polarized with ${\bf n}_\gamma=\pm{\bm\epsilon}_3$, which corresponds to a polarization 4-vector with $\epsilon_\LCperp=\{1,0\}$ and $\{0,1\}$, i.e. parallel and perpendicular to the field, as expected. Consider now instead an initial electron that was polarized along the laser propagation, ${\bf n}=\hat{\bf k}$. We again sum over the spin of the final-state electrons, but we want to know the difference in the probability between a positron polarized up or down along $\hat{\bf k}$. The only term that contributes to this difference is $\hat{\bf k}\cdot\hat{\bf R}_{\gamma3}^{\rm BW}\cdot\hat{\bf R}_{\gamma0}^{\rm C}\cdot\hat{\bf k}$ and the relevant polarization states of the intermediate photon are ${\bf n}_\gamma=\pm{\bm\epsilon}_2$, which correspond to left- and right-handed circular polarization.
So, for $\mathbb{P}({\bf n}_3=\hat{\bf k})-\mathbb{P}({\bf n}_3=-\hat{\bf k})$ we also do not need the general gluing prescription $\langle{\bf n}{\bf n}\rangle={\bf 1}$, but the two polarization states of the intermediate photon that we would have to sum over are ${\bf n}_\gamma=\pm{\bm\epsilon}_2$, while for $\mathbb{P}({\bf n}_3=\hat{\bf k})+\mathbb{P}({\bf n}_3=-\hat{\bf k})$ we need ${\bf n}_\gamma=\pm{\bm\epsilon}_3$. So, even if we are in a regime where one can replace $\langle{\bf n}{\bf n}\rangle$ with single spin/polarization sums, it can still be that one needs to use different bases for different quantities.
The general prescription $\langle{\bf n}{\bf n}\rangle={\bf 1}$, on the other hand, works for all cases.  

As noted in e.g.~\cite{Chen:2019vly,Seipt:2019ddd}, when trying to find set-ups to produce polarized fermion beams one is faced with the problem that the field points in different directions during its oscillations, e.g. for a linearly polarized, almost monochromatic laser the magnetic field direction $\hat{\bf B}$ flips between e.g. ${\bf e}_2$ and $-{\bf e}_2$, which means that these terms that could induce a polarization tend to average out when integrated over such a pulse. Note, though, that even if we drop all these terms we can still have nonzero matrix products: If we drop the terms proportional to $\hat{\bf B}$ (and ${\bf S}\cdot\hat{\bf B}$, which also involves the electric field direction) then $\hat{\bf R}_0^{\rm C}, \hat{\bf R}_1^{\rm C}, \hat{\bf R}_2^{\rm BW}, \hat{\bf R}_3^{\rm BW}\to{\bf 0}$ and $\hat{\bf R}_{\gamma0}^{\rm C},\hat{\bf R}_{\gamma1}^{\rm C},\hat{\bf R}_{\gamma2}^{\rm BW},\hat{\bf R}_{\gamma3}^{\rm BW}\propto{\bm\epsilon}_2\hat{\bf k}$. If we also average/sum over all the external fermion spins, or only consider fermions polarized along $\hat{\bf k}$, then any sequence of $3\times3$ matrices for the fermion spin must start and end with $\hat{\bf k}$, e.g. $\hat{\bf k}\cdot\dots\cdot\hat{\bf R}_{\gamma10}^{\rm C}\cdot\hat{\bf R}_{10}\cdot\hat{\bf k}$. Since we have already dropped terms with $\hat{\bf k}\hat{\bf B}$, which could otherwise couple the ${\bf e}_3$ with the ${\bf e}_1$ and ${\bf e}_2$ components, we see that also ${\bf 1}_\LCperp$ and ${\bm\sigma}_i^{(3)}$ drop out. So, the only $3\times3$ matrix that remains is ${\bf 1}_\LCpara=\hat{\bf k}\hat{\bf k}$. 
This means that the matrix multiplication reduces to a one-dimensional problem and one can simply replace the rule $\langle{\bf n}{\bf n}\rangle\to{\bf 1}$ for fermions with a sum over two basis vectors ${\bf n}=\pm\hat{\bf k}$. Note that this special basis is not along the magnetic field; it is along the propagation direction of the laser. This is the same spin basis as the one in the previous section for a circularly polarized laser. For the photon polarization there does not seem to be a simple basis (that works for all terms), because both ${\bm\epsilon}_2$ and $\hat{\bf E}\cdot{\bf S}\cdot\hat{\bf E}$ remain.
If no pairs are produced and if we sum over the polarization of the emitted photons, then the probability separates into two parts: $\langle\mathbb{R}\rangle\langle\mathbb{R}\rangle...\langle\mathbb{R}\rangle+{\bf n}_1\cdot{\bf R}_{10}\cdot{\bf R}_{10}...\cdot{\bf R}_{10}\cdot{\bf n}_0$, and if we average and sum over the spin of the initial and final electron then we only have $\langle\mathbb{R}\rangle\langle\mathbb{R}\rangle...\langle\mathbb{R}\rangle$ with no matrix multiplication or spin sums at all, which would therefore make the study of cascades much simpler.

\section{Mass operator}\label{MassOperatorSection}

So, far we have shown how to use the $\mathcal{O}(\alpha)$ Mueller matrices for nonlinear Compton and Breit-Wheeler as building blocks for higher-order tree-level diagrams. 
Now we will derive the $\mathcal{O}(\alpha)$ Mueller matrix for the electron mass loop (and later the photon polarization loop) and show how to use it as an additional building block for higher-order processes that include loops.
That such loops can be important for the generation of polarized electron beams in circularly polarized monochromatic lasers in the perturbative regime have been explained in~\cite{KotkinUseLoops,Kotkin:1997me}. Here we will study a general, pulsed plane wave with arbitrary polarization and in the nonlinear regime. 
The mass operator is also needed~\cite{BaierSokolovTernov} to derive the corrections to the Bargmann-Michel-Telegdi (BMT) equation~\cite{Bargmann:1959gz} for determining the time evolution of the spin of electrons in storage rings, i.e. for the Sokolov-Ternov effect~\cite{Sokolov:1963zn}. 
Spin effects due to the mass operator have also been studied in~\cite{Meuren:2011hv}.
The possibility that the spin of an electron can flip due to the loop was recently studied in~\cite{Ilderton:2020gno}, where it was shown that this effect is $\mathcal{O}(\alpha^2)$. To obtain the $\mathcal{O}(\alpha)$ Mueller matrix we have to consider general spin transitions.

\subsection{General results}

We present the derivation of the loop in Appendix-\ref{Derivation of loop}. To zeroth order we have $
\mathbb{P}^{(0)}=(1/2){\bf N}^{(1)}\cdot{\bf N}^{(0)}$. For the first order, we find for a general field
\be\label{loopP1nn}
\mathbb{P}^{\rm L}=\langle\mathbb{P}^{\rm L}\rangle+{\bf P}_0^{\rm L}\cdot{\bf n}_0+{\bf P}_1^{\rm L}\cdot{\bf n}_1+{\bf n}_1\cdot{\bf P}_{10}^{\rm L}\cdot{\bf n}_0 \;,
\ee
where
\be\label{PfromR}
\begin{split}
\{\langle\mathbb{P}^{\rm L}\rangle,{\bf P}_{0,1}^{\rm L},{\bf P}_{10}^{\rm L}\}=\frac{i\alpha}{4\pi b_0}&\int_0^1\ud s\int\frac{\ud^2\phi_{2,1}}{\theta}\exp\left\{\frac{ir\Theta}{2b_0}\right\} \\
\times&\{\langle\mathbb{R}^{\rm L}\rangle,{\bf R}_{0,1}^{\rm L},{\bf R}_{10}^{\rm L}\} \;,
\end{split}
\ee
where
\be
\langle\mathbb{R}^{\rm L}\rangle=-\frac{\kappa}{2}\left(\frac{2ib_0}{r\theta}+D_1+1\right)+1
\ee
or equivalently
\be
\langle\mathbb{R}^{\rm L}\rangle=\frac{\kappa}{4}({\bf a}(\phi_2)-{\bf a}(\phi_1))^2+1 \;,
\ee
\be
{\bf R}_0^{\rm L}={\bf R}_1^{\rm L}=-q\left({\bf 1}+\left[1+\frac{1}{s}\right]\hat{\bf k}\,{\bf X}\right)\cdot{\bf V} 
\ee
and
\be\label{RL10general}
\begin{split}
{\bf R}_{10}^{\rm L}=&\langle\mathbb{R}^{\rm L}\rangle{\bf 1} \\
-&\text{sign}(\theta)\left[\frac{q}{2}[{\bf Y}\hat{\bf k}-\hat{\bf k}{\bf Y}]-q\left[1+\frac{1}{s}\right]({\bf X}\cdot{\bf V}){\bm\sigma}_2\right] \;,
\end{split}
\ee
where $r=(1/s)-1$, $\kappa=(1/s)+s$, $q=1-s$ is the photon momentum fraction,
$D_1={\bf w}_1\cdot{\bf w}_2$, ${\bf Y}={\bf w}_2-{\bf w}_1$ and ${\bf X}$, ${\bf V}$ and ${\bm\sigma}$ are defined in~\eqref{XVdefinition} and~\eqref{sigmaDefinition}. 
We can also write this as a 4D Mueller-matrix with ${\bf N}=\{1,{\bf n}\}$,
\be
\mathbb{P}^{\rm L}=\frac{1}{2}{\bf N}_1\cdot{\bf M}^{\rm L}\cdot{\bf N}_0 \;,
\ee
where
\be\label{MLgeneral}
{\bf M}^{\rm L}=2\langle\mathbb{P}^{\rm L}\rangle{\bf 1}^{(4)}+2[{\bf e}_0{\bf P}_0^{\rm L}+{\bf P}_0^{\rm L}{\bf e}_0]+{\bf M}^{\rm L}_{\rm rot} \;,
\ee
where ${\bf 1}^{(4)}$ is a 4D unit matrix, ${\bf e}_0=\{1,{\bf 0}\}$ (e.g. ${\bf e}_0\cdot{\bf N}=1$), the difference between the 4D and 3D versions of ${\bf P}_0^{\rm L}$ is ${\bf P}_0^{\rm L}=\{0,{\bf P}_0^{\rm L}\}$ (i.e. ${\bf e}_0\cdot{\bf P}_0^{\rm L}=0$), and the $4\times4$ matrix ${\bf M}^{\rm L}_{\rm rot}$ is obtained from the $3\times3$ matrix ${\bf P}_{10}^{\rm L}-\mathbb{P}^{\rm L}{\bf 1}^{(3)}$ by adding zeroes (${\bf e}_0\cdot{\bf M}_{\rm rot}^{\rm L}={\bf M}_{\rm rot}^{\rm L}\cdot{\bf e}_0={\bf0}$). ${\bf M}_{\rm rot}^{\rm L}$ only has off-diagonal terms and, as we will show below, it leads to rotation of the Stokes vector. The reason for pulling out a factor of $1/2$ is because the sum over the spin of an intermediate electron state gives a factor of $2$, e.g. gluing together two Mueller matrices gives $(1/2){\bf M}\cdot2{\bf 1}\cdot(1/2){\bf M}=(1/2){\bf M}\cdot{\bf M}$, so any sequence of Mueller matrices will have an overall factor of $1/2$. Also, since $\mathbb{P}^{(0)}=(1/2){\bf N}^{(1)}\cdot{\bf N}^{(0)}$ the zeroth-order Mueller matrix is simply ${\bf 1}^{(4)}$. 

The first thing to note is that $\langle\mathbb{R}^{\rm L}\rangle$ and ${\bf R}_0^{\rm L}$ are exactly identical to the corresponding quantities in nonlinear Compton scattering~\cite{Dinu:2018efz,Dinu:2019pau} but with opposite overall sign\footnote{Actually, when we in this section compare with Compton scattering we only compare with the terms that remain after summing over the polarization of the emitted photon, i.e. $\langle\mathbb{P}^{\rm C}\rangle$, ${\bf P}_{0}^{\rm C}$, ${\bf P}_{1}^{\rm C}$ and ${\bf P}_{10}^{\rm C}$, and for these terms the sum over polarization just gives an overall factor of 2. To avoid having factors of 2 everywhere we absorb it into the definition of these terms. So, when we in this section write e.g. $\langle\mathbb{P}^{\rm C}\rangle$ it should be understood that this includes an overall factor of 2 compared to e.g.~\cite{Dinu:2019pau}.}
\be
\langle\mathbb{R}^{\rm L}\rangle=-\langle\mathbb{R}^{\rm C}\rangle \qquad
{\bf R}_0^{\rm L}=-{\bf R}_0^{\rm C} \;.
\ee
 This has to be because by summing over all possible final states, the probability has to be $1$, i.e. the loop has to exactly cancel the probability of single nonlinear Compton scattering. Since this should happen regardless of which initial state one starts with, this means that $\langle\mathbb{R}^{\rm L}\rangle$ and ${\bf R}_0$ should be the same as in the nonlinear Compton case. This cancellation is also what ensures that inclusive probabilities are infrared finite for unipolar fields~\cite{Dinu:2012tj,Ilderton:2012qe} and is important for expectation values describing radiation reaction~\cite{Ilderton:2013tb,Ilderton:2013dba}.

The second thing to note is that, since the zeroth order amplitude vanishes for two orthogonal spins, the loop cannot contribute to the spin-flip at $\mathcal{O}(\alpha)$~\cite{Ilderton:2020gno}, i.e. $\mathbb{P}^{\rm L}=0$ for ${\bf n}_1=-{\bf n}_0$. That this should hold for arbitrary initial spin ${\bf n}_0$ and arbitrary field is ensured by the fact that
\be
({\bf P}_0^{\rm L}-{\bf P}_1^{\rm L})\cdot{\bf n}_0=0 \qquad
\langle\mathbb{R}^{\rm L}\rangle-{\bf n}_0\cdot{\bf P}_{10}^{\rm L}\cdot{\bf n}_0=0 \;.
\ee
So, $\langle\mathbb{P}^{\rm L}\rangle$, ${\bf P}_0$,  ${\bf P}_1$ and the diagonal ${\bf 1}$ part of ${\bf P}_{10}$ could have been guessed from our results in~\cite{Dinu:2018efz,Dinu:2019pau} for Compton scattering.

Of course, this does not mean that the loop is not important, because, as expected from unitarity, for ${\bf n}_1\ne-{\bf n}_0$ it is in general on the same order of magnitude as Compton scattering. Moreover, the loop also contains terms that cannot be obtained from Compton scattering. These are the off-diagonal terms in ${\bf P}_{10}^{\rm L}$. To obtain these off-diagonal elements of the Mueller matrix we need to consider general ${\bf n}_0$ and ${\bf n}_1\ne\pm{\bf n}_0$.  

We find, though, that there is a relation between the $\theta$ integrands for the loop's off-diagonal terms and its diagonal terms, and hence with Compton scattering, given by
\be\label{iSignRelation}
({\bf R}_{10}^{\rm L}-\langle\mathbb{R}^{\rm L}\rangle{\bf 1})_{jk}=i\text{ sign}(\theta){\bf R}^{\rm L}_{0i}\varepsilon_{ijk} \;,
\ee 
where $\varepsilon_{ijk}$ is the Levi-Civita tensor with $\varepsilon_{123}=1$, ${\bf R}^{\rm L}_{0i}={\bf e}_i\cdot{\bf R}_0^{\rm L}=-{\bf e}_i\cdot{\bf R}_0^{\rm C}$ and there is a sum over $i=1,2,3$. If we rewrite the $\theta$ integral in~\eqref{PfromR} as an integral over only $\theta>0$ and define
\be
{\bf Q}:=\frac{i\alpha}{2\pi b_0}\int_0^1\ud s\int\ud\sigma\int_0^\infty\frac{\ud\theta}{\theta}\exp\left\{\frac{ir\Theta}{2b_0}\right\}
{\bf R}_0^{\rm L} \;,
\ee
then
\be\label{ReImRelation}
{\bf P}_0^{\rm L}=\text{Re}{\bf Q} \qquad
({\bf P}_{10}^{\rm L}-\langle\mathbb{P}^{\rm L}\rangle{\bf 1})_{jk}=-\text{Im}{\bf Q}_i\varepsilon_{ijk} \;.
\ee
Thus, the off-diagonal loop terms are given by the imaginary part of a quantity whose real part gives ${\bf P}_0^{\rm L}=-{\bf P}_0^{\rm C}$.

We always integrate over the transverse momenta, and we showed in~\cite{Dinu:2019pau} that it is possible to perform these integrals for each $\mathcal{O}(\alpha)$ step separately before gluing them together. In contrast, the longitudinal momentum $s$ integrals are in general intertwined. For example, if an electron emits two photons then the electron has a lower longitudinal momentum in the second step. So, for e.g. Compton scattering one cannot in general integrate the $\mathcal{O}(\alpha)$ steps separately before one glues together sequences of them.
However, for the loop we can of course always perform the $s$ integral before inserting the loop into a cascade diagram. This means, for example, that even though $\langle\mathbb{P}^{\rm L}\rangle(s)=-\langle\mathbb{P}^{\rm C}\rangle(s)$ these two terms might not cancel if they are inserted into a general cascade, because the total $s$ integrand is different for Compton scattering because the later steps depend on how much of the longitudinal momentum that was emitted, while the in- and outgoing momenta are the same in a loop step.

However, if we restrict to a single step (and integrate over $s$), and if we sum over the polarization of the emitted photon for the contribution from Compton scattering, i.e. we do not observe this photon, then the probability that the electron starts with Stokes vector ${\bf n}_0$ and end up with ${\bf n}_1$ is given by
\be\label{generalProbSum}
\begin{split}
\mathbb{P}=&\mathbb{P}^{(0)}+\mathbb{P}^{(1)}=\mathbb{P}^{(0)}+\mathbb{P}^{\rm L}+\mathbb{P}^{\rm C}=\frac{1}{2}(1+{\bf n}_1\cdot{\bf n}_0) \\
&+({\bf P}_1^{\rm L}+{\bf P}_1^{\rm C})\cdot{\bf n}_1+{\bf n}_1\cdot({\bf P}_{10}^{\rm L}+{\bf P}_{10}^{\rm C})\cdot{\bf n}_0 \;.
\end{split}
\ee
This equation is exact, i.e. the contributions from $\langle\mathbb{P}^{\rm L}\rangle$ and $\langle\mathbb{P}^{\rm C}\rangle$, and ${\bf P}_0^{\rm L}$ and ${\bf P}_0^{\rm C}$ cancel in any regime. Note that $\mathbb{P}$ only depends on the initial spin ${\bf n}_0$ via terms that also depend on the final spin ${\bf n}_1$, and $\sum_{{\bf n}_1=\pm{\bf n}_r}\mathbb{P}=1$.
In the following we will show that there is in general also a partial cancellation in the remaining terms. We will see below that in some regimes the ${\bf P}_1^{\rm L}+{\bf P}_1^{\rm C}$ term is negligible, and then the only change is due to the ${\bf P}_{10}^{\rm L}+{\bf P}_{10}^{\rm C}$ term. The off-diagonal terms of this matrix leads to a rotation of ${\bf n}$, while the diagonal terms lead to a change of the degree of polarization. However, since the probability should not become negative or larger than 1 for ${\bf n}_1=\pm{\bf n}_0$ and ${\bf n}_0^2=1$, these diagonal terms have to be negative, so that $\mathbb{P}=(1/2)(1\pm[1-\alpha|...|])$. So, before the interaction the probability is equal to 0 and 1 for ${\bf n}_1=\pm{\bf n}_0$ (with ${\bf n}_0^2=1$), but afterwards there is no direction ${\bf n}_1$ that gives $\mathbb{P}=1$, and, hence, these negative diagonal elements lead to a lower degree of polarization. Thus, if one wants to increase the degree of polarization, then the ${\bf P}_1^{\rm L}+{\bf P}_1^{\rm C}$ term should not be negligible. This can of course also be seen if we start with an unpolarized particle ${\bf n}_0\to{\bf 0}$ but want a polarized outgoing particle.

\subsection{Circular polarization}\label{loop circular polarization}

If one starts with an unpolarized electron, then the spin of the outgoing electron is determined by $\langle\mathbb{R}^{\rm L}\rangle$ and ${\bf R}_1^{\rm L}$. ${\bf R}_1^{\rm L}$ is similar to and tends to cancel parts of the corresponding quantity in nonlinear Compton~\cite{Dinu:2018efz,Dinu:2019pau} 
\be
{\bf R}_1^{\rm C}=q\left(\frac{\bf 1}{s}+\left[1+\frac{1}{s}\right]\hat{\bf k}\,{\bf X}\right)\cdot{\bf V} \;,
\ee
but we see that in general they do not cancel each other exactly. However, there is a cancellation of the leading order of the soft-photon part ($q\ll1$, $s\sim1$). Moreover, as mentioned above, for a long pulse with circular polarization only the term proportional to ${\bf X}\cdot{\bf V}$ in ${\bf R}_1$ contributes to leading order. Since this term is exactly the same as in the Compton case (but with opposite sign), this means that to leading order in the pulse length, the loop cancels the contribution from nonlinear Compton for ${\bf P}_1$. 
Note that, while the first term in ${\bf R}_1$ only gives a small contribution because it is linear in the field and therefore averages out upon performing the $\phi$ integrals, this is not the case for the ${\bf X}\cdot{\bf V}$ term. So, if one just considers the change in spin due to photon emission, then one would find a significant effect for electrons polarized along the laser propagation, ${\bf n}\propto\hat{\bf k}$. However, the resulting electron beam actually has a  much lower polarization because the loop cancels this effect to leading order. This cancellation for circularly polarized fields is expected from the monochromatic case at $\mathcal{O}(a_0^2)$ in~\cite{KotkinUseLoops,Kotkin:1997me}.  

However, although ${\bf P}_1^{\rm L}$ and ${\bf P}_1^{\rm C}$ cancel, the terms that depend on both the initial and final Stokes vectors remain, so for circular polarization~\eqref{generalProbSum} reduces to
\be\label{circProbSum}
\mathbb{P}=\frac{1}{2}(1+{\bf n}_1\cdot{\bf n}_0)+{\bf n}_1\cdot({\bf P}_{10}^{\rm L}+{\bf P}_{10}^{\rm C})\cdot{\bf n}_0 \;,
\ee
where ${\bf P}_{10}^{\rm C}$ is given by~\eqref{QC10circ}. For the ${\bf 1}$ part of the loop contribution~\eqref{RL10general} we have the same integral as in $\langle\mathbb{P}^{\rm C}\rangle$, so this part is given by minus~\eqref{aveQCcirc}. The ${\bf Y}\hat{\bf k}-\hat{\bf k}{\bf Y}$ part of~\eqref{RL10general} does not contribute to leading order. For the remaining part we have 
\be
{\bf P}_{10}^{\rm L}=-\langle\mathbb{P}^{\rm C}\rangle{\bf 1}+i{\bm\sigma_2^{(3)}}\frac{\alpha\mathcal{T}}{4\pi b_0}\int_0^1\ud s\int\ud u\; \tilde{\bf R}_{10}^{\rm L} \;,
\ee
where
\be
\begin{split}
\tilde{\bf R}_{10}^{\rm L}=&2\text{Re}ia_0^2(u)q\left(1+\frac{1}{s}\right)\int_0^\infty\frac{\ud\theta}{\theta} \\
&\times\left(\frac{\cos\theta-1}{\theta}+\frac{\sin\theta}{2}\right)\exp\left\{\frac{ir}{2b_0}\Theta\right\} \;,
\end{split}
\ee
where $\Theta$ is given by~\eqref{MLMF}. We can see that this term can be important even without actually evaluating it, because the contribution from Compton scattering ${\bf P}_{10}^{\rm C}$ in~\eqref{QC10circ} is a diagonal matrix, so if we start with e.g. ${\bf n}_0=\{1,0,0\}$ and calculate the probability that ${\bf n}_1=\{0,1,0\}$ then only the loop contributes (thanks to the ${\bm\sigma_2^{(3)}}$ term). 

\begin{figure}
\includegraphics[width=\linewidth]{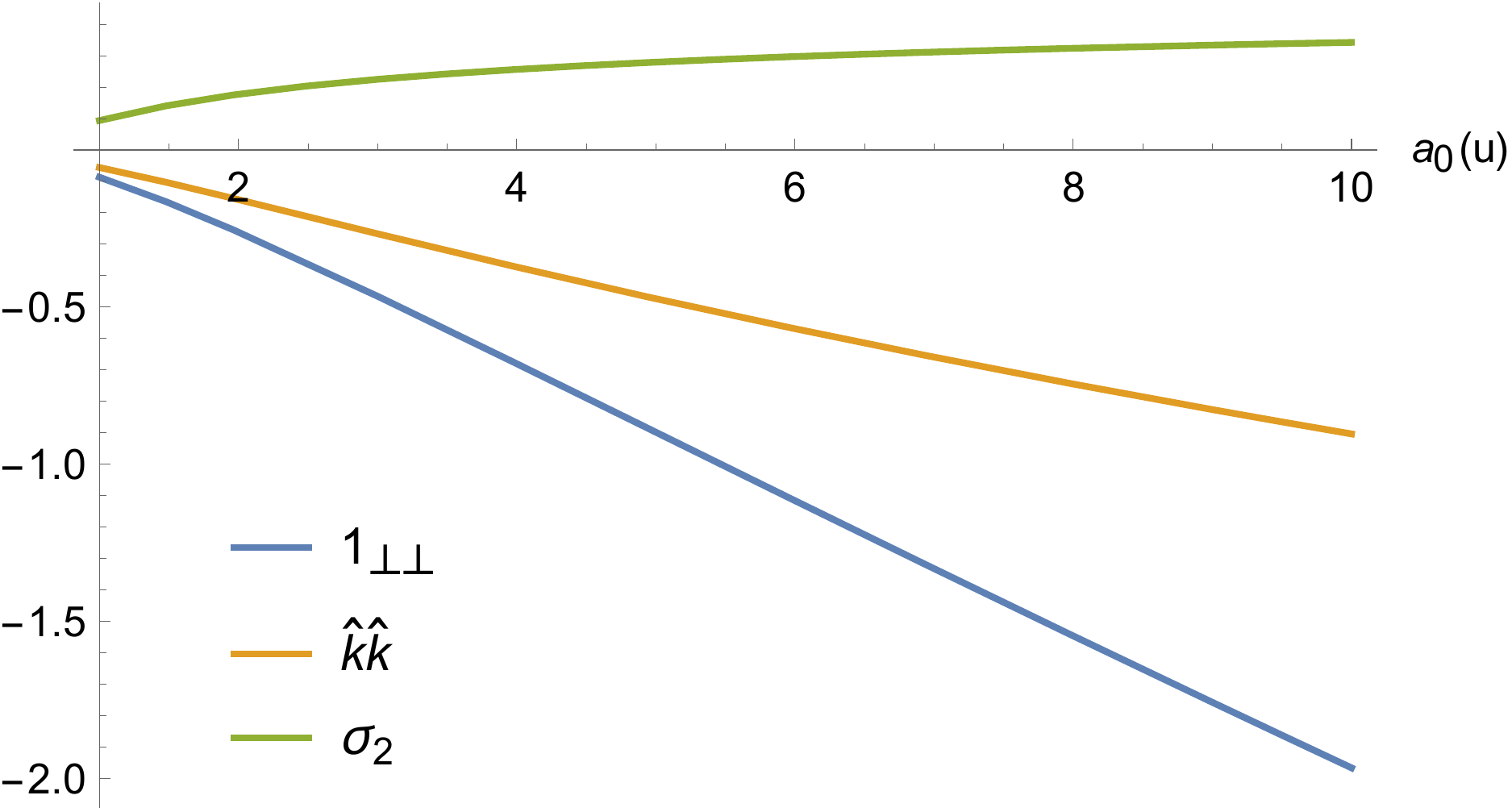}
\caption{The integrated probability for a circularly polarized field in LMF, where $\mathbb{P}$ is given by~\eqref{circProbSum} and ${\bf P}_{10}^{\rm L}+{\bf P}_{10}^{\rm C}=\frac{\alpha\mathcal{T}}{2\pi}\int\ud u(R_{\perp}{\bf 1}_\perp+R_\parallel\hat{\bf k}\hat{\bf k}+R_2i{\bm\sigma}_2^{(3)})$, where $R_i(a_0(u),b_0)$. The plot shows $R_\perp$, $R_\parallel$ and $R_2$ at $b_0=1/2$.}
\label{sintcircfig}
\end{figure}

In Fig.~\ref{sintcircfig} we plot these results for a circularly polarized field in LMF. We see that the diagonal terms in ${\bf P}_{10}^{\rm L}+{\bf P}_{10}^{\rm C}$ are negative, which they have to be as explained after~\eqref{generalProbSum}.
We are considering $\mathcal{O}(\alpha)$ here integrated over the longitudinal momentum of the emitted photon. As expected from the nonlinear Compton case~\cite{Dinu:2013hsd}, we can perform the longitudinal momentum $s$ integrals in terms of cosine and sine integrals, as explained in Appendix.~\ref{The final momentum integral}. These $s$ integrals can be performed for any field shape and polarization. For a circularly polarized field we can further approximate the effective mass, which appears in the argument of the cosine/sine integrals, as in~\eqref{MLMF}, and then we can perform the $\theta$ integral numerically. It turns out that this way, i.e. performing the $s$ integral analytically and then the $\theta$ integral numerically, is actually more convenient than first performing the $\theta$ integral analytically in terms of sums of Bessel functions and then the $s$ integral numerically. Also the former approach works for general fields (with the full effective mass, of course), while the latter only works for circularly polarized field in LMF. However, when going beyond $\mathcal{O}(\alpha)$ one might not be able to perform the $s$ integrals, since they couple nontrivially because of the recoil due to photon emission.

\subsection{Locally constant field approximation}\label{LCF loop and C}

For large $a_0$ we can rescale $\theta\to\theta/a_0$ and expand to leading order in $1/a_0$ with $\chi=a_0b_0$ kept fixed. This gives the LCF approximation. We find
\be\label{LCFPfromRloop}
\{\langle\mathbb{P}^{\rm L}\rangle,{\bf P}_0^{\rm L},\dots\}=\frac{\alpha}{2}\int\frac{\ud\sigma}{b_0}\int_0^1\!\ud s\,\{\langle\hat{\mathbb{R}}^{\rm L}\rangle,\hat{\bf R}_0^{\rm L},\dots\} \;,
\ee
where 
%\be
%\langle\mathbb{P}^{(1)}\rangle=\frac{\alpha}{2}\int_0^1\ud s\int\frac{\ud\sigma}{b_0}\left(\text{Ai}_1(\xi)+\kappa\frac{\text{Ai}'(\xi)}{\xi}\right) \;,
%\ee 
\be\label{LCFRLoopaveDefinition}
\langle\hat{\mathbb{R}}^{\rm L}\rangle=\text{Ai}_1(\xi)+\kappa\frac{\text{Ai}'(\xi)}{\xi} \;,
\ee 
\be\label{LCFRLoop0Definition}
\hat{\bf R}_0^{\rm L}=-q\frac{\text{Ai}(\xi)}{\sqrt{\xi}}\hat{\bf B} \;,
\ee
\be\label{LCFRLoop10Definition}
\hat{\bf R}_{10}^{\rm L}=\langle\hat{\mathbb{R}}^{\rm L}\rangle{\bf 1}+q\frac{\text{Gi}(\xi)}{\sqrt{\xi}}(\hat{\bf E}\,\hat{\bf k}-\hat{\bf k}\,\hat{\bf E}) \;,
\ee
where $\xi=(r/\chi(\sigma))^{2/3}$ with $\chi(\sigma)=b_0|a'(\sigma)|$, $\text{Gi}(\xi)$ is the Scorer Gi function~\cite{DLMF}
and $\hat{\bf E}$ and $\hat{\bf B}$ are the local electric- and magnetic-field direction as defined in~\eqref{EBdefinition}.
%\be
%\{{\bf P}_1,{\bf P}_1^{\rm C}\}=\frac{\alpha}{2}\int_0^1\ud s\int\frac{\ud\sigma}{b_0}q\left\{1,-\frac{1}{s}\right\}\frac{\text{Ai}(\xi)}{\sqrt{\xi}}{\bf e}(\sigma) \;,
%\ee
Given the above discussion about the exact expressions, \eqref{LCFRLoopaveDefinition} and~\eqref{LCFRLoop0Definition} are of course exactly the same as~\eqref{LCFRCaveDefinition} and~\eqref{LCFRC0Definition} except for the opposite sign. 

In~\eqref{LCFRLoop10Definition}, on the other hand, we find a term with $\text{Gi}$, which does not appear in any of the expressions in Sec.~\ref{LCF building blocks} for the spin and polarization of nonlinear Compton. One might nevertheless have guessed this term from the general relation in~\eqref{ReImRelation}, because $\text{Ai}$ and $\text{Gi}$ give the real and imaginary parts of the following integral
\be
\frac{\text{Ai}(\xi)+i\text{Gi}(\xi)}{\sqrt{\xi}}=\frac{1}{\pi}\int_0^\infty\ud\tau\exp\left\{i\xi^{3/2}\left(\tau+\frac{\tau^3}{3}\right)\right\} \;.
\ee 
Note that this $\hat{\bf E}\,\hat{\bf k}-\hat{\bf k}\,\hat{\bf E}$ term is the only one that couples Stokes vectors parallel to $\hat{\bf E}$ and $\hat{\bf k}$, so if we have a linearly polarized field and an initial Stokes vector that is parallel to the laser propagation direction, ${\bf n}_0=\pm\hat{\bf k}$, then the loop is necessary for the probability that the final Stokes vector is parallel to the electric field, ${\bf n}_1=\pm\hat{\bf E}$, while Compton scattering does not contribute to this.

The $\text{Gi}$ function appears in the $\mathcal{O}(\alpha^2)$ results in~\cite{Ilderton:2020gno} for spin flip. In order to compare with those results, consider ${\bf N}_0=\{1,{\bf n}_0\}$ and ${\bf N}_f=\{1,-{\bf n}_0\}$ with ${\bf n}^2=1$. The Mueller matrix has a matrix structure as in~\eqref{MLgeneral} with ${\bf P}_0^{\rm L}\propto\hat{\bf B}$ and ${\bf M}_{\rm rot}^{\rm L}\propto\hat{\bf E}\hat{\bf k}-\hat{\bf k}\hat{\bf E}$. At $\mathcal{O}(\alpha^2)$ the probability is given by $\mathbb{P}_{\rm flip}^{{\rm L}(2)}=(1/2){\bf N}_f\cdot (T_\sigma/2){\bf M}^{\rm L}\cdot{\bf M}^{\rm L}\cdot{\bf N}_0$. We consider for simplicity a linearly polarized field, then the only terms that contribute to spin flip are the ones proportional to $({\bf e}_0\hat{\bf B}+\hat{\bf B}{\bf e}_0)^2={\bf e}_0{\bf e}_0+\hat{\bf B}\hat{\bf B}$ and $(\hat{\bf E}\hat{\bf k}-\hat{\bf k}\hat{\bf E})^2=-(\hat{\bf E}\hat{\bf E}+\hat{\bf k}\hat{\bf k})$. The lightfront time ordering is trivial and simply gives an overall factor of $(T_\sigma/2)\to1/2$. We find
\be\label{spinFlipSecondO}
\begin{split}
\mathbb{P}_{\rm flip}^{{\rm L}(2)}&={\bf N}_f\cdot\bigg[\left(\frac{\alpha}{2b_0}\int\ud\sigma\int_0^1\ud q\;q\frac{\text{Ai}(\xi)}{\sqrt{\xi}}\right)^2({\bf e}_0{\bf e}_0+\hat{\bf B}\hat{\bf B}) \\
&-\left(\frac{\alpha}{2b_0}\int\ud\sigma\int_0^1\ud q\;q\frac{\text{Gi}(\xi)}{\sqrt{\xi}}\right)^2(\hat{\bf E}\hat{\bf E}+\hat{\bf k}\hat{\bf k})\bigg]\cdot{\bf N}_0 \;.
\end{split}
\ee
From this we can see that $\mathbb{P}_{\rm flip}^{{\rm L}(2)}=0$ if ${\bf n}=\pm\hat{\bf B}$, while $\mathbb{P}_{\rm flip}^{{\rm L}(2)}\ne0$ if $\hat{\bf E}\cdot{\bf n}\ne0$ or $\hat{\bf k}\cdot{\bf n}\ne0$. $\mathbb{P}_{\rm flip}^{{\rm L}(2)}$ is independent of $\nu$ if ${\bf n}=\cos\nu\hat{\bf E}+\sin\nu\hat{\bf k}$, e.g. the loop contribution is the same for spin polarized along the electric field or the laser propagation direction. For ${\bf n}=\pm\hat{\bf k}$ we find agreement with Eq.~(20) in~\cite{Ilderton:2020gno}.
So, although there is no spin flip at $\mathcal{O}(\alpha)$, the $\mathcal{O}(\alpha)$ Mueller matrix ${\bf M}^{\rm L}$ contains the necessary information to obtain $\mathbb{P}_{\rm flip}^{{\rm L}(2)}$ from ${\bf M}^{\rm L}\cdot{\bf M}^{\rm L}$. In Appendix~\ref{glueloop} we show that this holds in general, for arbitrary field shape and polarization.    

For what follows, it is natural to combine the diagonal part of $\hat{\bf R}_{10}^{\rm L}$, $\langle\hat{\mathbb{R}}^{\rm L}\rangle{\bf 1}=-\langle\hat{\mathbb{R}}^{\rm C}\rangle{\bf 1}$, with ${\bf R}_{10}^{\rm C}$ from Compton scattering~\eqref{LCFRC10Definition} (which only has diagonal terms)
\be
\langle\hat{\mathbb{R}}^{\rm L}\rangle{\bf 1}+{\bf R}_{10}^{\rm C}=-(\kappa-2)\left(\text{Ai}_1(\xi){\bf 1}_\LCpara-\frac{\text{Ai}'(\xi)}{\xi}{\bf 1}_\LCperp\right) \;.
\ee
Since $\kappa-2=q^2/s>0$, $\text{Ai}_1>0$ and $\text{Ai}'<0$, these diagonal terms are all negative.

For an oscillating field, ${\bf P}_1$ and ${\bf P}_1^{\rm C}$ tend to average out (each term separately) because $\hat{\bf B}(\sigma)$ changes direction. It has been realized in recent literature that one can prevent this by choosing asymmetric fields, e.g. two-colored fields~\cite{Chen:2019vly,Seipt:2019ddd}. Here we will study the $\sigma$ integrand as a function of $\chi(\sigma)$, so the following results are relevant for a general (e.g. asymmetric) field shape.

Thus, the probability terms in LCF can be expressed as
\be\label{P1LCFJ}
{\bf P}_1=\frac{\alpha}{2}\int\frac{\ud\sigma}{b_0}\hat{\bf B}(J_1^{\rm C}+J_1^{\rm L})
\ee
and
\be\label{P10LCFJ}
{\bf P}_{10}=\frac{\alpha}{2}\int\frac{\ud\sigma}{b_0}(J_\perp{\bf 1}_\perp+J_\parallel{\bf 1}_\parallel+J_r[\hat{\bf E}\,\hat{\bf k}-\hat{\bf k}\,\hat{\bf E}]) \;.
\ee

\begin{figure}
\includegraphics[width=\linewidth]{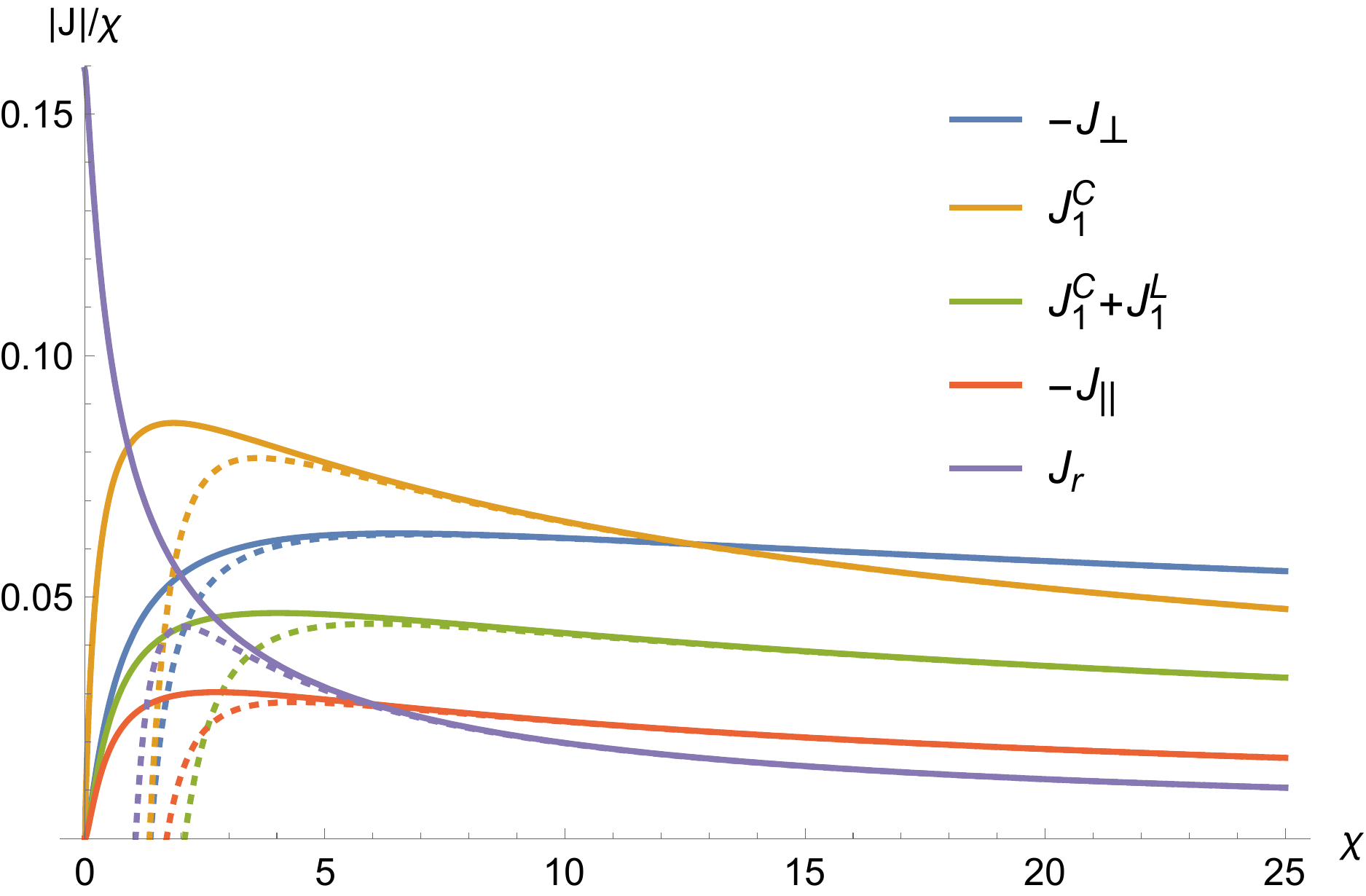}
\caption{The different contributions to the Mueller matrix for the initial and final spin of the electron, in LCF. The dashed lines show the large-$\chi$ expansion arbitrarily truncated at $\chi^{-2}$.}
\label{spinMatLCFfig}
\end{figure}

We consider first the $\chi\ll1$ and $\chi\gg1$ expansions. A simple way to obtain these is to first calculate the Mellin transform with respect to $\chi$, as explained in Appendix~\ref{Mellin-transform-section}. 
For the first few terms we obtain for the low-energy expansion
\be\label{J1Clow}
J_1^{\rm C}=\frac{\sqrt{3}}{4}\chi^2-2\chi^3+\mathcal{O}(\chi^4)
\ee
\be
J_1^{\rm L}=-\frac{\sqrt{3}}{4}\chi^2+3\chi^3+\mathcal{O}(\chi^4)
\ee
\be
J_\perp=-\frac{5\sqrt{3}}{8}\chi^3+\mathcal{O}(\chi^4)
\ee
\be
J_\parallel=-\frac{35}{24\sqrt{3}}\chi^3+\mathcal{O}(\chi^4)
\ee
\be
J_r=\frac{\chi}{2\pi}+\mathcal{O}(\chi^3\ln\chi) \;.
\ee
Note that the contributions from Compton scattering and the loop to $J_1$ cancel to leading order. We also see that the rotational term $J_r$ is the only term that contribute to the overall leading order.
For large $\chi$ we find
\be\label{J1Clarge}
J_1^{\rm C}=\frac{2\Gamma\left[\frac{4}{3}\right]\chi^{1/3}}{3^{2/3}}-\sqrt{3}-\frac{4\Gamma\left[-\frac{1}{3}\right]}{9\times3^{1/3}\chi^{1/3}}+\mathcal{O}(\chi^{-1}) 
\ee
\be\label{J1Llarge}
J_1^{\rm L}=-\frac{\Gamma\left[\frac{1}{3}\right]\chi^{1/3}}{9\times3^{2/3}}-\frac{2\Gamma\left[-\frac{1}{3}\right]}{27\times3^{1/3}\chi^{1/3}}+\mathcal{O}(\chi^{-1})
\ee
\be
J_\perp=-\frac{2\Gamma\left[\frac{2}{3}\right]\chi^{2/3}}{9\times3^{1/3}}+\frac{1}{2}+\frac{20\Gamma\left[-\frac{2}{3}\right]}{27\times3^{2/3}\chi^{2/3}}+\mathcal{O}(\chi^{-1})
\ee 
\be
J_\parallel=-\frac{1}{3}\left[\ln\left[\frac{\chi}{\sqrt{3}}\right]-\gamma_{\rm E}\right]+\frac{1}{2}-\frac{20\Gamma\left[\frac{1}{3}\right]}{9\times3^{2/3}\chi^{2/3}}+\mathcal{O}(\chi^{-1})
\ee
\be
J_r=\frac{\Gamma\left[\frac{4}{3}\right]\chi^{1/3}}{9\times3^{1/6}}+\frac{\Gamma\left[\frac{5}{3}\right]}{3^{11/6}\chi^{1/3}}+\mathcal{O}(\chi^{-1}\ln\chi) \;.
\ee
The appearance of fractional powers like $\chi^{1/3}$ can mean a slow convergence~\cite{Seipt:2020diz,Mironov:2020gbi}, but with the Mellin transform it is in any case easy to obtain higher orders in these expansions. In Fig.~\ref{spinMatLCFfig} we have arbitrarily truncated the large-$\chi$ expansion at $\chi^{-2}$ for all terms (in some terms the correction is merely suppressed as $\mathcal{O}(\chi^{-7/3})$).
From these expansions we see that $J_1^{\rm C}$ and $J_1^{\rm L}$ do not cancel each other beyond the small-$\chi$ limit. However, they continue to be on the same order of magnitude for arbitrary $\chi$. In the asymptotically large-$\chi$ limit we have $J_1^{\rm C}\sim-6J_1^{\rm L}$, but around the maximum they are closer and $J_1^{\rm C}+J_1^{\rm L}\sim J_1^{\rm C}/2$. Thus, in the LCF regime we find that the loop is numerically important for any value of $\chi$.   

\subsection{LCF + low-energy approximation}\label{compare Sokolov-Ternov}

In the low-$\chi$ limit we can sum up the $\alpha$ expansion explicitly. For this it is convenient to use 4D Mueller matrices. We can write the $3\times3$ matrix ${\bf P}_{10}$ as a $4\times4$ matrix by just setting ${\bf P}_{10}^{00}={\bf P}_{10}^{0i}={\bf P}_{10}^{i0}=0$, with $i=1,2,3$. The 3D vector ${\bf P}_1$ becomes a $4\times4$ matrix by replacing $\hat{\bf B}\to\hat{\bf B}\hat{\bf e}_0$, where $\hat{\bf e}_0=\{1,0,0,0\}$ and $\hat{\bf e}_0\cdot\hat{\bf B}=0$.
We can then write the contribution from Compton scattering and the loop in terms of a 4D Mueller matrix as
\be
\mathbb{P}^{\rm C}+\mathbb{P}^{\rm L}=\frac{1}{2}{\bf N}_1\cdot\int\ud\sigma{\bf m}(\sigma)\cdot{\bf N}_0 \;,
\ee
where 
\be
\begin{split}
{\bf m}=\alpha\frac{\chi}{b_0}\bigg[&\chi^2\hat{\bf B}\hat{\bf e}_0-\frac{5\sqrt{3}}{8}\chi^2{\bf 1}_\LCperp-\frac{35}{24\sqrt{3}}\chi^2{\bf 1}_\LCpara \\
&+\frac{1}{2\pi}(\hat{\bf E}\,\hat{\bf k}-\hat{\bf k}\,\hat{\bf E})\bigg] \;.
\end{split}
\ee
For a linearly polarized field we have
\be\label{MuellerLCFlowLin}
{\bf m}=\alpha\frac{\chi}{b_0}\begin{pmatrix}0&0&0&0\\ 0&-\frac{5\sqrt{3}}{8}\chi^2&0&\frac{1}{2\pi}\\ \chi^2&0&-\frac{5\sqrt{3}}{8}\chi^2&0\\ 0&-\frac{1}{2\pi}&0&-\frac{35}{24\sqrt{3}}\chi^2 \end{pmatrix} \;.
\ee
According to the gluing method, one can approximate the $\mathcal{O}(\alpha^2)$ term as\footnote{There is an overall factor of $2$ because a sum over the spin in the intermediate electron state is replaced by an spin average, $\sum_{\bf n}\to2\langle...\rangle$, and according to the gluing prescription one should replace $\langle{\bf n}{\bf n}\rangle\to{\bf 1}$, which is equivalent to multiplying the 4D Mueller matrices.}
\be
\mathbb{P}^{(2)}\approx\frac{1}{2}{\bf N}_f\cdot\int\ud\sigma_2\int^{\sigma_2}\ud\sigma_1{\bf m}(\sigma_2)\cdot{\bf m}(\sigma_1)\cdot{\bf N}_0 \;.
\ee
In general the Mueller matrices would be connected also via the integrals over the longitudinal momenta, but here in the low-$\chi$ limit they are only connected via (lightfront) time ordering. So, the sum over all orders in $\alpha$ gives a time-ordered exponential
\be\label{LCFresumLowChi}
\mathbb{P}=\sum_{j=0}^\infty\mathbb{P}^{(j)}=\frac{1}{2}{\bf N}_f\cdot T_\sigma\exp\left\{\int^\sigma\ud\sigma'{\bf m}(\sigma')\right\}\cdot{\bf N}_0 \;,
\ee
where $T_\sigma$ stands for lightfront time ordering. Although we have derived these results with the goal of predicting what happens after the electron has left the pulse, i.e. for $\sigma\to\infty$, we have written $\mathbb{P}(\sigma)$ as a function of a finite $\sigma$ since this allows us to obtain a differential equation for a ${\bf n}(\sigma)$, which might be simpler to solve, even if one is only interested in $\sigma\to\infty$. So, let
\be
{\bf N}(\sigma)=T_\sigma\exp\left\{\int^\sigma\ud\sigma'{\bf m}(\sigma')\right\}\cdot{\bf N}_0 \;,
\ee 
then
\be
{\bf N}'(\sigma)={\bf m}(\sigma)\cdot{\bf N}(\sigma) \;.
\ee
The first element is conserved, so ${\bf N}(\sigma)=\{1,{\bf n}(\sigma)\}$, and for the remaining 3D part we have
\be\label{diffEqStokes}
\begin{split}
n_1'=&\Omega n_3-\frac{n_1}{T} \\
n_2'=&\frac{1}{T}\left(\frac{8}{5\sqrt{3}}-n_2\right) \\
n_3'=&-\Omega n_1-\frac{7}{9}\frac{n_3}{T}  \;,
\end{split}
\ee
where 
\be
\frac{1}{T}=\frac{5\sqrt{3}}{8}\alpha\frac{\chi^3}{b_0} \qquad \Omega=\eta\frac{\chi}{b_0} \qquad \eta=\frac{\alpha}{2\pi} \;.
\ee
Eq.~\eqref{diffEqStokes} agrees with Eq.~(3.24) in~\cite{BaierSokolovTernov}, which describes the time evolution of the Stokes vector in a magnetic field and for a high-energy particle. This is expected since a general field appears as a crossed (plane wave) field for a high-energy particle ($\chi$ can still be small even if $\gamma\gg1$). This is a nontrivial check of our gluing method as well as many of its building blocks. This is encouraging since this is a regime where the dominant contribution comes from low-energy photons, which one might otherwise have expected to be challenging for a gluing/incoherent product approach (cf.~\cite{DiPiazza:2017raw,DiPiazza:2018bfu,Ilderton:2018nws} for the break-down of LCF for soft photons). To understand this one should note that some problems due to soft (or infrared divergent) photons are either absent or can be expected to be less severe thanks to the inclusion of the loop, because of the soft-photon cancellation between the loop and Compton scattering.

The solution to~\eqref{diffEqStokes} for a constant field can be found in~\cite{BaierSokolovTernov}, and in particular the solution for $n_2$ can be written down immediately. However, it might be difficult to obtain a simple differential equation away from the low-$\chi$ regime, because in addition to time ordering one also has ``(longitudinal) momentum ordering'' due to Compton scattering steps. So, in Appendix~\ref{LCFlowSol} we instead take a step back and calculate~\eqref{LCFresumLowChi} directly. For an initially polarized particle, the $\Omega$ terms lead to rotation.
For an initially unpolarized particle the probability to observe ${\bf n}$ in the final state is given by
\be\label{STunPolToPol}
\mathbb{P}=\frac{1}{2}\left[1+\frac{8}{5\sqrt{3}}{\bf n}\cdot\hat{\bf B}\left(1-\exp\left\{-\frac{5\sqrt{3}}{8}\alpha\int\frac{\ud\sigma}{b_0}\chi^3\right\}\right)\right] \;.
\ee
The maximum probability is achieved with ${\bf n}=\hat{\bf B}$.
Since we have absorbed $e$ into the definition of the background field, for $e<0$ $\hat{\bf B}$ is actually anti-parallel to the magnetic field, so electrons will polarize anti-parallel to the magnetic field, which is well known.
While the integrand in the exponent is small, if the pulse is sufficiently long then the exponential becomes small and one approaches the upper limit for the induced polarization of the electron beam, namely $8/(5\sqrt{3})\approx0.92$~\cite{Sokolov:1963zn,BaierSokolovTernov}. However, the pulse would have to be very long to compensate for $\alpha\chi^3/b_0\ll1$ (there is of course also the problem that the field polarization would in general oscillate).

The maximum polarization can be obtained directly from the Mueller matrix in~\eqref{MuellerLCFlowLin} without finding the complete solution. One just has to notice that ${\bf m}\cdot\{1,0,8/(5\sqrt{3}),0\}={\bf 0}$, so ${\bf N}=\{1,0,8/(5\sqrt{3}),0\}$ is an eigenvector\footnote{The other three eigenvectors have vanishing first element ${\bf N}=\{0,{\bf n}\}$.} of the Mueller matrix with zero as eigenvalue. This means that applying further Mueller matrices will not change this Stokes vector. We can also see from the differential equation~\eqref{diffEqStokes} that this corresponds to $\ud{\bf n}/\ud\sigma=0$. Thus, $8/(5\sqrt{3})$ represents the maximum degree of polarization.    

\subsection{LCF for larger $\chi$}\label{LCF larger chi}

\begin{figure}
\includegraphics[width=\linewidth]{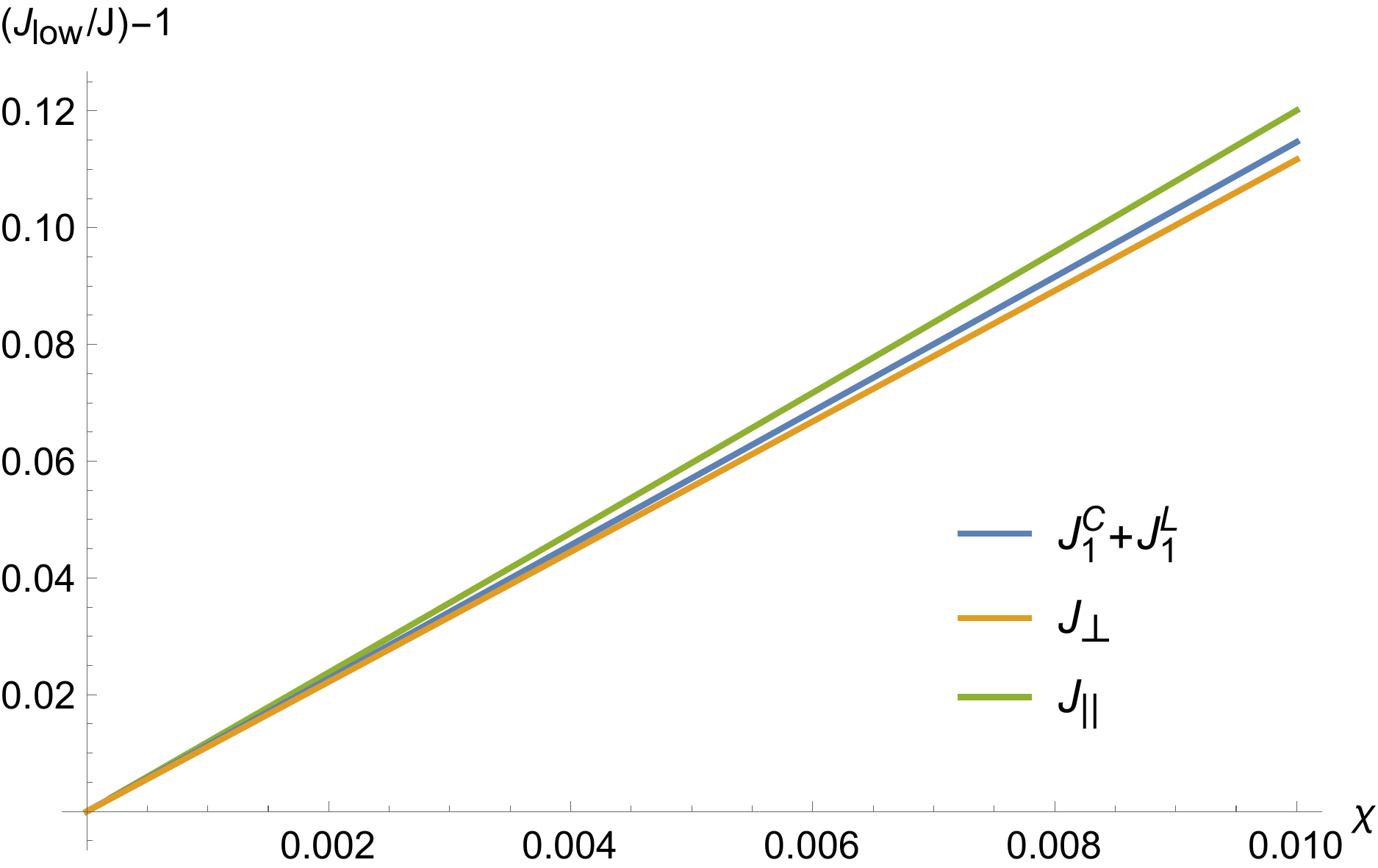}
\caption{The relative error of the leading order in the small-$\chi$ expansion. The relative error for $J_r$ is smaller and is therefore not shown.}
\label{relerlowFig}
\end{figure}

We have just shown that at leading order in $\chi\ll1$ we can explicitly resum the $\alpha$ expansion and we recover the results in~\cite{BaierSokolovTernov}. The next question then is how small $\chi$ has to be in order for these results to give a good approximation. In Fig.~\ref{relerlowFig} we see that the relative error for the individual terms is already $\sim10\%$ at $\chi\sim0.01$. So, one might expect significant corrections even if $\chi$ is quite small. At larger $\chi$ Fig.~\ref{spinMatLCFfig} shows that the rotational term $J_r$ decreases while the other terms first increase and then slowly decrease. So, while the rotation is the dominant effect at $\chi\ll1$, at larger $\chi$ one can expect that rotation and damping become on the same order of magnitude. 

\begin{figure}
\includegraphics[width=\linewidth]{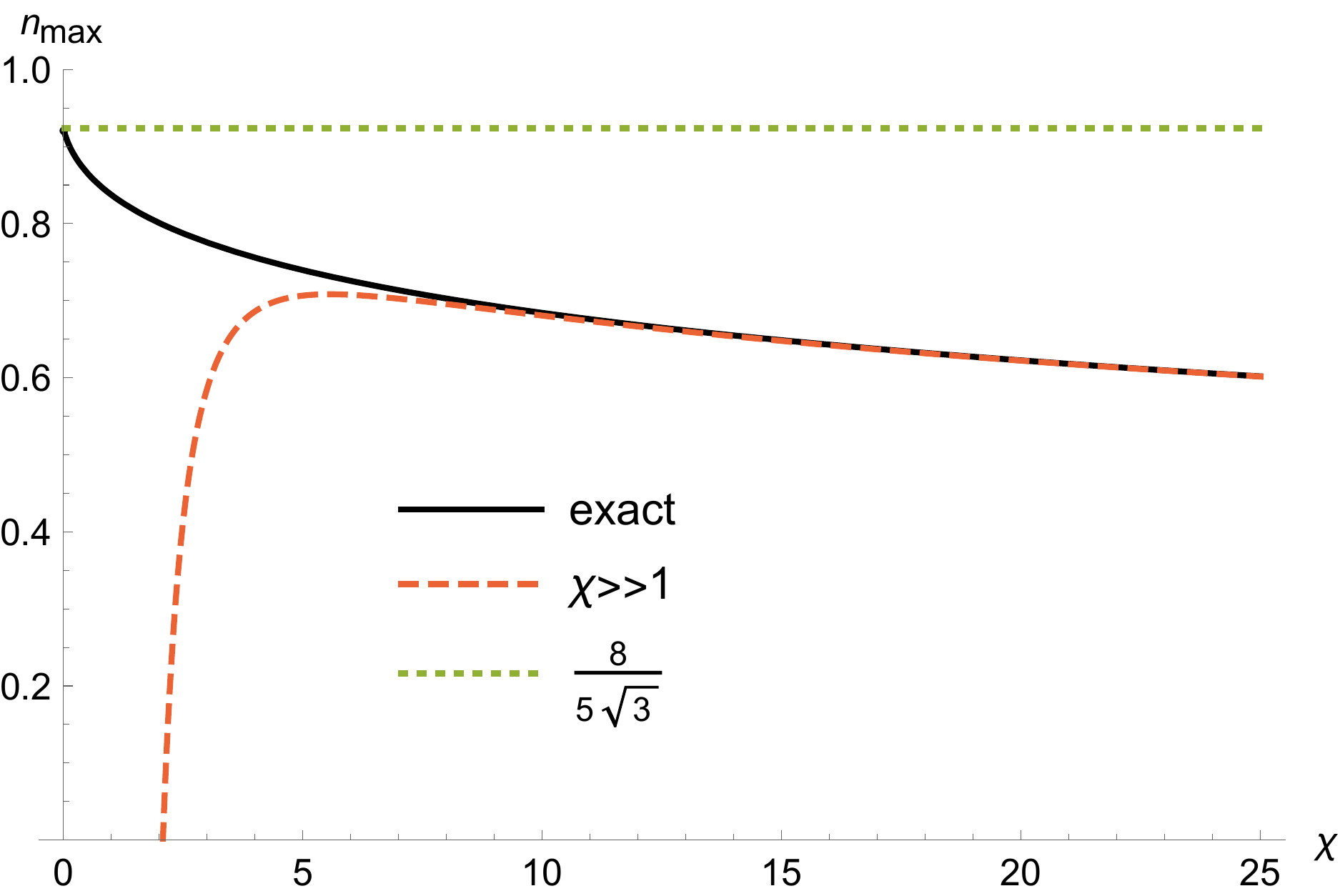}
\caption{The estimate in~\eqref{nmax} of the maximum degree of polarization as a function of $\chi$. The $\chi\gg1$ line is obtained from the large-$\chi$ expansions in Fig.~\ref{spinMatLCFfig}, without re-expanding the resulting ratio. The leading order in the $\chi\gg1$ limit (shown in~\eqref{lowhighnmax}) is larger than one for $\chi\lesssim40$, so cannot be used here.}
\label{maxPolFig}
\end{figure}

At larger $\chi$ one would in general expect it to be necessary to include the recoil on the electrons due to the emission of photons, i.e. radiation reaction. This would mean that we can no longer perform the integral over the longitudinal momentum of the emitted photon for each ${\bf M}^{\rm C}$ separately. So, in general one would need a numerical treatment. However, while waiting for such numerical results, we can try to go ahead and use the $s$ integrated results~\eqref{P1LCFJ} and~\eqref{P10LCFJ} anyway, hoping that it will at least give a decent idea of the scaling. We can in general find an eigenvector of the Mueller matrix ${\bf m}$ with eigenvalue zero, given by 
\be\label{nmax}
{\bf N}=\{1,0,n_{\rm max},0\} \qquad n_{\rm max}=-\frac{J_1^{\rm C}+J_1^{\rm L}}{J_\LCperp}\;,
\ee
where the ${\bf e}_2$ component points along the magnetic field.
This suggests a maximum degree of polarization given by $n_{\rm max}$. For $\chi\ll1$ and $\chi\gg1$ we have
\be\label{lowhighnmax}
n_{\rm max}(\chi\ll1)\approx\frac{8}{5\sqrt{3}}  \quad
n_{\rm max}(\chi\gg1)\approx\frac{5\Gamma\left[\frac{1}{3}\right]}{2\Gamma\left[\frac{2}{3}\right](3\chi)^{1/3}} \;.
\ee
The $\chi\ll1$ limit has already been discussed. The $\chi\gg1$ limit agrees with~\cite{Lobanov1980}. However, the exact result converges very slowly to this leading order, which can be seen from ${\rm NLO}/{\rm LO}\sim-2.4/\chi^{1/3}$. In Fig.~\ref{maxPolFig} we show that $n_{\rm max}$ is a monotonically decreasing function of $\chi$, so the low-$\chi$/Sokolov-Ternov result is the overall maximum. It would be interesting to check these results with a numerical treatment that includes radiation reaction.

\subsection{Low energy limit}\label{loop low energy limit}

In this subsection we will consider the low-energy limit, i.e. small $b_0$. Here we are interested in the probability integrated over all the momenta and summed over the polarization of the emitted photon (for the contribution from Compton scattering), so the starting point is~\eqref{generalProbSum} with the terms expressed in terms of cosine/sine integrals as in Appendix~\ref{The final momentum integral}. For ${\bf P}_{10}^{\rm C}$ and for the $\langle\hat{\mathbb{R}}^{\rm L}\rangle{\bf 1}$ part of ${\bf P}_{10}^{\rm L}$ we obtain the leading order by expanding the integrand for large $\varphi$. We find that these two terms cancel to leading order. 
In the remaining part of ${\bf P}_{10}^{\rm L}$ we rescale $\theta\to2b_0\theta$ and then expand the integrand in $b_0$. Performing the resulting $\theta$ integral gives (omitting the argument of ${\bf a}(\sigma)$)
\be
{\bf P}_{10}^{\rm L}+{\bf P}_{10}^{\rm C}\approx\frac{\alpha}{2\pi}\int\ud\sigma\frac{1}{2}[{\bf a}',\hat{\bf k}]=:\frac{\alpha}{2\pi}\int\ud\sigma\frac{\bf F}{2}=:\frac{\bm\delta}{2} \;,
\ee
and if we choose the constant part of the potential so that ${\bf a}(-\infty)=0$ then
\be
{\bm\delta}=\frac{\alpha}{2\pi}[{\bf a}(\infty),\hat{\bf k}] \;.
\ee
We also find that ${\bf P}_1^{\rm L}+{\bf P}_1^{\rm C}$ vanishes to leading order. Hence, the only terms that remain in the low energy/classical limit are terms that come from the loop. We see that the field has to be unipolar to have a nonzero change in the low-energy limit. So, from~\eqref{generalProbSum} we finally find
\be
\mathbb{P}=\frac{1}{2}\left(1+{\bf n}_1\cdot[{\bf 1}+{\bm\delta}]\cdot{\bf n}_0\right) \;.
\ee 
From this we can read off the new spin state
\be
{\bf n}_f=[{\bf 1}+{\bm\delta}]\cdot{\bf n}_0 \;.
\ee
Note that, since ${\bm\delta}$ is an antisymmetric matrix, the new Stokes vector is a unit vector to the order of $\alpha$ that we are working with, i.e. ${\bf n}_f^2=1+\mathcal{O}(\alpha^2)$. Note also that the change in the Stokes vector is orthogonal to the initial vector, i.e. $\pm{\bf n}_0\cdot{\bm\delta}\cdot{\bf n}_0=0$, which holds for an arbitrary ${\bf n}_0$ because ${\bm\delta}$ is antisymmetric.

So far we have considered the probability at $\mathcal{O}(\alpha)$. In~\cite{Dinu:2018efz,Dinu:2019pau} we showed how the Mueller matrices for nonlinear Compton and Breit-Wheeler can be glued together to form approximations of higher-order processes. We show in Appendix~\ref{glueloop} that the obvious generalization of the gluing method to processes with loops is indeed correct. Hence, we should replace (no sum over $j$)
\be
\sum{\bf n}_j{\bf n}_j\to2\langle{\bf n}_j{\bf n}_j\rangle\to2{\bf 1}
\ee
for each intermediate Stokes vector ${\bf n}_j$. In the low-energy limit we have (cf.~\eqref{generalProbSum})
\be
\mathbb{P}^{(1)}=\mathbb{P}^{\rm L}+\mathbb{P}^{\rm C}\approx\frac{1}{2}{\bf n}_1\cdot\frac{\alpha}{2\pi}\int\ud\sigma{\bf F}\cdot{\bf n}_0 \;,
\ee
so at $\mathcal{O}(\alpha^2)$ the gluing prescription applied to the intermediate Stokes vector ${\bf n}_1$ gives
\be
\mathbb{P}^{(2)}\approx\frac{1}{2}{\bf n}_2\cdot\left(\frac{\alpha}{2\pi}\right)^2\int\ud\sigma_2\int^{\sigma_2}\!\ud\sigma_1{\bf F}(\sigma_2)\cdot{\bf F}(\sigma_1)\cdot{\bf n}_0 \;,
\ee
where the restriction to $\sigma_1<\sigma_2$ comes from demanding that the second step (either photon emission or a loop) happens after the first step. Continuing in this way, we find the expansion of a $\sigma$-ordered exponential. So, in this low-energy limit we can resum all the $\alpha$ orders, and we find that the probability to go from ${\bf n}_0$ to ${\bf n}$ is given by
\be
\mathbb{P}=\frac{1}{2}(1+{\bf n}\cdot{\bf n}_f) \;,
\ee
where
\be\label{resummedStokes}
{\bf n}_f=T\exp\Big\{\frac{\alpha}{2\pi}\int{\bf F}\Big\}\cdot{\bf n}_0 \;.
\ee

In general one would only expect the gluing/product approach to give an approximation, but 
we will now show that this low-energy limit agrees exactly with the solution of the BMT equation~\cite{Bargmann:1959gz}
\be\label{BMTeq}
\frac{\ud\alpha^\mu}{\ud\tau}=\left[(1+\mu)F^{\mu\nu}-\mu\pi^\mu\pi_\sigma F^{\sigma\nu}\right]\alpha_\nu \;,
\ee
where $\alpha^\mu$ is the spin 4-vector related to the Stokes vector as in~\eqref{Stokes4vec}, and $\mu$ is the anomalous magnetic moment. The momentum $p_\mu$ in~\eqref{alphaidefin} should be replaced by the time-dependent momentum, and since we know that in the classical limit this is to leading order given by $\pi(\phi)$, we now have the following $\phi$-dependent basis
\be
\alpha_\mu^{(1,2)}(\phi)=-\delta_\mu^{1,2}-\frac{\pi_{1,2}}{kp}k_\mu
\quad
\alpha_\mu^{(3)}(\phi)=\pi_\mu-\frac{1}{kp}k_\mu \;.
\ee
Inserting $\alpha_\mu=n_i\alpha^{(i)}_\mu$ into~\eqref{BMTeq}, using $\dot{\alpha}^{(i)\mu}=F^{\mu\nu}\alpha^{(i)}_\nu$ to cancel the first term on the right-hand side of~\eqref{BMTeq}, $\alpha_\mu^{(i)}\alpha^{(j)\mu}=-\delta_{ij}$ to project onto a single $\dot{n}_i$, $\alpha^{(i)}\pi=0$ to cancel the last term in~\eqref{BMTeq}\footnote{We have projected with the three vectors $\alpha^{(i)}$. The last term is needed since $\pi[F-\mu\pi\pi F]=0$ means that there is zero overlap with $\pi_\mu$.}, $\dot{f}=kp\ud f/\ud\phi$, and finally $\alpha^{(i)}F\alpha^{(j)}=-kp[{\bf a}'_i\hat{\bf k}_j-\hat{\bf k}_i{\bf a}'_j]$ we find that the BMT equation reduces to
\be
\frac{\ud{\bf n}}{\ud\phi}=\mu[{\bf a}'\hat{\bf k}-\hat{\bf k}{\bf a}']\cdot{\bf n}=:\mu{\bf F}\cdot{\bf n} \;,
\ee  
and hence the solution is given by
%\be
%\begin{split}
%{\bf n}=&T\exp\Big\{\mu\int_{\phi_0}^\phi{\bf F}\Big\}{\bf n}(\phi_0)
%=\bigg\{{\bf 1}+\mu\int^\phi{\bf F} \\
%&+\mu^2\int^\phi\!\ud\phi_2\int^{\phi_2}\!\ud\phi_2{\bf F}(\phi_2)\cdot{\bf F}(\phi_1)+\dots\bigg\}\cdot{\bf n}_0 \;.
%\end{split}
%\ee
\be
{\bf n}=T\exp\Big\{\mu\int_{\phi_0}^\phi{\bf F}\Big\}{\bf n}_0 \;,
\ee
where ${\bf n}_0={\bf n}(\phi_0)$.
This is exactly the same as the vector~\eqref{resummedStokes} that gives the maximum probability ($\mathbb{P}=1$ in this case).

For linear polarization we have ${\bf F}(\phi)=a'(\phi)\hat{\bf F}$, where $\hat{\bf F}=\hat{\bf E}\hat{\bf k}-\hat{\bf k}\hat{\bf E}$ is a constant matrix and $\hat{\bf E}$ is as before a unit vector pointing in the electric-field direction. Since $\hat{\bf F}^2=-(\hat{\bf E}\hat{\bf E}+\hat{\bf k}\hat{\bf k})$ and $\hat{\bf F}^3=-\hat{\bf F}$ we find (assuming $a(\phi_0)=0$) 
\be\label{BMTsoln}
{\bf n}=\left\{\hat{\bf B}\hat{\bf B}+(\hat{\bf E}\hat{\bf E}+\hat{\bf k}\hat{\bf k})\cos[\mu a(\phi)]+\hat{\bf F}\sin[\mu a(\phi)]\right\}\cdot{\bf n}_0 \;,
\ee
where $\hat{\bf B}$ is again a unit vector in the magnetic-field direction. For example, if we start with ${\bf n}_0=\hat{\bf k}$ then ${\bf n}=\hat{\bf k}\cos[\mu a]+\hat{\bf E}\sin[\mu a]$.
Here we have considered the leading order in the low-energy expansion, where the degree of polarization is constant, ${\bf n}^2(\phi)=1$. At higher orders, also the degree of polarization can change, as in~\eqref{STunPolToPol} and~\cite{BaierSokolovTernov} in the LCF regime.

\subsection{Electrons with negligible recoil}\label{Electrons with negligible recoil}

We have now seen how the higher orders can be resummed into a time-ordered exponential in the low-energy limit. In general it is not possible to obtain such compact results, because, after an electron has emitted a photon carrying a significant fraction of its longitudinal momentum, the second step has effectively a different $b_0$ (although we still use $b_0$ for the initial momentum). However, if we consider the probability that the final electron has a longitudinal momentum close to the initial one, then any photon that is emitted must be soft and so all the longitudinal momentum integrals for the Compton steps are restricted to small values of the photon momentum $q$. For a field with ${\bf a}(\infty)={\bf a}(-\infty)$ there is no IR divergence~\cite{Dinu:2012tj,Ilderton:2012qe}, and in~\cite{DiPiazza:2017raw,DiPiazza:2018bfu} it has been shown explicitly that the longitudinal momentum spectrum, $\mathbb{P}^{\rm C}(s_1=s_0-q_1)$ in our notation, has a finite, constant soft-photon limit $q_1\to0$. So, the momentum integral in each Compton step has to leading order a constant integrand, and is therefore simply proportional to the length of the integration interval, which is smaller (or equal if only one photon is emitted) than the difference between the final and initial electron momentum, $1-s_f\ll1$. Hence, the contribution from Compton scattering can be made small by choosing the final electron momentum to be very close to the initial momentum. The contribution from loops, on the other hand, is not restricted at all by this, and we still have the same longitudinal momentum integrals. Thus, in this limit of negligible recoil, we can neglect Compton scattering but still have a nontrivial spin effect due to the loop. This simplifies the calculation tremendously, as we can again resum the $\alpha$ expansion into a time-ordered exponential. 

In terms of a 4D Mueller matrix we have
\be\label{4DMuellerLoop}
\mathbb{P}^{\rm L}=\frac{1}{2}{\bf N}_1\cdot{\bf M}^{\rm L}\cdot{\bf N}_0=\frac{1}{2}{\bf N}_1\cdot\int\ud\sigma{\bf m}(\sigma)\cdot{\bf N}_0 \;,
\ee
According to the gluing prescription, the $\mathcal{O}(\alpha^2)$ term is approximately given by
\be
\mathbb{P}^{(2)}\approx\frac{1}{2}{\bf N}_f\cdot\int\ud\sigma_2\int^{\sigma_2}\ud\sigma_1{\bf m}(\sigma_2)\cdot{\bf m}(\sigma_1)\cdot{\bf N}_0 \;,
\ee 
and similarly for higher orders. Thus,
\be\label{loopResum}
\mathbb{P}=\sum_{j=0}^\infty\mathbb{P}^{(j)}\approx\frac{1}{2}{\bf N}_f T_\sigma\exp\left\{\int\ud\sigma{\bf m}\right\}{\bf N}_0 \;.
\ee

Things simplify further if we assume a linearly polarized field. With ${\bf a}(\phi)=a(\phi){\bf e}_1$, the Mueller matrix in~\eqref{MLgeneral} simplifies to
\be\label{MLlinear}
\begin{split}
{\bf M}^{\rm L}=&2\langle\mathbb{P}^{\rm L}\rangle{\bf 1}^{(4)}+2P_0^{\rm L}[{\bf e}_0{\bf e}_2+{\bf e}_2{\bf e}_0] \\
&+M_{\rm rot}^{\rm L}[{\bf e}_1{\bf e}_3-{\bf e}_3{\bf e}_1]=:\int\ud\sigma{\bf m}^{\rm L}(\sigma) \;,
\end{split}
\ee
where $P_0^{\rm L}$ and $M_{\rm rot}^{\rm L}$ are obtained by matching with~\eqref{MLgeneral}. We do not have to choose a field shape or calculate $P_0^{\rm L}$ and $M_{\rm rot}^{\rm L}$ to see that ${\bf m}(\sigma)$ in~\eqref{MLlinear} is written as the sum of three matrices, and each matrix commutes with itself and the other matrices at different $\sigma$. The time-ordering hence becomes trivial and we find
\be\label{noRecoilLinGen}
\begin{split}
\mathbb{P}
=&\frac{1}{2}{\bf N}_f \cdot e^{-2\langle\mathbb{P}^{\rm C}\rangle}\bigg[[{\bf e}_0{\bf e}_0+{\bf e}_2{\bf e}_2]\cosh(2P_0^{\rm C}) \\
&-[{\bf e}_0{\bf e}_2+{\bf e}_2{\bf e}_0]\sinh(2P_0^{\rm C})+[{\bf e}_1{\bf e}_1+{\bf e}_3{\bf e}_3]\cos M_{\rm rot}^{\rm L} \\
&+[{\bf e}_1{\bf e}_3-{\bf e}_3{\bf e}_1]\sin M_{\rm rot}^{\rm L}\bigg]\cdot{\bf N}_0 \;,
\end{split}
\ee 
where we have used $\langle\mathbb{P}^{\rm L}\rangle=-\langle\mathbb{P}^{\rm C}\rangle$ and $P_0^{\rm L}=-P_0^{\rm C}$ to write loop contributions in terms of Compton terms. 
Eq.~\eqref{noRecoilLinGen} can be seen as a generalization of Eq.~(14) in~\cite{Meuren:2011hv} to arbitrary electron polarization (including an initially unpolarized electron). Considering spins that are not parallel or anti-parallel, ${\bf n}_f\ne\pm{\bf n}_0$, also allows us to see an effect already at $\mathcal{O}(\alpha)$.
If the initial particle is unpolarized, ${\bf N}_0={\bf e}_0$, then
\be\label{noRecoilLinUnPol}
\mathbb{P}=\frac{1}{2}{\bf N}_f \cdot e^{-2\langle\mathbb{P}^{\rm C}\rangle}\cosh(2P_0^{\rm C})[{\bf e}_0-\tanh(2P_0^{\rm C}){\bf e}_2] \;,
\ee
which means that the electron tends to become polarized parallel to $\pm{\bf e}_2$, i.e. parallel (or anti-parallel) to the magnetic field, as one might expect. Recall that in this fermion-loop section we have absorbed a factor of $2$ into $\langle\mathbb{P}^{\rm C}\rangle$ and $P_0^{\rm C}$ to account for a trivial factor of $2$ coming from summing these photon-polarization independent terms over the polarization of the emitted photon. So, $2\langle\mathbb{P}^{\rm C}\rangle\pm2P_0^{\rm C}$ gives the total probability of Compton scattering by an initial electron with polarization ${\bf n}=\pm{\bf e}_2$ 
and summed over the polarization of the emitted photon and the spin of the final-state electron, where the latter gives the (explicit) factor of $2$. Note that~\eqref{noRecoilLinUnPol} comes solely from the loop, but it is written in terms of the Compton scattering probabilities in order to show that the induced electron polarization is a consequence of the fact that the probability to emit a photon is higher for ${\bf n}=\text{sign}(P_0){\bf e}_2$, so in the forward direction there will be more electrons with ${\bf n}=-\text{sign}(P_0){\bf e}_2$. 

It might at first seem natural to find LCF approximations of the above results. However, this can be problematic because in this section we have assumed that Compton scattering can be neglected, which is justified if the Compton spectrum is bounded for low photon momentum, but in the LCF regime the spectrum is IR divergent (see~\cite{Dinu:2012tj,Ilderton:2012qe,DiPiazza:2017raw,DiPiazza:2018bfu} for a comparison of IR in LCF and non-LCF). Although it happens to be an integrable singularity, it might anyway lead to a too large contribution from soft photons, which would mean that Compton scattering is not negligible. Also the formation length might become too large (compared to the pulse length) for the gluing approach (at least if the loop and Compton scattering are considered separately).  
However, the rotational term from the loop has no counterpart in Compton scattering, so it makes more sense to consider it separately. The LCF approximation of this term is given by~\eqref{LCFRLoop10Definition}
\be
M_{\rm rot}^{\rm L,LCF}=\alpha\int\frac{\ud\sigma}{b_0}\int_0^1\ud s\, q\frac{\text{Gi}(\xi)}{\sqrt{\xi}}
\ee  
and if we identify this with $\mu a(\phi)$ in the solution~\eqref{BMTsoln} to the BMT equation, then we find a field-dependent anomalous magnetic moment that agrees with the literature~\cite{BaierSokolovTernov,Li:2018fcz,Ilderton:2020gno} (this is immediately clear by comparing with Eq.~(25) in~\cite{Ilderton:2020gno}). To leading order in $\chi\ll1$ this reduces to the usual $\mu=\alpha/(2\pi)$. Note though that the difference between $\mu(\chi)$ and $\alpha/(2\pi)$ can be expected to be on the same order of magnitude as the other, non-rotational terms. 

We can find very similar expressions for a circularly polarized field in the LMF approximation. Here the Mueller matrix can be written as
\be\label{MLcirc}
\begin{split}
{\bf M}^{\rm L}=&2\langle\mathbb{P}^{\rm L}\rangle{\bf 1}^{(4)}+2P_0^{\rm L}[{\bf e}_0{\bf e}_3+{\bf e}_3{\bf e}_0] \\
&+M_{\rm rot}^{\rm L}[{\bf e}_1{\bf e}_2-{\bf e}_2{\bf e}_1] \;,
\end{split}
\ee
where $P_0^{\rm L}$ and $M_{\rm rot}^{\rm L}$ are obtained by matching with~\eqref{MLgeneral} (we use the same notation, but $P_0^{\rm L}$ and $M_{\rm rot}^{\rm L}$ are of course different from the linear case~\eqref{MLlinear}).
This has exactly the same matrix form as in~\eqref{MLlinear}, except that ${\bf e}_2\leftrightarrow{\bf e}_3$, which means that the spin structure of the probability is obtained by making the replacement ${\bf e}_2\leftrightarrow{\bf e}_3$ in~\eqref{noRecoilLinGen} or~\eqref{noRecoilLinUnPol}. So, the rotational terms lead to spin rotation in the ${\bf e}_1,{\bf e}_2$ plane, i.e. the plane that contains the rotating field polarization, and (e.g.) an unpolarized initial particle will tend to become polarized in the $\pm{\bf e}_3$ direction, i.e. parallel to the laser propagation. As mentioned above, the oscillation of a circularly rotating field does not lead to an averaging out of the induced spin polarization due to the loop or Compton scattering separately, in contrast to a linearly oscillating field. We saw above that there is nevertheless a cancellation between the loop and Compton scattering for such a field. However, in this subsection we are in a regime where Compton scattering is negligible, so it cannot cancel the loop contribution. Thus, if we select those electrons that have kept most of the initial momentum, then a circularly polarized field may lead to electron polarization.

\section{Polarization operator}\label{PolarizationOperatorSection}

In this section we will consider the polarization dependence of the polarization operator. The photon polarization is given by~\eqref{epsilonDefinition}. As in the fermion case, we describe the initial state by a wave packet
\be
|\text{in}\rangle=\int\ud\tilde{l}f\epsilon^\mu a_\mu^\dagger|0\rangle \qquad \int\ud\tilde{l}|f|^2=1 \;,
\ee 
where the annihilation and creation operators satisfy
\be
[a_\mu(l),a_\nu^\dagger(l')]=-2l_\LCm\bar{\delta}(l-l')L_{\mu\nu} \;,
\ee
where in the lightfront gauge we have
\be\label{Lmunu}
L_{\mu\nu}=g_{\mu\nu}-\frac{k_\mu l_\nu+l_\mu k_\nu}{kl}  \;.
\ee
The amplitude for an initial photon with momentum $l_\mu$ and polarization $\epsilon_\mu$ to a final photon with $l'_\mu$ and polarization $\epsilon'_\mu$ is given by
\be
2l_\LCm\bar{\delta}(l'-l)M:=\langle0|\bar{\epsilon}'a' U\epsilon a^\dagger|0\rangle \;,
\ee
where $U$ is the evolution operator. 
While the momentum is conserved, the polarization can change and the probability for this is given by $\mathbb{P}=|M|^2$. The leading order amplitude $M_0$ is given by the same expression as for the mass operator~\eqref{M0}, with $\rho_0$ and $\lambda_0$ for the initial photon and $\rho_1$ and $\lambda_1$ for the final photon. The calculation of $M_1$ is similar e.g. to~\cite{Dinu:2013gaa}, e.g. renormalization of the UV divergence leads to the subtraction of the field-independent part. So, we simply state the results.

The probability takes the same form as for the mass operator~\eqref{loopP1nn} and~\eqref{PfromR}, where now $r=\frac{1}{s}+\frac{1}{1-s}$ and 
\be
\langle\mathbb{R}^{\rm L}\rangle=-\frac{\kappa}{2}\left(\frac{2ib_0}{r_2\theta}+D_1+1\right)-1 \;,
\ee
\be\label{R0polarop}
{\bf R}_{1,k}^{\rm L}={\bf R}_{0,k}^{\rm L}={\bf w}_1\!\cdot\!\left({\bf S}_k+\frac{\kappa}{2}\delta_{k2}{\bm\sigma}_2\right)\!\cdot\!{\bf w}_2  \;,
\ee
\be\label{R10polarop}
\begin{split}
{\bf R}_{10,ij}^{\rm L}&=\langle\mathbb{R}^{\rm L}\rangle\delta_{ij} \\
&+i\text{sign}(\theta){\bf w}_1\!\cdot\!\left({\bm\sigma}_1\varepsilon_{1ij}+{\bm\sigma}_3\varepsilon_{3ij}+\frac{\kappa}{2}{\bm\sigma}_2\varepsilon_{2ij}\right)\!\cdot\!{\bf w}_2
\end{split} \;,
\ee
where $\kappa=\frac{s}{1-s}+\frac{1-s}{s}$, ${\bf S}_k=\delta_{k1}{\bm\sigma}_1+\delta_{k3}{\bm\sigma}_3$, and $\varepsilon_{ijk}$ is the Levi-Civita symbol with $\varepsilon_{123}=1$. 
It is again easy to check that the loop contribution at $\mathcal{O}(\alpha)$ vanishes for ${\bf n}_1=-{\bf n}_0$, but it is in general nonzero.
As expected, $\langle\mathbb{R}^{\rm L}\rangle$ and ${\bf R}_0^{\rm L}$ are, apart from the overall sign, exactly the same as in the nonlinear Breit-Wheeler case. As in the electron mass operator case, this follows from the fact that the sum of the probabilities of all possible final states at $\mathcal{O}(\alpha)$ has to be $1$, so by summing these loop results over the final polarization the result has to exactly cancel the probability of nonlinear Breit-Wheeler summed over the spin of the electron-positron pair, and this should happen for arbitrary initial polarization. 
Then the diagonal $\langle\mathbb{R}^{\rm L}\rangle{\bf 1}$ term in ${\bf R}_{10}^{\rm L}$ and the fact that ${\bf R}_1^{\rm L}={\bf R}_0^{\rm L}$ ensure that the loop gives no contribution to spin flip ${\bf n}_1=-{\bf n}_0$ at $\mathcal{O}(\alpha)$. So, one could have guessed these terms from the corresponding results for Breit-Wheeler pair production. The off-diagonal terms in ${\bf R}_{10}^{\rm L}$ cannot be obtained directly from $\mathbb{P}^{\rm BW}$. However, we can immediately see that we have the same relation between the off-diagonal terms ${\bf R}_{10,ij}^{\rm L}-\langle\mathbb{R}\rangle^{\rm L}\delta_{ij}$ and ${\bf R}_0^{\rm L}$ as in~\eqref{iSignRelation} for the electron mass-operator loop.   

If no pairs are created, then we can resum the sum of products of Mueller matrices as in~\eqref{loopResum}. The restriction to only loop diagrams is less restrictive in the case of polarization loops, because pair production is a threshold process which is exponentially suppressed at low energies, while in the electron case we have to e.g. restrict ourselves to electrons with negligible recoil to be able to neglect photon emission. 

We write the first-order result in terms of a 4D Mueller matrix as in~\eqref{4DMuellerLoop}.
The $\langle\mathbb{R}^{\rm L}\rangle\delta_{ij}$ term in ${\bf R}_{10}^{\rm L}$ combines with $\langle\mathbb{R}\rangle$ to form a term proportional to the 4D unit matrix ${\bf 1}^{(4)}$, which hence commutes with all the other contributions and can be separated from the time-ordered exponential. Since $2\langle\mathbb{R}^{\rm L}\rangle=-4\langle\mathbb{R}^{\rm BW}\rangle$, this part becomes $e^{-4\langle\mathbb{P}^{\rm BW}\rangle}$, where $4\langle\mathbb{P}^{\rm BW}\rangle$ gives the probability of nonlinear Breit-Wheeler pair production, summed over the spins of the fermions and averaged over the polarization of the photon. 

The remaining part of ${\bf m}$ simplifies for a linearly polarized field. With ${\bf a}=a(\phi){\bf e}_1$ we have
\be
{\bf R}_1^{\rm L}={\bf R}_0^{\rm L}=w_1w_2{\bm\epsilon}_3
\ee 
and
\be
{\bf R}_{10}^{\rm L}-\langle\mathbb{R}^{\rm L}\rangle{\bf 1}^{(3)}=i\text{sign}(\theta)w_1w_2({\bm\epsilon}_1{\bm\epsilon}_2-{\bm\epsilon}_2{\bm\epsilon}_1) \;,
\ee
where ${\bf w}_1=w_1{\bf e}_1$. Hence, the Mueller matrix separates into three simple and mutually commuting matrices
\be
{\bf m}=...({\bm\epsilon}_0{\bm\epsilon}_3+{\bm\epsilon}_3{\bm\epsilon}_0)+...({\bm\epsilon}_1{\bm\epsilon}_2-{\bm\epsilon}_2{\bm\epsilon}_1)+...{\bf 1}^{(4)} \;.
\ee 
The time ordering becomes trivial and using $({\bm\epsilon}_0{\bm\epsilon}_3+{\bm\epsilon}_3{\bm\epsilon}_0)^2={\bm\epsilon}_0{\bm\epsilon}_0+{\bm\epsilon}_3{\bm\epsilon}_3$ and $({\bm\epsilon}_1{\bm\epsilon}_2-{\bm\epsilon}_2{\bm\epsilon}_1)^2=-({\bm\epsilon}_1{\bm\epsilon}_1+{\bm\epsilon}_2{\bm\epsilon}_2)$ we find
\be
\begin{split}
\mathbb{P}
=&\frac{1}{2}{\bf N}_f \cdot e^{-4\langle\mathbb{P}^{\rm BW}\rangle}\bigg[[{\bm\epsilon}_0{\bm\epsilon}_0+{\bm\epsilon}_3{\bm\epsilon}_3]\cosh\nu \\
&+[{\bm\epsilon}_0{\bm\epsilon}_3+{\bm\epsilon}_3{\bm\epsilon}_0]\sinh\nu+[{\bm\epsilon}_1{\bm\epsilon}_1+{\bm\epsilon}_2{\bm\epsilon}_2]\cos\varphi \\
&+[{\bm\epsilon}_1{\bm\epsilon}_2-{\bm\epsilon}_2{\bm\epsilon}_1]\sin\varphi\bigg]\cdot{\bf N}_0 \;,
\end{split}
\ee 
where
\be
\begin{split}
\nu&=2{\bm\epsilon}_3\cdot{\bf P}_0^{\rm L}=-4{\bm\epsilon}_3\cdot{\bf P}_\gamma^{\rm BW} \\
&=\frac{i\alpha}{2\pi b_0}\int_0^1\ud s\int\frac{\ud^2\phi_{2,1}}{\theta}w_1w_2\exp\left\{\frac{ir\Theta}{2b_0}\right\} \;,
\end{split}
\ee
where we have a factor of $4$ in the first line because ${\bf P}_\gamma^{\rm BW}$ gives the dependence of the Breit-Wheeler probability on the Stokes vector of the photon but for definite spins of the electron and positron, so summing over their spins gives a trivial factor of $4$ for this term, and
\be
\begin{split}
\varphi&=2{\bm\epsilon}_1\cdot{\bf P}^{\rm L}_{10}\cdot{\bm\epsilon}_2 \\
&=\frac{i\alpha}{2\pi b_0}\int_0^1\ud s\int\frac{\ud^2\phi_{2,1}}{\theta}i\text{sign}(\theta)w_1w_2\exp\left\{\frac{ir\Theta}{2b_0}\right\} \;.
\end{split}
\ee 
In particular, for an initially unpolarized photon, ${\bf N}_0={\bm\epsilon}_0$, we have 
\be
\mathbb{P}=\frac{1}{2}{\bf N}_f \cdot\left[{\bm\epsilon}_0+{\bm\epsilon}_3\tanh\nu\right]\cosh\nu e^{-4\langle\mathbb{P}^{\rm BW}\rangle} \;,
\ee
so it will tend to become polarized with ${\bf n}\propto{\bm\epsilon}_3$. Recall that $+{\bm\epsilon}_3$ and $-{\bm\epsilon}_3$ correspond, respectively, to a polarization 4-vector with $\epsilon_\LCperp=\{1,0\}$ and $\epsilon_\LCperp=\{0,1\}$, i.e. parallel and perpendicular to the field polarization. However, note that the pair production probability for a photon with ${\bf n}=\pm{\bm\epsilon}_3$ is given by $4\langle\mathbb{P}^{\rm BW}\rangle\mp\nu$, so $\nu$ cannot be larger than $4\langle\mathbb{P}^{\rm BW}\rangle$. So, if $\tanh\nu$ is not small then $e^{-4\langle\mathbb{P}^{\rm BW}\rangle}$ will be significantly smaller than $1$. In other words, the price for this induced polarization is that a significant fraction of the initial photons will decay into pairs.  

In the LCF regime we find~\eqref{LCFPfromRloop} with
\be
\hat{\bf R}_0^{\rm L}=-\frac{\text{Ai}'(\xi)}{\xi}\hat{\bf E}\cdot({\bm\epsilon}_1{\bm\sigma}_1^{(3)}+{\bm\epsilon}_3{\bm\sigma}_3^{(3)})\cdot\hat{\bf E} \;,
\ee
\be
\hat{\bf R}_{10,ij}^{\rm L}=\langle\hat{\mathbb{R}}^{\rm L}\rangle{\bf 1}+\frac{\text{Gi}'(\xi)}{\xi}\hat{\bf E}\cdot\left({\bm\sigma}_1^{(3)}\varepsilon_{1ij}+{\bm\sigma}_3^{(3)}\varepsilon_{3ij}\right)\cdot\hat{\bf E} \;,
\ee
while $\langle\hat{\mathbb{R}}^{\rm L}\rangle$ and $\hat{\bf R}_0^{\rm L}$ are simply obtained from the corresponding expressions for nonlinear Breit-Wheeler. As a curiosity, we note that the Scorer-Gi function appear in $\hat{\bf R}_{10}^{\rm L}$ both in the electron case~\eqref{LCFRLoop10Definition} and in the polarization operator. One could again have expected this from the general relation in~\eqref{ReImRelation}, because
\be
\frac{\text{Ai}'(\xi)+i\text{Gi}'(\xi)}{\xi}=\frac{1}{\pi}\int_0^\infty\ud\tau\; i\tau\exp\left\{i\xi^{3/2}\left(\tau+\frac{\tau^3}{3}\right)\right\} \;.
\ee
To compare with the literature, we consider for simplicity a linearly polarized field with $\hat{\bf E}={\bf e}_1$.
We have 
\be\label{MuellerPhotonLCF}
\begin{split}
{\bf m}=\frac{\alpha}{b_0}\int_0^1\!\ud s\bigg[&\langle\mathbb{R}^{\rm L}\rangle{\bf 1}^{(4)}-\frac{\text{Ai}'(\xi)}{\xi}({\bm\epsilon}_0{\bm\epsilon}_3+{\bm\epsilon}_3{\bm\epsilon}_0) \\
&+\frac{\text{Gi}'(\xi)}{\xi}({\bm\epsilon}_1{\bm\epsilon}_2-{\bm\epsilon}_2{\bm\epsilon}_1)\bigg] \;.
\end{split}
\ee
In the LCF regime, \eqref{MuellerPhotonLCF} is already written as the sum of three mutually commuting matrices which give
\be
\nu=-\alpha\int\frac{\ud\sigma}{b_0}\int_0^1\ud s\frac{\text{Ai}'(\xi)}{\xi}
\ee
and
\be
\varphi=\alpha\int\frac{\ud\sigma}{b_0}\int_0^1\ud s\frac{\text{Gi}'(\xi)}{\xi} \;.
\ee
This agrees with Eq.~(11) in~\cite{Bragin:2017yau} and Eq.~(81) in~\cite{Ritus1985}. 
For low $\chi$, $\langle\mathbb{P}^{\rm BW}\rangle$ and $\nu$ become exponentially suppressed, while
\be\label{LCFvarphiLow}
\varphi=-\frac{\alpha b_0}{30\pi}\int\ud\sigma a'^2(\sigma) \;.
\ee
and hence
\be
\begin{split}
\mathbb{P}=\frac{1}{2}{\bf N}_f\cdot\bigg[&{\bf 1}^{(4)}+[{\bm\epsilon}_1{\bm\epsilon}_1+{\bm\epsilon}_2{\bm\epsilon}_2](\cos\varphi-1) \\
&+[{\bm\epsilon}_1{\bm\epsilon}_2-{\bm\epsilon}_2{\bm\epsilon}_1]\sin\varphi\bigg]\cdot{\bf N}_0 \;.
\end{split}
\ee
The second term shows that the probability to flip polarization, ${\bf n}=\pm{\bm\epsilon}_1\to\mp{\bm\epsilon}_1$, is to leading order given by $\mathbb{P}_{\rm flip}\approx\varphi^2/4$. This agrees with Eq.~(41) in~\cite{Dinu:2013gaa} (for the factors of $2$, note that ${\bf n}=\pm{\bm\epsilon}_1$ corresponds to a polarization 4-vector with $\epsilon_\LCperp=(1/\sqrt{2})\{1,\pm1\}$). Note that we have obtained $\mathbb{P}_{\rm flip}$ by gluing together the Mueller matrices for $\mathbb{P}^{\rm L}=\mathcal{O}(\alpha)$ (two Mueller matrices for the leading order). Hence, $\mathbb{P}^{\rm L}$ contains the necessary information to obtain via the gluing approach the full spin-flip probability at $\mathcal{O}(\alpha^2)$, even though (a single factor of) $\mathbb{P}^{\rm L}$ vanishes for spin flip. 
The flip probability has a quadratic scaling $\mathbb{P}_{\rm flip}\sim(\alpha b_0)^2$.
However, an important point made in~\cite{Bragin:2017yau} is that there are terms that have linear scaling $\sim\varphi$. These correspond to the off-diagonal terms of the Mueller matrix.

Thus, in the low-$\chi$ limit only rotational terms remain, i.e. the degree of polarization is constant. This can be compared with the low-energy limit of the propagation of an electron through the laser, where one also finds that only rotational terms contribute to leading order. However, in that case the non-rotational terms are only suppressed by a higher power (and give the Sokolov-Ternov effect), while for the propagation of a photon the non-rotational terms are exponentially suppressed.

We can also find simple results for a circularly polarized field in the LMF regime. From~\eqref{QBWgamma} we have
\be
{\bf R}_1={\bf R}_0\propto{\bm\epsilon}_2
\ee
and from the similarity between~\eqref{R0polarop} and~\eqref{R10polarop} we can immediately see that only the $\varepsilon_{2ij}$ part of the rotational term remain, i.e.
\be
{\bf R}_{10}-\langle\mathbb{R}^{\rm L}\rangle{\bf 1}^{(3)}\propto {\bm\epsilon}_3{\bm\epsilon}_1-{\bm\epsilon}_1{\bm\epsilon}_3 \;.
\ee
So, we have essentially the same matrix structure as in the linear case, we just have to replace ${\bm\epsilon}_1\to{\bm\epsilon}_3$, ${\bm\epsilon}_3\to{\bm\epsilon}_2$ and ${\bm\epsilon}_2\to{\bm\epsilon}_1$. Thus, we find
\be
\begin{split}
\mathbb{P}
=&\frac{1}{2}{\bf N}_f \cdot e^{-4\langle\mathbb{P}^{\rm BW}\rangle}\bigg[[{\bm\epsilon}_0{\bm\epsilon}_0+{\bm\epsilon}_2{\bm\epsilon}_2]\cosh\nu \\
&+[{\bm\epsilon}_0{\bm\epsilon}_2+{\bm\epsilon}_2{\bm\epsilon}_0]\sinh\nu+[{\bm\epsilon}_1{\bm\epsilon}_1+{\bm\epsilon}_3{\bm\epsilon}_3]\cos\varphi \\
&+[{\bm\epsilon}_3{\bm\epsilon}_1-{\bm\epsilon}_1{\bm\epsilon}_3]\sin\varphi\bigg]\cdot{\bf N}_0 \;,
\end{split}
\ee 
where $\nu=2{\bm\epsilon}_2\cdot{\bf P}_0^{\rm L}$ and $\varphi=2{\bm\epsilon}_3\cdot{\bf P}^{\rm L}_{10}\cdot{\bm\epsilon}_1$.

In the low-energy limit, $b_0\gg1$, the pair-production probability becomes exponentially suppressed, and consequently $\langle\mathbb{R}^{\rm L}\rangle$ and ${\bf R}_1^{\rm L}$ too become exponentially suppressed. In contrast, the $\text{sign}(\theta)$-part of ${\bf R}^{\rm L}_{10}$ only leads to a power-law scaling and is therefore much less suppressed. We obtain the leading order by rescaling $\theta\to b_0\theta$ and performing the resulting $\theta$ and $s$ integrals. We find
\be\label{MuellerPolLow}
\begin{split}
\mathbb{P}=&\frac{1}{2}{\bf N}_{1i}\bigg(\delta_{ij} \\
&-\frac{\alpha b_0}{30\pi}\int\ud\sigma{\bf a}'(\sigma)\cdot[{\bm\sigma}_1\varepsilon_{1ij}+{\bm\sigma}_3\varepsilon_{3ij}]\cdot{\bf a}'(\sigma)\bigg){\bf N}_{0j} \\
=&\frac{1}{2}{\bf N}_{1i}\bigg(\delta_{ij}-\frac{\alpha b_0}{30\pi}\int\ud\sigma\, a'^2(\sigma){\bm\epsilon}_{\rm E,k}(\sigma)\varepsilon_{ijk}\bigg){\bf N}_{0j} \;,
\end{split}
\ee
where ${\bm\epsilon}_{\rm E}$ corresponds to polarization parallel to the local electric field, see~\eqref{photonStokesE} (and the Levi-Civita tensor has a trivial zeroth component, $\varepsilon_{ijk}{\bf N}_k=\varepsilon_{ijk}{\bf n}_k$).
Eq.~\eqref{MuellerPolLow} holds for arbitrary (e.g. elliptical) polarization of the background field. For a linearly polarized field we recover~\eqref{LCFvarphiLow}, which was obtained from the low-$\chi$ limit of the LCF approximation, while~\eqref{MuellerPolLow} holds even if $a_0$ is not large. The reason for this is that reducing $b_0$ or increasing $a_0$ both lead to dominant contribution from small $\theta$. 
We note again that we have a nonzero result already at $\mathcal{O}(\alpha)$ because we are considering a general polarization transition.
From $(\cos\mu,\sin\mu)\cdot{\bm\sigma}_1\cdot(\cos\mu,\sin\mu)=\sin(2\mu)$ and $(\cos\mu,\sin\mu)\cdot{\bm\sigma}_3\cdot(\cos\mu,\sin\mu)=\cos(2\mu)$ we see that the $\sigma$ integral in~\eqref{MuellerPolLow} tends to average to zero for a circularly rotating field, but we can at least see that for this term to be nonzero the Stokes vector of either the initial or the final probe photon needs to have a nonzero ${\bm\epsilon}_2$ component, i.e. the probe photon should have a nonzero degree of circular polarization, as pointed out in~\cite{Bragin:2017yau} (in the LCF regime).

\section{Conclusions}\label{conclusions}

We have studied the spin and polarization dependences in the $\mathcal{O}(\alpha)$ processes in a plane wave (nonlinear/nonperturbative in the field), i.e. the tree processes $e^{\LCm}\to e^{\LCm}+\gamma$ (or $e^{\LCp}\to e^{\LCp}+\gamma$) and $\gamma\to e^\LCm+e^\LCp$, and the $\mathcal{O}(\alpha)$ loop contributions to $e^{\LCm}\to e^{\LCm}$ and $\gamma\to\gamma$ (i.e. the cross-term between the zeroth- and first-order amplitude terms). We have allowed for arbitrary field polarization and arbitrary spin and polarization of the scattering particles. The dependence of the probability on the spin/polarization of any incoming or outgoing particle is expressed in terms of Stokes vectors ${\bf N}=\{1,{\bf n}\}$ and Mueller matrices ${\bf M}$. 

We have calculated all elements of these Mueller matrices. These include diagonal and off-diagonal terms that describe e.g. spin flip in any direction and spin rotation. 
There are several reasons for considering completely general spin transitions: 

1) The off-diagonal, rotational terms can be much larger than the non-rotational terms. This is the case e.g. for the spin precession of low-energy electrons, or for vacuum birefringence~\cite{Bragin:2017yau}. Thus, considering the full Mueller matrix can lead to a larger signal.   

2) Even if one does not measure the spin/polarization of the initial and final particles, one has to consider spin sums of intermediate particles in order to approximate higher orders with sequences of first-order processes. 
A spin sum on the amplitude level becomes a double spin sum on the probability level, and one cannot always find a basis where these double sums reduce to single sums. In such cases we can still use the Mueller-matrix approach.

In~\cite{Dinu:2019pau} we derived the Mueller matrices for the tree processes in the most general case. Already at $\mathcal{O}(\alpha^2)$, these general results give a huge simplification compared to an exact calculation. But there are important special cases where one can derive even simpler expressions. So, in this paper we have derived LMF and LCF approximations approximations for the Mueller matrices. LMF and LCF are of course well-used methods, so some elements of these Mueller matrices correspond to quantities that have been obtained before, but expressed in different ways, i.e. not as Mueller matrices. Thus, in addition to providing the building blocks needed for general cascades, the full Mueller matrices also complement the literature by allowing completely general spin transitions.
 
Having these approximations of the Mueller matrices is of course useful in practice. For example, here we have shown that the LMF approximation agrees very well with the exact results for trident, which is encouraging for studying higher order processes, for which an exact treatment would be impossible and a full Mueller-matrix approach potentially more time consuming than necessary.  
However, even without a numerical evaluation, these approximate Mueller matrices also show us which spin/polarization states that are important and under which conditions one could use single spin sums instead of the Mueller-matrix approach. Here we have shown that one spin basis may reduce the double sums to single sums for one part (e.g. the spin average), while a different basis may do the same for another part (e.g. the spin difference).

In this paper we have also derived the full Mueller matrices for the first-order loops $e^{\LCm}\to e^{\LCm}$ and $\gamma\to\gamma$, $\mathbb{P}^{\rm L}=(1/2){\bf N}_1\cdot{\bf M}^{\rm L}\cdot{\bf N}_0$. Since these come from the cross term between the zeroth and the first-order amplitudes, and since the zeroth order vanishes for two orthogonal spin states, one finds that $\mathbb{P}^{\rm L}$ vanishes for spin flip, ${\bf n}_1=-{\bf n}_0$ (cf.~\cite{Ilderton:2020gno}). However, ${\bf M}^{\rm L}$ is of course nonzero and in general $\mathbb{P}^{\rm L}$ is on the same order of magnitude as nonlinear Compton or Breit-Wheeler, as can be expected from unitarity. In fact, the loop contribution tends to cancel parts of nonlinear Compton, either partially or completely.  
Also, the loop contains off-diagonal/rotational terms that are not present in nonlinear Compton. And, importantly, we have shown that, although $\mathbb{P}^{\rm L}$ vanishes for spin flip, ${\bf M}^{\rm L}$ nevertheless contains all the spin/polarization information needed in order to approximate a general higher-order cascade process. For example, spin flip can be obtained from ${\bf M}^{\rm L}\cdot{\bf M}^{\rm L}$ or from higher-order products ${\bf M}^{\rm L}\cdot{\bf M}^{\rm L}\dots{\bf M}^{\rm L}$.  

For photons that travel through the laser field without pair production, we have resummed the sum of products of ${\bf M}^{\rm L}$ into a time-ordered exponential of ${\bf M}^{\rm L}$. We have found simple expressions for a general linearly polarized field, and in LCF we find agreement with the results in~\cite{Bragin:2017yau}, which were obtained with a different approach. We have also found similar results for a circularly polarized field in LMF.
For an electron traveling through the laser one in general has to consider photon emission and the loop. Due to radiation reaction, the product of Mueller matrices are not only time ordered, but also occur at a different longitudinal momentum\footnote{The transverse momenta can and have all been integrated at each step separately.}, which makes the general case challenging for an analytical approach. However, for low-energy electrons we can to leading order neglect the recoil, which allows us to resum the series in ${\bf M}^{\rm L}+{\bf M}^{\rm C}$ into a time-ordered exponential. We have found agreement with the solution to the BMT equation and with the extra terms~\cite{BaierSokolovTernov} due to the Sokolov-Ternov effect.

These time-ordered resummations are nontrivial checks of the general gluing/Mueller-matrix approach, and clearly illustrate the importance of loops; indeed the nontrivial part of the BMT equation comes only from the loop. By restricting ourselves to final-state electrons that have lost only a negligible fraction of their longitudinal momenta, we have also been able to obtain time-ordered exponentials for higher-energy electrons. 
However, for the general (and potentially most important) cases one would need to resort to a numerical treatment. It would, in particular, be interesting to use the Mueller-matrix approach to study the generation of polarized particle beams due to the interaction with the laser, and to compare with other, numerical (PIC) approaches.

\acknowledgements

G.~T. thanks Anton Ilderton for useful comments on a draft of this paper.

\appendix
\section{Bessel functions in LMF}\label{Bessel function in LMF}

In this section we will show how to rewrite the $\theta$ integrals that appear in LMF for circular polarization. All components are expressed in terms of three integrals,
\be
\mathcal{J}_0=\frac{i}{2\pi}\int\frac{\ud\theta}{\theta}\exp\left\{\frac{ir}{2b_0}\Theta\right\} \;,
\ee  
where $r=(1/s_1)-(1/s_0)$ for photon emission and $r=(1/s_2)+(1/s_3)$ for pair production,
\be
\begin{split}
\mathcal{J}_1=&\frac{i}{2\pi}\int\frac{\ud\theta}{\theta}\left(\frac{2ib_0}{r\theta}+1+{\bf w}_1\!\cdot\!{\bf w}_2\right)\exp\left\{\frac{ir}{2b_0}\Theta\right\} \\
=&-2a_0^2(u)\frac{i}{2\pi}\int\frac{\ud\theta}{\theta}\sin^2\left(\frac{\theta}{2}\right)\exp\left\{\frac{ir}{2b_0}\Theta\right\}
\end{split}
\ee
and
\be
\begin{split}
\mathcal{J}_2=&\frac{1}{2\pi}\int\frac{\ud\theta}{\theta}{\bf w}_1\!\cdot\!i{\bm\sigma}_2^{(3)}\!\cdot\!{\bf w}_2\exp\left\{\frac{ir}{2b_0}\Theta\right\} \\
=&-\frac{a_0^2(u)}{2\pi}\int\frac{\ud\theta}{\theta}\left(\sin\theta-\frac{4}{\theta}\sin^2\frac{\theta}{2}\right)\exp\left\{\frac{ir}{2b_0}\Theta\right\} \;,
\end{split}
\ee
where $\Theta$ is given by~\eqref{MLMF}. As expected from the literature (see e.g.~\cite{Ivanov:2004fi,Ivanov:2004vh,Heinzl:2020ynb}), we can perform the $\theta$ integrals in terms of Bessel functions.
In order to use some well-known formulas for Bessel functions, we first have to simplify the $\theta$ dependence of the exponent. We do this by introducing new integrals over $p_1$ and $p_2$. We rewrite $1$ in the $\theta$ integrand as one of the following components\footnote{The transverse momentum integrals which we have performed in order to arrive at the final expressions for $\mathbb{P}$ and ${\bf P}$ in~\cite{Dinu:2019pau} actually have similar Gaussian forms, but it is not necessary to go back and undo those integrals.}
%\be
%\int\ud^2p\frac{c_2}{\pi}\left\{1,2c_2p_2^2,\frac{4}{3}c_2^2p_2^4\right\}e^{-c_2(p_1^2+p_2^2)-c_1p_1-\frac{c_1^2}{4c_2}} \;,
%\ee 
\be
\{1,1\}=\int\ud^2p\frac{c_2}{\pi}\left\{1,2c_2p_2^2\right\}e^{-c_2(p_1^2+p_2^2)-c_1p_1-\frac{c_1^2}{4c_2}} \;,
\ee 
where
\be
c_1=-i\sin\frac{\theta}{2} \qquad c_2=-\frac{ir\theta}{8ra_0^2(u)} \;.
\ee
We choose the first and second component for terms in the integrand of $\mathcal{J}_i$ proportional to $1/\theta$ and $1/\theta^2$, respectively.
The point of doing this is that now the $\theta$ part of the exponent is much simpler, and we can use the Jacobi-Anger expansion~\cite{BesselDLMF}
\be
\exp\left\{ip_1\sin\frac{\theta}{2}\right\}=\sum_{m=-\infty}^\infty e^{-\frac{im\theta}{2}}J_{-m}(p_1) \;,
\ee
where $J$ is the Bessel function. All the odd order vanish because they give antisymmetric $p_1$ integrals, so we replace $m\to2n$.
Next we change to cylidrical integration variables, $p_1=p\cos\nu$ and $p_2=p=\sin\nu$. We have three different $\nu$ integrals, which can be performed using e.g. the tabulated integrals in~\cite{BesselDLMF}, giving
\be
\int_0^{2\pi}\frac{\ud\nu}{\pi} J_{2n}(p\cos\nu)=2J_n^2 
\ee
and
\be
\int_0^{2\pi}\frac{\ud\nu}{\pi}\sin^2\nu J_{2n}(p\cos\nu)=J_n^2-J_{n-1}J_{n+1} \;,
\ee
%and
%\be
%\begin{split}
%\int_0^{2\pi}&\frac{\ud\nu}{\pi}\sin^4\nu J_{2n}(p\cos\nu) \\
%&=\frac{3}{4}J_n^2-J_{n-1}J_{n+1}-\frac{1}{4}J_{n-2}J_{n+2} \;,
%\end{split}
%\ee
where the suppressed arguments on the right-hand side are $p/2$. The $\theta$ integrals are now trivial and give delta functions, which we use to perform the $p$ integral. We can simplify the result using recurrence relations~\cite{BesselDLMF} between $J_{n-1}$, $J_n$ and $J_{n+1}$.
We find
\be\label{J0Bessel}
\mathcal{J}_0=\sum_n J_n^2(z) \;,
\ee
\be\label{J1Bessel}
\mathcal{J}_1=\frac{a_0^2(u)}{2}\sum_n[J_{n+1}^2(z)+J_{n-1}^2(z)-2J_n^2(z)]
\ee
and
%\be
%\mathcal{J}_2=2\pi\sum_n\left(\frac{b_0z}{r}-\frac{na_0^2(u)}{z}\right)J_n(z)[J_{n-1}(z)-J_{n+1}(z)] \;,
%\ee
\be\label{J2Bessel}
\mathcal{J}_2=\frac{1}{2}\sum_n\left(\frac{b_0z^2}{nr}-a_0^2(u)\right)[J_{n-1}^2(z)-J_{n+1}^2(z)] \;,
\ee
where the argument of the Bessel functions is
\be
z=a_0(u)\frac{r}{b_0}\left[\frac{2b_0n}{r}-1-a_0^2(u)\right]^\frac{1}{2} \;.
\ee
This implies a minimum $n$,
\be
n\geq\frac{r}{2b_0}[1+a_0^2(u)] \;.
\ee

These three Bessel-function combinations also appear in~\cite{Ivanov:2004fi,Ivanov:2004vh}. To compare with the results in~\cite{Ivanov:2004fi,Ivanov:2004vh} for the spin/polarization structure, it is important to recall that we have integrated over all transverse momenta. So, our results can be compared with section 4.3 in~\cite{Ivanov:2004fi}, and we have checked that we have agreement for the terms there that have been written out explicitly, which correspond to our $\langle\mathbb{R}\rangle^{\rm C}$, ${\bf R}_0^{\rm C}$, ${\bf R}_1^{\rm C}$, ${\bf R}_\gamma^{\rm C}$, ${\bf R}_{\gamma0}^{\rm C}$ and ${\bf R}_{01}^{\rm C}$. So, although we have taken a rather different approach, using in particular spin and polarization bases that are common in lightfront quantization and which are especially convenient when dealing with plane-wave backgrounds, we can nevertheless compare with previous treatments of spin and polarization~\cite{Ivanov:2004fi,Ivanov:2004vh}.
Spin and polarization effects in Compton scattering and Breit-Wheeler pair production in a circularly polarized laser have also been studied in~\cite{Tsai:1992ek}.

\section{Derivation of loop}\label{Derivation of loop}

As in~\cite{Dinu:2018efz} we use a basis that is common in the lightfront quantization formalism \cite{Kogut:1969xa,Brodsky:1997de,Heinzl:2000ht,Neville:1971uc},
\begin{align}
	\gamma^0=&
	\begin{pmatrix}
		0&0&1&0 \\
		0&0&0&1 \\
		1&0&0&0 \\
		0&1&0&0
	\end{pmatrix}
	&
	\gamma^1=&
	\begin{pmatrix}
		0&0&0&1 \\
		0&0&1&0 \\
		0&-1&0&0 \\
		-1&0&0&0
	\end{pmatrix}
	\nonumber\\
	\gamma^2=&
	\begin{pmatrix}
		0&0&0&-i \\
		0&0&i&0 \\
		0&i&0&0 \\
		-i&0&0&0
	\end{pmatrix}
	&
	\gamma^3=&
	\begin{pmatrix}
		0&0&1&0 \\
		0&0&0&-1 \\
		-1&0&0&0 \\
		0&1&0&0
	\end{pmatrix} 
\; ,
\end{align}
\be
u_{\scriptscriptstyle\uparrow}=\frac{1}{\sqrt{2p_\LCm}}\begin{pmatrix}1\\0\\2p_\LCm\\-p_1-ip_2 \end{pmatrix} \qquad
u_{\scriptscriptstyle\downarrow}=\frac{1}{\sqrt{2p_\LCm}}\begin{pmatrix}p_1-ip_2\\2p_\LCm\\0\\1 \end{pmatrix} \;.
\ee
A general spinor is given by
\be
u=\cos\left(\frac{\rho}{2}\right)u_{\scriptscriptstyle\uparrow}+\sin\left(\frac{\rho}{2}\right)e^{i\lambda}u_{\scriptscriptstyle\downarrow} \;,
\ee
which corresponds to a Stokes vector as in~\eqref{Stokes3D} and to the following mode operator
\be
\bar{b}=\cos\left(\frac{\rho}{2}\right)\bar{b}_{\scriptscriptstyle\uparrow}+\sin\left(\frac{\rho}{2}\right)e^{i\lambda}\bar{b}_{\scriptscriptstyle\downarrow} \;,
\ee
where the mode operators are normalized according to $\{b_r(q),\bar{b}_{r'}(q)\}=2p_\LCm\bar{\delta}(q-q')\delta_{rr'}$ with $\bar{\delta}(...)=(2\pi)^3\delta_{\LCm,\LCperp}(...)$. 
Although we will not consider any nontrivial wave-packet effects here, it is still convenient to start with an electron in an initial state given by a wave packet
\be
|\text{in}\rangle=\int\!\ud\tilde{p}f\bar{b}|0\rangle 
\qquad
\int\ud\tilde{p}|f|^2=1 \;,
\ee
where $\ud\tilde{p}=\theta(p_\LCm)\ud p_\LCm\ud^2p_\LCperp/(2p_\LCm(2\pi)^3)$.
The amplitude for the no-emission process is given by
\be
\langle0|b^{(1)}U\bar{b}^{(0)}|0\rangle=:2p_\LCm\bar{\delta}(p'-p)M \;,
\ee
where $p_\mu$ and $p'_\mu$ are the momenta of the initial and final electron, respectively, and $U$ is the evolution operator. While the momentum is conserved, the spin can change. The probability for this is given by
\be
\mathbb{P}=\int\ud\tilde{p}'\left|\int\ud\tilde{p} f 2p_\LCm\bar{\delta}(p'-p)M\right|^2 \;.
\ee
With a sharply peaked wave packet, this simplifies to
\be
\mathbb{P}=|M|^2 \;.
\ee
At zeroth order we have
\be\label{M0}
M_0=\cos\left[\frac{\rho_1}{2}\right]\cos\left[\frac{\rho_0}{2}\right]+\sin\left[\frac{\rho_1}{2}\right]\sin\left[\frac{\rho_0}{2}\right]e^{i(\lambda_0-\lambda_1)} \;,
\ee
(note that $M_0=0$ for two orthogonal spins, e.g. for $\lambda_1=\lambda_0$ and $\rho_1=\rho_0+\pi$) and
\be
\mathbb{P}^{(0)}=\frac{1}{2}{\bf N}^{(1)}\cdot{\bf N}^{(0)} \;,
\ee
where ${\bf N}^{(i)}$ are the 4D Stokes vectors obtained by substituting $\rho_i$ and $\lambda_i$ into~\eqref{Stokes3D} and~\eqref{Stokes4D}. So, at zeroth order the Mueller matrix is simply given by the identity matrix, as expected.

The calculation of $\mathcal{O}(\alpha)$ is similar to the double nonlinear Compton case, as described in the appendix of~\cite{Dinu:2018efz}. One can use either the standard covariant approach or the lightfront-quantization approach. There are two terms in the amplitude. One comes from the instantaneous part of the lightfront Hamiltonian, and contributes to e.g. double Compton scattering. However, in this case, it only gives a background-field-independent term. Since the effect of renormalization is to subtract the field-independent part, only the non-instantaneous part of the lightfront Hamiltonian gives a nontrivial contribution\footnote{For more details about this renormalization, see~\cite{BaierRenorm}}.
Thus we find
\be
\begin{split}
M_1=&\frac{\pi\alpha}{kp}\int\ud\tilde{l}\frac{\theta(kP)}{kP}L_{\mu\nu}\int\ud\phi_2\ud\phi_1\theta(\theta_{21}) \\
\times& e^{il(x_1-x_2)}\bar{\varphi}_p(\phi_2)\varphi_P(\phi_2)\bar{\varphi}_P(\phi_1)\varphi_p(\phi_1) \\
\times&\bar{u}^{(1)}\bar{K}_p(\phi_2)\gamma^\mu K_P(\phi_2)(\slashed{P}+1)\bar{K}_P(\phi_1)\gamma^\nu K_p(\phi_1)u^{(0)} \;,
\end{split}
\ee  
where $l_\mu$ and $P_{\LCm,\LCperp}=(p-l)_{\LCm,\LCperp}$ are the momenta of the intermediate photon and electron, respectively, $L_{\mu\nu}$ is given by~\eqref{Lmunu}, and the scalar and spinor parts of the Volkov solution are given by
\be
\varphi=\exp\left\{-i\left(px+\int^{kx}\frac{2ap-a^2}{2kp}\right)\right\}  
\ee
and
\be
K=1+\frac{\slashed{k}\slashed{a}}{2kp} \qquad
\bar{K}=1-\frac{\slashed{k}\slashed{a}}{2kp} \;.
\ee
For the first-order probability we have
\be
\mathbb{P}^{\rm L}=2\text{Re}M_0\bar{M}_1 \;.
\ee
The zeroth order amplitude can be expressed as
\be
M_0=\frac{1}{2}\bar{u}^{(1)}u^{(0)} \;,
\ee
and then we can express the spin dependence in terms of the Stokes vectors right from the start by using
\be\label{uubWithN}
u_\alpha\bar{u}_\beta={\bf N}\cdot{\bm\Omega}_{\alpha\beta} \;,
\ee
where
\be
\begin{split}
{\bm\Omega}_{\alpha\beta}:=\frac{1}{2}&\{u_{\scriptscriptstyle\uparrow}\bar{u}_{\scriptscriptstyle\uparrow}+u_{\scriptscriptstyle\downarrow}\bar{u}_{\scriptscriptstyle\downarrow},u_{\scriptscriptstyle\downarrow}\bar{u}_{\scriptscriptstyle\uparrow}+u_{\scriptscriptstyle\uparrow}\bar{u}_{\scriptscriptstyle\downarrow}, \\
&i(u_{\scriptscriptstyle\downarrow}\bar{u}_{\scriptscriptstyle\uparrow}-u_{\scriptscriptstyle\uparrow}\bar{u}_{\scriptscriptstyle\downarrow}),u_{\scriptscriptstyle\uparrow}\bar{u}_{\scriptscriptstyle\uparrow}-u_{\scriptscriptstyle\downarrow}\bar{u}_{\scriptscriptstyle\downarrow}\}_{\alpha\beta} \;.
\end{split}
\ee
The spinors $u$ are ordinary spinors with 4 elements (normalized as $\bar{u}u=2$), but if we restrict to the 2D space spanned by the electron spinors then ${\bm\Omega}$ acts as the vector of the Pauli matrices $\{{\bf1},{\bm\sigma}_1,{\bm\sigma}_2,{\bm\sigma}_3\}$.
%\be
%\begin{split}
%\bar{u}_\alpha^{(1)}u_\gamma^{(1)}=N^{(0)}\cdot&\frac{1}{2}\{\bar{u}_{\scriptscriptstyle\uparrow}u_{\scriptscriptstyle\uparrow}+\bar{u}_{\scriptscriptstyle\downarrow}u_{\scriptscriptstyle\downarrow},\bar{u}_{\scriptscriptstyle\downarrow}u_{\scriptscriptstyle\uparrow}+\bar{u}_{\scriptscriptstyle\uparrow}u_{\scriptscriptstyle\downarrow}, \\
%&-i(\bar{u}_{\scriptscriptstyle\downarrow}u_{\scriptscriptstyle\uparrow}-\bar{u}_{\scriptscriptstyle\uparrow}u_{\scriptscriptstyle\downarrow}),\bar{u}_{\scriptscriptstyle\uparrow}u_{\scriptscriptstyle\uparrow}-\bar{u}_{\scriptscriptstyle\downarrow}u_{\scriptscriptstyle\downarrow}\}_{\alpha\gamma} \;.
%\end{split}
%\ee

In simplifying $\mathbb{P}^{\rm L}=2\text{Re }M_0\bar{M}_1$ we use for example
\be
2\text{Re}\int_0^\infty\ud\theta\frac{i}{\theta}\left(e^{ic\Theta}-e^{ic\theta}\right)=i\int_{-\infty}^\infty\frac{\ud\theta}{\theta}e^{ic\Theta} \;,
\ee
where in the second expression the integration contour is equivalent to $\theta\to\theta+i\epsilon$ with $\epsilon>0$. The reason for writing it like this rather than with factors of $\partial\Theta/\partial\theta$ as in~\cite{Dinu:2013gaa} is that we want to compare with the results in~\cite{Dinu:2019pau} for nonlinear Compton and Breit-Wheeler.

\section{Gluing together loops}\label{glueloop}

In~\cite{Dinu:2019pau} we showed how to glue together the probabilities of nonlinear Compton and Breit-Wheeler pair production for tree-level diagrams. The outcome is that a higher-order diagram is obtained by multiplying the first-order Mueller matrices. The obvious generalization to diagrams with loops is that the Mueller matrix describing the first-order loop contribution, i.e. $2\text{Re} M_0\bar{M}_1$, should also be multiplied in the same way. This is the case, but the proof is somewhat longer than the tree-level case. So, we will show this in this section.

For comparison, let us first recall how the Mueller-matrix multiplication emerges in tree-level diagrams. For such diagrams there are no coherent diagrams (in the sense made clear below), and an intermediate electron has a spin sum given by
\be\label{intermediateSpinSum}
\begin{split}
&|\bar{B}_\beta(\slashed{p}+1)_{\beta\alpha}A_\alpha|^2=|\bar{B}_\beta\sum_r u_\beta\bar{u}_\alpha A_\alpha|^2 \\
&=\sum_{rr'}u_\alpha\bar{u}'_\beta\bar{B}_\alpha B_\beta u'_\gamma\bar{u}_\delta \bar{A}_\gamma A_\delta \;,
\end{split}
\ee 
where the spin sums are over e.g. $r=\uparrow,\downarrow$, $A$ describes the steps that lead to this intermediate state and $B$ describes all the subsequent steps. We can for any combination of the two spins $r$ and $r'$ write
\be
u_\alpha\bar{u}'_\beta={\bf N}\cdot{\bm\Omega}_{\alpha\beta} \qquad
u'_\alpha\bar{u}_\beta={\bf N}^*\cdot{\bm\Omega}_{\alpha\beta} \;.
\ee
The double sum over $r$ and $r'$ corresponds to a single sum over 4 different Stokes vectors ${\bf N}$. If we sum over $r=\uparrow,\downarrow$ then we have ${\bf N}=\{1,0,0,\pm1\}$ and ${\bf N}=\{0,1,\pm i,0\}$. This gives
\be\label{intermediateSpinSum2}
\eqref{intermediateSpinSum}=\sum_{\bf N} {\bf N}_i{\bf N}_j^* \bar{B}{\bm\Omega}^i B \bar{A}{\bm\Omega}^j A=2(\bar{B}{\bm\Omega}B)\cdot(\bar{A}{\bm\Omega}A) \;.
\ee
This should be compared with the probability that the first steps, represented by $A$, lead to a final-state particle with a real Stokes vector ${\bf N}$, which can be expressed as
\be
|\bar{u}A|^2={\bf N}\cdot(\bar{A}{\bm\Omega}A) \;,
\ee
and the probability that an initial particle with a real ${\bf N}$ leads to the steps represented by $B$, i.e.
\be
|\bar{B}u|^2={\bf N}\cdot(\bar{B}{\bm\Omega}B) \;.
\ee
By comparing with~\eqref{intermediateSpinSum2} we see that we should: express the probability of producing the intermediate state as if it were a final state with Stokes vector ${\bf N}$ as $\mathbb{P}={\bf N}\cdot{\bf A}$, which gives ${\bf A}$; express the probability of the subsequent steps happening as if the intermediate state were an initial state with Stokes vector ${\bf N}$ as $\mathbb{P}={\bf N}\cdot{\bf B}$, which gives ${\bf B}$; the probability for the whole process is then given by $\mathbb{P}=2{\bf B}\cdot{\bf A}$. The factor of 2 can be seen as a consequence of the fact that there are two orthogonal spin state, but it should be noted that $2{\bf B}\cdot{\bf A}$ comes from a double spin sum on the probability level, which can in general not be expressed as a single spin sum. The fact that there are no other overall factors is shown in~\cite{Dinu:2019pau}.
Since this factorization happens for all intermediate particles, the total probability can be expressed as a sequence of first-order Mueller matrices. If we are only interested in a single fermion line and if we sum over the polarization of the emitted photons, then it is convenient to write the probability of nonlinear Compton as $\mathbb{P}^{\rm C}=(1/2){\bf N}_1\cdot{\bf M}^{\rm C}\cdot{\bf N}_0$, because then the factors of $1/2$ cancel against the factors of $2$ from the spin sum, and the total probability is simply given by $(1/2){\bf N}_f\cdot{\bf M}^{\rm C}\cdot{\bf M}^{\rm C}\dots{\bf M}^{\rm C}\cdot{\bf N}_0$.

Now we turn to loops. In lightfront-time ordered perturbation theory the first order amplitude is given by
\be
\begin{split}
2p_\LCm\bar{\delta}(p'-p)M_1=&\langle0|b'(-)\int\ud x^\LCp_2\ud x^\LCp_1\theta(x^\LCp_2-x^\LCp_1) \\
&\times H_1(x^\LCp_2)H_1(x^\LCp_1)\bar{b}|0\rangle \;,
\end{split}
\ee
where $H_1$ is the non-instantaneous part of the lightfront Hamiltonian. There are of course several different loops at $\mathcal{O}(\alpha^2)$, but here we focus on only the one that is expected to give the leading order for long pulses or intense fields, i.e. the one that can be thought of as $\sim M_1 M_1$. More precisely, this part is obtained by inserting the projection operator
\be
\sum_r\int\ud\tilde{P}\bar{b}(P,r)|0\rangle\langle0|b(P,r)
\ee
between $H_1(x^\LCp_4)H_1(x^\LCp_3)$ and $H_1(x^\LCp_2)H_1(x^\LCp_1)$, which gives
\be
M_2=\sum_r M_1(r_f\leftarrow r)M_1(r\leftarrow r_0) \;,
\ee
where the spin sum $r$ is over any two orthogonal spin states, $r_0$ ($r_f$) is the arbitrary initial (final) spin state, and where the product $M_1M_1$ has the following lightfront-time ordering. The initial time ordering $\theta(x^\LCp_3-x^\LCp_2)$ already gives a separation into a second step that happens at a later lightfront time than the first step, but to leading order we can replace this by $\theta(\sigma_{43}-\sigma_{21})$, where $\sigma_{ij}=(\phi_i+\phi_j)/2$, which treats $\phi_1$ and $\phi_2$ (and $\phi_3$ and $\phi_4$) symmetrically, and which allows us to perform the integrals over $\theta_{ij}=\phi_i-\phi_j$ for each step separately.  

To perform the matrix calculations it is convenient to express everything in a 2D space rather than with the 4D spinors. For this we write an arbitrary spinor as
\be\label{usuperuu}
u=c_{\scriptscriptstyle\uparrow} u_{\scriptscriptstyle\uparrow}+c_{\scriptscriptstyle\downarrow}u_{\scriptscriptstyle\downarrow} \quad\to\quad {\bf u}=\{c_{\scriptscriptstyle\uparrow},c_{\scriptscriptstyle\downarrow}\} \;.
\ee
The Stokes vector is now ${\bf N}^a={\bf u}^*\cdot{\bm\sigma}^i\cdot{\bf u}$, where $a=0,...,3$, ${\bm\sigma}^0={\bf1}$ and ${\bm\sigma}^{1,2,3}$ are the usual $2\times2$ Pauli matrices. Now we can write
\be
M_0={\bf u}_f^*\cdot{\bf u}_0 \quad M_1=:{\bf u}_f^*\cdot{\bf w}\cdot{\bf u}_0
\quad M_2={\bf u}_f^*\cdot\frac{T}{2}{\bf w}\cdot{\bf w}\cdot{\bf u}_0 \;.
\ee
The higher-order terms can be expressed in a similar fashion, so we can resum them into a time-ordered exponential
\be
M=\sum_{n=0}^\infty={\bf u}_f^* Te^{\bf w}{\bf u}_0 \;.
\ee
Using
\be
{\bf u}_i{\bf u}_j^*=\frac{1}{2}{\bf N}^a{\bm\sigma}_{ij}^a \;,
\ee
where $i,j=1,2$ and with a sum over $a=0,...,3$, we can write the probability as
\be
\mathbb{P}=|M|^2=\frac{1}{2}{\bf N}_f\cdot{\bf M}\cdot{\bf N}_0 \;,
\ee
where the Mueller matrix is given by
\be\label{MuellerTrace}
{\bf M}_{ba}=\frac{1}{2}\text{tr}\left[{\bm\sigma}^bTe^{\bf w}{\bm\sigma}^a \bar{T}e^{\bar{\bf w}}\right] \;,
\ee
where $\bar{T}$ means anti-time-ordering. In order to simplify this we restrict the lightfront-time $\sigma$ integrals from $\int^\infty$ to $\int^\sigma$ and then we take the derivative with respect to $\sigma$,
\be\label{derivativeOfM}
\frac{\ud{\bf M}_{ba}}{\ud\sigma}=\frac{1}{2}\text{tr}\left[(\bar{\bf w}'{\bm\sigma}^b+{\bm\sigma}^b{\bf w}')Te^{\bf w}{\bm\sigma}^a \bar{T}e^{\bar{\bf w}}\right] \;.
\ee
The idea is that this derivative should be given by the first-order Mueller matrix, which is obtained by expanding~\eqref{MuellerTrace} to first order in ${\bf w}\propto\alpha$,
\be
{\bf M}_{ba}^1=\frac{1}{2}\text{tr}\left[(\bar{\bf w}{\bm\sigma}^b+{\bm\sigma}^b{\bf w}){\bm\sigma}^a\right]  \;.
\ee
Since any $2\times2$ matrix can be written as a sum of the four Pauli matrices, with coefficients obtained using $\text{tr}{\bm\sigma}^a{\bm\sigma}^b=2\delta_{ab}$, we can write 
\be
(\bar{\bf w}'{\bm\sigma}^b+{\bm\sigma}^b{\bm w}')_{ij}=\frac{\ud{\bf M}_{bc}^1}{\ud\sigma}{\bm\sigma}^c_{ij} \;,
\ee
and substituting this into~\eqref{derivativeOfM} gives the desired result
\be
\frac{\ud{\bf M}}{\ud\sigma}=\frac{\ud{\bf M}^1}{\ud\sigma}\cdot{\bf M} \;.
\ee
Thus, the total Mueller matrix is given by the time-ordered exponential of the first-order Mueller matrix,
\be
{\bf M}=T e^{{\bf M}_1} \;.
\ee

So far in this section we have considered the loop correction to the electron line. However, the corresponding calculations for the series of polarization loops for the photon line are basically the same. For example, instead of~\eqref{uubWithN} we have
\be
\epsilon_\mu\bar{\epsilon}'_\nu={\bf N}\cdot{\bm\Omega}_{\mu\nu} \;,
\ee
where 
\be
\begin{split}
{\bm\Omega}_{\mu\nu}=&\frac{1}{2}\Big\{\epsilon_\mu^{(1)}\bar{\epsilon}_\nu^{(1)}+\epsilon_\mu^{(2)}\bar{\epsilon}_\nu^{(2)},\epsilon_\mu^{(2)}\bar{\epsilon}_\nu^{(1)}+\epsilon_\mu^{(1)}\bar{\epsilon}_\nu^{(2)}, \\
&i(\epsilon_\mu^{(2)}\bar{\epsilon}_\nu^{(1)}-\epsilon_\mu^{(1)}\bar{\epsilon}_\nu^{(2)}),\epsilon_\mu^{(1)}\bar{\epsilon}_\nu^{(1)}-\epsilon_\mu^{(2)}\bar{\epsilon}_\nu^{(2)}\Big\}
\end{split} \;,
\ee
where $\epsilon_\mu^{(1)}$ and $\epsilon_\mu^{(2)}$ are the two (lightfront gauge) polarization vectors with $\epsilon_\LCperp^{(1)}=\{1,0\}$ and $\epsilon_\LCperp^{(2)}=\{0,1\}$. For $\epsilon'=\epsilon$ given by~\eqref{epsilonDefinition} we have ${\bf N}=\{1,{\bf n}\}$ with ${\bf n}$ as in~\eqref{Stokes3D}. For an intermediate photon we can replace double polarization sums $\sum_{\epsilon,\epsilon'}$ with a single sum over 4 Stokes vectors, e.g. ${\bf N}=\{1,0,0,\pm1\}$ and ${\bf N}=\{0,1,\pm i,0\}$, just as in the fermion case. And instead of~\eqref{usuperuu} we have $\epsilon_\mu=c_1\epsilon_\mu^{(1)}+c_2\epsilon_\mu^{(2)}$. The rest of the calculations is the same, and therefore the conclusion is also the same, i.e. one should express the polarization dependence of the loop at $\mathcal{O}(\alpha)$, $2\text{Re }M_0\bar{M}_1$, in terms of a Mueller matrix ${\bf M}$ and then higher orders can be approximated by a time-ordered product of a sequence of Mueller matrices.

Since the Mueller matrix for the loop is constructed from the $\mathcal{O}(\alpha)$ probability $\mathbb{P}^{\rm L}$, and since $\mathbb{P}^{\rm L}=0$ for spin flip, it might not be obvious how the Mueller-matrix approach can describe spin flip. To explain this we consider $\mathcal{O}(\alpha^2)$. For a general spin transition there are two contributions, which we can express as 
\be
|M_1|^2=\frac{1}{2}{\bf N}_f^b\frac{1}{2}\text{tr}[{\bm\sigma}^b{\bf w}{\bm\sigma}^a\bar{\bf w}]{\bf N}_0^a
\ee 
and
\be
2\text{Re}M_0\bar{M}_2=\frac{1}{2}{\bf N}_f^b\frac{1}{2}\text{tr}\bigg[{\bm\sigma}^b{\bm\sigma}^a\frac{\bar{T}}{2}\bar{\bf w}\bar{\bf w}+{\bm\sigma}^b\frac{T}{2}{\bf w}{\bf w}{\bm\sigma}^a\bigg]{\bf N}_0^a \;.
\ee
The sum of these two gives the $\mathcal{O}(\alpha^2)$ part of the Mueller-matrix resummation~\eqref{MuellerTrace}. For spin flip, ${\bf N}_0=\{1,{\bf n}\}$ and ${\bf N}_f=\{1,-{\bf n}\}$, we have $M_0=0$, so $2\text{Re}M_0\bar{M}_2=0$. Thus,
\be
\mathbb{P}_{\rm flip}^{{\rm L}(2)}=|M_1|^2=\frac{1}{2}{\bf N}_f\cdot\frac{T}{2}{\bf M}^{\rm L}\cdot{\bf M}^{\rm L}\cdot{\bf N}_0 \;,
\ee
so the Mueller-matrix approach can handle spin flip even though the one-loop contribution $\mathbb{P}_{\rm flip}^{{\rm L}(1)}=0$. In fact, while the higher-order amplitudes have been approximated as $M_n\sim M_1\dots M_1$, $M_1$ is exact, so the Mueller-matrix approach actually gives the exact spin-flip probability at $\mathcal{O}(\alpha^2)$. Note that, while ${\bf M}^{\rm L}$ contains all the information needed to describe spin flip, the converse is not true; knowing $|M_1|^2$ is not enough to find the full Mueller matrix.

\subsection{The final momentum integral}\label{The final momentum integral}

If no parameter is large or small we can in general not approximate the $\phi$ integrals as in e.g. LCF or LMF. However,
just like in the nonlinear-Compton case~\cite{Dinu:2013hsd}, we can perform the last remaining momentum integral in terms of sine and cosine integrals for arbitrary pulse shape. In fact, in a cascade we would not be able to integrate the probability of Compton scattering over the longitudinal momentum before gluing together the steps, but each loop has an independent longitudinal momentum integral which mean that we can perform all the momentum integrals in the loop before gluing together. 
We find
\be
\begin{split}
\{\langle\mathbb{P}^{\rm L}\rangle,{\bf P}_{0,1}^{\rm L},{\bf P}_{10}^{\rm L}\}=\frac{\alpha}{2\pi b_0}\int\ud\sigma&\int_0^\infty\frac{\ud\theta}{\theta}\\
\times&\{\langle\hat{\mathbb{R}}^{\rm L}\rangle,\hat{{\bf R}}_{0,1}^{\rm L},\hat{{\bf R}}_{10}^{\rm L}\} \;,
\end{split}
\ee
where
\be
\begin{split}
\langle\hat{\mathbb{R}}^{\rm L}\rangle=&-\frac{1}{4}({\bf a}(\phi_2)-{\bf a}(\phi_1))^2\left(\frac{\varphi}{2}+\left[1-\frac{\varphi^2}{2}\right]\mathcal{S}(\varphi)\right) \\
&-\left(1-\frac{\theta}{\Theta}\frac{\ud\Theta}{\ud\theta}\right)\varphi\mathcal{C}(\varphi) \;,
\end{split}
\ee
\be
\begin{split}
\hat{\bf R}_1^{\rm L}=&-i\bigg\{{\bf 1}\left[\frac{1}{2}+\frac{\varphi^2}{2}\mathcal{C}-\varphi\mathcal{S}\right] \\
&+\hat{\bf k}\,{\bf X}\left[-\frac{1}{2}+\left(1+\frac{\varphi^2}{2}\right)\mathcal{C}\right]\bigg\}\cdot{\bf V} \;,
\end{split}
\ee
\be
\begin{split}
\hat{\bf R}_{10}^{\rm L}=&\langle\hat{\mathbb{R}}^{(1)}\rangle{\bf 1}+\frac{1}{2}[{\bf Y}\hat{\bf k}-\hat{\bf k}{\bf Y}]\left[\varphi\mathcal{C}-\frac{\varphi}{2}(1-\varphi\mathcal{S})\right]\\
&-({\bf X}\cdot{\bf V}){\bm\sigma}_2\left[\left(1+\frac{\varphi^2}{2}\right)\mathcal{S}-\frac{\varphi}{2}\right] \;,
\end{split}
\ee
and for Compton scattering we have
\be
\begin{split}
\hat{\bf R}_1^{\rm C}=&i\bigg\{{\bf 1}[-1+\mathcal{C}+\varphi\mathcal{S}] \\
&+\hat{\bf k}\,{\bf X}\left[-\frac{1}{2}+\left(1+\frac{\varphi^2}{2}\right)\mathcal{C}\right]\bigg\}\cdot{\bf V} \;,
\end{split}
\ee
\be
\begin{split}
\hat{\bf R}_{10}^{\rm C}&=-{\bf X}\hat{\bf k}[\mathcal{S}-\varphi\mathcal{C}]+\hat{\bf k}{\bf X}\left[\varphi\mathcal{C}-\frac{\varphi}{2}(1-\varphi\mathcal{S})\right] \\
&+\frac{1}{2}\hat{\bf k}\hat{\bf k}\left(1-\frac{\theta}{\Theta}\frac{\ud\Theta}{\ud\theta}\right)\left[\left(1-\frac{\varphi^2}{2}\right)\mathcal{S}-2\varphi\mathcal{C}+\frac{\varphi}{2}\right] \\
&+\left[1-\frac{\theta}{\Theta}\frac{\ud\Theta}{\ud\theta}+\frac{1}{2}({\bf a}(\phi_2)-{\bf a}(\phi_1))^2\right] \\
&\times\left\{{\bf 1}_\LCperp\varphi\mathcal{C}+\frac{1}{2}\hat{\bf k}\hat{\bf k}\left[\frac{\varphi}{2}+\left(1-\frac{\varphi^2}{2}\right)\mathcal{S}\right]\right\} \;,
\end{split}
\ee
(the factor of $i$ cancels against the factor of $i$ in $V$) where
\be
\mathcal{S}(\varphi)=\sin(\varphi)\text{Ci}(\varphi)-\cos(\varphi)\text{si}(\varphi)
\ee
and
\be
\mathcal{C}(\varphi)=-\cos(\varphi)\text{Ci}(\varphi)-\sin(\varphi)\text{si}(\varphi) \;,
\ee
where $\text{Ci}$ and $\text{si}(\varphi)=\text{Si}(\varphi)-\frac{\pi}{2}$ are cosine and sine integrals (see~\cite{DLMF}) with argument $\varphi=\Theta/(2b_0)$.

\section{Series expansions from the Mellin transform}\label{Mellin-transform-section}

A simple way to obtain the $\chi\ll1$ and $\chi\gg1$ expansions in LCF is to first calculate the Mellin transform~\cite{RitusMellin,Lobanov1980} with respect to $\chi$, defined by
\be
\tilde{J}(S)=\int_0^\infty\ud\chi\;\chi^{S-1}J(\chi) \;.
\ee
It turns out to be convenient to rescale the variable of the transform $S\to2t$. We first change variables in $J(\chi)$ from $s_1$ to $\xi=(r/\chi)^{2/3}$, $r=(1/s_1)-1$. Then we change order of integration, and first integrate over $\chi$. This leads in general to a simpler $\xi$ integral, which can also be performed explicitly.  
For these two integrals over $\chi$ and $\xi$ to be convergent, one finds a condition on $t$ on the form $t_1<\text{Re }t<t_2$, where $t_1$ and $t_2$ are two constants. For example, for $J_1^{\rm C}$ we find $-1<\text{Re }t<-1/6$.
The inverse is given by
\be
J(\chi)=\int_\gamma\frac{\ud S}{2\pi i}\chi^{-S}\tilde{J}(S) \;,
\ee
where the integration path $\gamma$ starts at $t_0-i\infty$, ends at $t_0+i\infty$ and goes through the real axis in the interval $t_1<\text{Re }t<t_2$. 
For all terms we find that $\tilde{J}$ can be expressed explicitly in term of $\Gamma$ functions and $\text{csc}(2\pi t)$ (which could also be written in terms of two $\Gamma$ functions). For example, for $J_1^C$ we have
\be
\tilde{J}_1^{\rm C}(S=2t)=-\frac{(1+2t)\text{csc}(2\pi t)}{4\times3^{2t-1/2}\chi^{2t}}\Gamma\left[-\frac{1}{6}-t\right]\Gamma\left[\frac{1}{6}-t\right] \;.
\ee
This means that it is simple to find the poles and the corresponding residues. All poles lie on the real axis, and we can deform the integration contour such that it encloses either $t<t_1$ counterclockwise, or $t>t_2$ clockwise; the small- and large-$\chi$ expansions are obtained from the first and second choice, respectively. In this way it is straightforward to obtain any number of terms in these expansions.

\section{Solution in LCF + $\chi\ll1$ regime}\label{LCFlowSol}

In this section we will calculate~\eqref{LCFresumLowChi} directly without first turning it into a differential equation. Of course, in general we would also not be able to find an exact resummation (exact at the level of the gluing approach, that is), but we would have sums of sequences of Mueller matrices, so this calculation could still give some relevant insights. Let us first separate the total Mueller matrix into four parts. In this 4D space we have  
\be
(\hat{\bf B}\hat{\bf e}_0)^2=\hat{\bf B}\hat{\bf e}_0\cdot{\bf 1}_\LCperp=\hat{\bf B}\hat{\bf e}_0\cdot{\bf 1}_\LCpara=\hat{\bf B}\hat{\bf e}_0\cdot(\hat{\bf E}\,\hat{\bf k}-\hat{\bf k}\,\hat{\bf E})={\bf 0} \;,
\ee
which means the $\hat{\bf B}\hat{\bf e}_0$ part of ${\bf m}$ can only appear in the first step. In the 3D formulation, this means that a term with ${\bf P}_1$ can only appear in the first step. Contrast this with the general case where one can have e.g. terms with (omitting all the arguments) $({\bf P}_0\cdot{\bf P}_{10}\dots{\bf P}_{10}\cdot{\bf P}_1)({\bf P}_0\cdot{\bf P}_{10}\dots{\bf P}_{10}\cdot{\bf P}_1)$, where a matrix multiplication can start at one step (with ${\bf P}_1$) and then end (with ${\bf P}_0$) at a an intermediate step, and then a new sequence of matrix products can start at a later step (with a second factor of ${\bf P}_1$). However, this is not possible here since, after integrating over all the momenta (which we can do independently at each step since we are in the low-$\chi$ regime where we can neglect radiation reaction), there is no ${\bf P}_0$ (and no $\langle\mathbb{P}\rangle$) term in the sum of the loop and Compton scattering. So, after a matrix product has started with a factor of ${\bf P}_1$ or the initial Stokes vector ${\bf n}_0$ it cannot end at any intermediate step, and since we have the same number of indices at each step (in contrast to a general cascade where the number of spin/polarization vectors increases with the production of particles) we find that ${\bf P}_1$ can only appear in the first step.  
So, we have two different contributions: one with a factor of $\hat{\bf B}\hat{\bf e}_0$ in the first step and the other with no factor of $\hat{\bf B}\hat{\bf e}_0$. 

For the first contribution we have $\hat{\bf B}\hat{\bf e}_0\cdot{\bf N}_0\propto\hat{\bf B}$, which means that the rotation part, $\hat{\bf E}\,\hat{\bf k}-\hat{\bf k}\,\hat{\bf E}$, drops out and we are left with a trivial matrix multiplication,  
\be
\begin{split}
{\bf N}_{\rm u}&=\int^\sigma\!\ud\sigma_1\,T_\sigma\exp\left\{\int_{\sigma_1}^\sigma\ud\sigma'{\bf m}(\sigma')\right\}\cdot\frac{\alpha\chi^3(\sigma_1)}{b_0}\hat{\bf B}\hat{\bf e}_0\cdot{\bf N}_0 \\
&=\frac{8}{5\sqrt{3}}\hat{\bf B}\int^\sigma\frac{\ud\sigma_1}{T(\sigma_1)}\exp\left\{-\int_{\sigma_1}^\sigma\frac{\ud\sigma'}{T(\sigma')}\right\} \\
&=\frac{8}{5\sqrt{3}}\hat{\bf B}\left(1-\exp\left\{-\int_{-\infty}^\infty\!\frac{\ud\sigma'}{T(\sigma')}\right\}\right) \;,
\end{split}
\ee
where in the last line we have taken the limit $\sigma\to\infty$ for an electron that has left the pulse.
This part does not depend on the initial Stokes vector ${\bf n}_0$, and it is the only nontrivial contribution for an unpolarized initial particle, ${\bf n}_0={\bf 0}$. 

For the second contribution we first note that the $\hat{\bf B}\hat{\bf B}$ part commutes with the rest of the Mueller matrix. For the rest of the Mueller matrix we write
\be
\begin{split}
&\alpha\frac{\chi}{b_0}\bigg[-\frac{5\sqrt{3}}{8}\chi^2\hat{\bf E}\hat{\bf E}-\frac{35}{24\sqrt{3}}\chi^2\hat{\bf k}\hat{\bf k} 
+\frac{1}{2\pi}(\hat{\bf E}\,\hat{\bf k}-\hat{\bf k}\,\hat{\bf E})\bigg] \\
&=c(\hat{\bf E}\hat{\bf E}+\hat{\bf k}\hat{\bf k})+{\bf m}_r \;,
\end{split}
\ee  
and then we choose the constant $c$ such that ${\bf m}_r^2\propto \hat{\bf E}\hat{\bf E}+\hat{\bf k}\hat{\bf k}$, which gives $c=-8/(9T)$ and ${\bf m}_r=\Omega(-\delta \hat{\bf E}\hat{\bf E}+\delta\hat{\bf k}\hat{\bf k}+\hat{\bf E}\hat{\bf k}-\hat{\bf k}\hat{\bf E})$, where $\delta=1/(9T\Omega)\propto\chi^2\ll1$. Hence, we have now separated the Mueller matrix (minus the $\hat{\bf B}\hat{\bf e}_0$ part) into mutually commuting matrices, and, since we are assuming a linearly polarized field, $\hat{\bf B}\hat{\bf B}$, $\hat{\bf E}\hat{\bf E}$ and $\hat{\bf k}\hat{\bf k}$ also commute at different lightfront times, so the time ordering for these parts becomes unnecessary. So, if ${\bf N}_0=\{1,c\hat{\bf B}\}$ with some constant $-1<c<1$, then 
\be
\begin{split}
{\bf N}_{\rm p}=&\exp\left\{-\int\frac{\ud\sigma}{T}\hat{\bf B}\hat{\bf B}\right\}\cdot{\bf N}_0 \\
=&\left[{\bf 1}+\hat{\bf B}\hat{\bf B}\left[\exp\left\{-\int\frac{\ud\sigma}{T}\right\}-1\right]\right]\cdot{\bf N}_0 \;.
\end{split}
\ee 
However, if ${\bf N}_0$ also has components along $\hat{\bf E}$ or $\hat{\bf k}$, then we also need ${\bf m}_r$, and, at the moment, ${\bf m}_r$ does not commute with itself at different ligthfront times. So, let us for simplicity consider a constant field. Then, from ${\bf m}_r^2=-\Omega(1-\delta^2)(\hat{\bf E}\hat{\bf E}+\hat{\bf k}\hat{\bf k})$ we see that the corresponding exponential separates into 
\be
\begin{split}
{\bf N}_{\rm p}=&\bigg\{\cos\left[\Delta\sigma\Omega\sqrt{1-\delta^2}(\hat{\bf E}\hat{\bf E}+\hat{\bf k}\hat{\bf k})\right] \\
&+(-\delta \hat{\bf E}\hat{\bf E}+\delta\hat{\bf k}\hat{\bf k}+\hat{\bf E}\hat{\bf k}-\hat{\bf k}\hat{\bf E})\frac{\sin\left[\Delta\sigma\Omega\sqrt{1-\delta^2}\right]}{\sqrt{1-\delta^2}}\bigg\} \\
&\times\exp\left\{-\frac{\Delta\sigma}{T}\left[\hat{\bf B}\hat{\bf B}+\frac{8}{9}(\hat{\bf E}\hat{\bf E}+\hat{\bf k}\hat{\bf k})\right]\right\}{\bf N}_0 \\
\approx&\left\{\cos\left[\Delta\sigma\Omega(\hat{\bf E}\hat{\bf E}+\hat{\bf k}\hat{\bf k})\right] +(\hat{\bf E}\hat{\bf k}-\hat{\bf k}\hat{\bf E})\sin\left[\Delta\sigma\Omega\right]\right\} \\
&\times\exp\left\{-\frac{\Delta\sigma}{T}\left[\hat{\bf B}\hat{\bf B}+\frac{8}{9}(\hat{\bf E}\hat{\bf E}+\hat{\bf k}\hat{\bf k})\right]\right\}{\bf N}_0 \;,
\end{split}
\ee
where we have used $\delta\ll1$ (we have already neglected such small terms). This part depends on the initial ${\bf n}_0$, but for sufficiently long pulses we have ${\bf N}_{\rm p}\to\{1,{\bf 0}\}$. If the initial particle is unpolarized, then we have ${\bf N}_{\rm p}=\{1,{\bf 0}\}$ (even for a inhomogeneous field).
These results for ${\bf N}_{\rm u}+{\bf N}_{\rm p}$ agree of course with~\cite{BaierSokolovTernov}.

\end{document}